\documentclass[a4paper,12pt,notitlepage]{article}
\usepackage{amsmath}
\usepackage{graphicx,lscape}
\usepackage{rotating}
\usepackage[english]{babel}
\usepackage[latin1]{inputenc}
\usepackage{latexsym}
\usepackage{amssymb}
\usepackage{epstopdf}
\usepackage{tikz}
\usepackage{amsfonts}
\usepackage{hyperref}
\usetikzlibrary{topaths,calc}

\newtheorem{algorithm}{Algorithm}

\newtheorem{assumption}{Assumption}

\evensidemargin\oddsidemargin
\input{tcilatex}

\tikzstyle{vertex} = [fill,shape=circle,node distance=80pt]
\tikzstyle{edge} = [fill,opacity=.5,fill opacity=.5,line cap=round, line join=round, line width=50pt]
\tikzstyle{elabel} =  [fill,shape=circle,node distance=30pt]

\pgfdeclarelayer{background}
\pgfsetlayers{background,main}

\begin{document}

\title{Teams:\\{\Large Heterogeneity, Sorting, and Complementarity}\footnote{Presented at the World Congress of the Econometric Society, invited session ``Frontiers in Modern Econometrics'', August 17 2020. I thank Manuel Arellano, Eric Auerbach, Philipp Kircher, Thibaut Lamadon, Jack Light, Elena Manresa, Charles Manski, Eleonora Patacchini, Alessandra Voena, Martin Weidner, and audiences at various venues for comments. I acknowledge support from NSF grant number SES-1658920.}}
\author{Stéphane Bonhomme \\
University of Chicago}%\\$\quad$\\$\quad$\\\\$\quad$\\PRELIMINARY DRAFT}
\maketitle

\begin{abstract}

How much do individuals contribute to team output? I propose an econometric framework to quantify individual contributions when only the output of their teams is observed. The identification strategy relies on following individuals who work in different teams over time. I consider two production technologies. For a production function that is additive in worker inputs, I propose a regression estimator and show how to obtain unbiased estimates of variance components that measure the contributions of heterogeneity and sorting. To estimate nonlinear models with complementarity, I propose a mixture approach under the assumption that individual types are discrete, and rely on a mean-field variational approximation for estimation. To illustrate the methods, I estimate the impact of economists on their research output, and the contributions of inventors to the quality of their patents.

\bigskip


\end{abstract}

\baselineskip21pt

\bigskip

\bigskip

\setcounter{page}{0}\thispagestyle{empty}

\newpage

\section{Introduction}

How to infer individual contributions when only team output is observed? This question is central to a number of areas, ranging from labor economics and the economics of innovation to sports and music. My goal in this chapter is twofold: I propose an econometric framework to estimate how team performance depends on its workers, and I apply it to study the production of economic research and the impact of individual inventors on the quality of their patents.  

I focus on three aspects of the relationship between workers and teams. The first one is worker \emph{heterogeneity}, which shapes team performance. The approach I propose relies on a simple intuition: when individuals work for different teams over time, the variation in team composition will be informative about which workers contribute the most to their teams and which ones contribute the least. Hence, the network structure of the data will be crucial in order to infer worker heterogeneity. 

The second aspect is the \emph{sorting} of workers into teams. Both heterogeneity and sorting contribute to the variation in output. I show in a model where production in additive in worker inputs that the two can be separately identified. The approach I rely on generalizes the ``plus-minus'' metric in US team sports (see Hvattum, 2019): intuitively, when a worker $i'$ joins a team and a worker $i$ leaves it, the difference in output identifies the difference between the types of $i'$ and $i$. More generally, I show how, in collaboration networks, worker types can be identified by using their participation in different teams, including when they work on their own.

The third aspect is the \emph{complementarity} between workers. The presence of complementarity is important for policies that influence the reallocation of workers across teams. Complementarities are also central to theories of sorting (e.g., Becker, 1973, Garicano and Rossi-Hansberg, 2006, Eeckhout and Kircher, 2018). To identify complementarities empirically, I build and estimate nonlinear models of team production, which rely on the assumption that worker types are discrete. This framework allows me to jointly document the role of heterogeneity, sorting and complementarity.

In the analysis I will draw connections with methods for matched employer-employee data (e.g., Abowd and Kramarz, 1999). In that literature, both firm types and worker types contribute to the wage outcome, which is observed at the worker level, and workers do not interact with one another inside the firm.\footnote{With some exceptions, such as the recent work by Herkenhoff \textit{et al.} (2018).} The worker-firm structure is thus a bipartite graph (Bonhomme, 2017). In contrast, in the collaboration networks I study here, workers interact with one another in the team. In addition, while I assume away team effects, so the team in my framework is simply a collection of workers, the outcome is only observed at the team level. Collaboration networks follow a hypergraph structure where teams vary in size (e.g., Turnbull \textit{et al.}, 2019, Lerner \textit{et al.}, 2019). Despite these differences, my framework will leverage insights from the matched employer-employee data literature. As an example, when production is additive in worker inputs, I will estimate worker types using linear regression, in the spirit of Abowd \textit{et al.} (1999).

There is a fourth important aspect in the relationship between workers and teams: the presence of \emph{other factors}, beyond worker types, that affect output. These other factors are lumped into the error term of the model. Their presence prevents one from recovering all worker types precisely, especially in data sets where the number of productions per worker is small and collaborations are sparse. As a result, regression-based estimates of sorting and heterogeneity may be biased. To correct the bias in additive models, I implement the method of Andrews \textit{et al.} (2008) developed for matched employer-employee data.\footnote{Kline \textit{et al.} (2020) generalize this approach to allow for unspecified heteroskedasticity. The nature of the bias is formally studied in Jochmans and Weidner (2020).} Bonhomme \textit{et al.} (2020) find using samples from various countries that the bias can be substantial. Similarly, when estimating additive models in networks of collaborations, I find that bias correction is important.

A main goal of this chapter is to propose methods that account for complementarity between workers. Bonhomme \textit{et al.} (2019, BLM) introduce a framework to allow for complementarity between workers and firms in matched employer-employee data. Here I build on this work to allow for complementarity between workers within a team. Like in BLM, I assume that heterogeneity takes a finite number of values. However, while BLM model \emph{firm} heterogeneity as discrete and estimate firm groups using the kmeans algorithm, in my setting estimates of \emph{worker} types are not sufficiently precise to follow a similar grouped fixed-effects approach. For this reason, I model the distribution of the discrete worker types using a random-effects approach.

The presence of unobserved worker heterogeneity in a network of teams makes estimation of nonlinear random-effects models challenging. Indeed, the likelihood function is non-separable, since the same worker may participate in multiple teams, and teams may contain multiple workers. To lower the computational cost, I use a mean-field variational method, which relies on a fully factored approximation. Variational estimators are widely used in networks and other complex data settings (Bishop, 2006, Blei \textit{et al.}, 2017). In stochastic blockmodels, which are closely related to the model I study, Bickel \textit{et al.} (2013) provide conditions for consistency. In my model I perform Monte Carlo simulations to probe the accuracy of the variational method in finite samples.\footnote{Codes to implement the methods are available on my \href{https://sites.google.com/site/stephanebonhommeresearch/}{\color{blue}{webpage}}.}

The framework can be extended in different ways. In particular, I consider adding a model of team formation on top of the model of team production. To do so, I implement a joint random-effects approach using a stochastic blockmodel as a simple statistical model of team formation. An interesting avenue for future work will be to combine the framework with economic models of team formation. It will also be important to understand under which conditions the methods I propose are robust to the team formation model being misspecified. 

I apply these methods to two empirical settings: economic research, and patents and innovation. Collaboration in scientific research is an extensively studied topic (e.g., Furman and Gaule, 2013). Aspects of the network of collaborations between economists have been studied by Goyal \textit{et al.} (2006), Fafchamps \textit{et al.} (2010), Hsieh \textit{et al.} (2018), and Anderson and Richards-Shubik (2019), among others. Here I focus on the impact of individual researchers and the shape of the production function. In the literature on patents and innovation, there is increasing interest in the role of individual inventors in innovation as an engine of endogenous growth (e.g., Akcigit \textit{et al.}, 2017, Bell \textit{et al.}, 2019, Pearce, 2019). Relative to this literature, I propose different measures of the type of an inventor -- that is, a measure of her ``quality'' -- which I infer from the network of patent collaborations using econometric techniques. 

I estimate the models using a sample of economists constructed by Ductor \textit{et al.} (2014), and a sample of inventors and their patents constructed by Akcigit \textit{et al.} (2016). In both samples, I find that an additive model of team production is useful to highlight the presence of heterogeneity and sorting. However, the additive model is only an imperfect tool to analyze these data. Indeed, for both economists and inventors, I find evidence of complementarity between high-type workers who collaborate in a team. These findings suggest that future research will need to account for complementarity and sorting when measuring individual contributions of researchers or inventors to their teams.

Related approaches have been proposed in the literature. Relative to the analysis of peer effects and spillovers (in particular, Arcidiacono \textit{et al.}, 2012, and Arcidiacono \textit{et al.}, 2017), a key difference in my context is that I only observe team output. In recent work, Devereux (2018) uses fixed-effects methods to estimate the link between individual productivity and team performance in tennis games. Most closely related to this chapter, Ahmadpoor and Jones (2019) use fixed-effects methods in nonlinear models to document heterogeneity and complementarity patterns with a focus on research, innovation and patenting.\footnote{Related methods have been used to estimate manager fixed-effects (e.g., Bertrand and Schoar, 2003).} While I build on this work, a specificity of my approach is that it explicitly accounts for the presence of other unobserved inputs in addition to the worker types, which motivates the use of new econometric methods. Moreover, estimates of nonlinear models with discrete types uncover novel forms of complementarity between workers. Another recent related article is Weidmann and Deming (2020), who design and analyze experiments to identify individual contributions to teamwork.

The outline of the chapter is as follows. I describe the framework in Section \ref{Model_sec}, and study additive and nonlinear models in Sections \ref{Additive_sec} and \ref{Complements_sec}, respectively. I apply the methods to study economic researchers in Section \ref{Research_sec}, and inventors and their patents in Section \ref{Inventors_sec}. Finally, I conclude in Section \ref{Conclu_sec}.

\section{Framework\label{Model_sec}}

Consider a set of \emph{nodes} $1,...,N$, which represent workers who collaborate and produce output in teams. Nodes are linked to each other by \emph{hyperedges} that represent the collaborations between workers. In a standard graph, an edge links two nodes $i$ and $i'$. In a hypergraph of collaborations, hyperedges can link a worker $i$ to herself when $i$ produces on her own, two workers $i$ and $i'$ who produce together, three workers $i,i',i''$ who produce together, and so on. I will refer to hyperedges as ``teams'', and denote the workers in a team $j$ of size $n_j=n$ as $(i_1(j),...,i_n(j))$. I denote the number of teams as $J$.

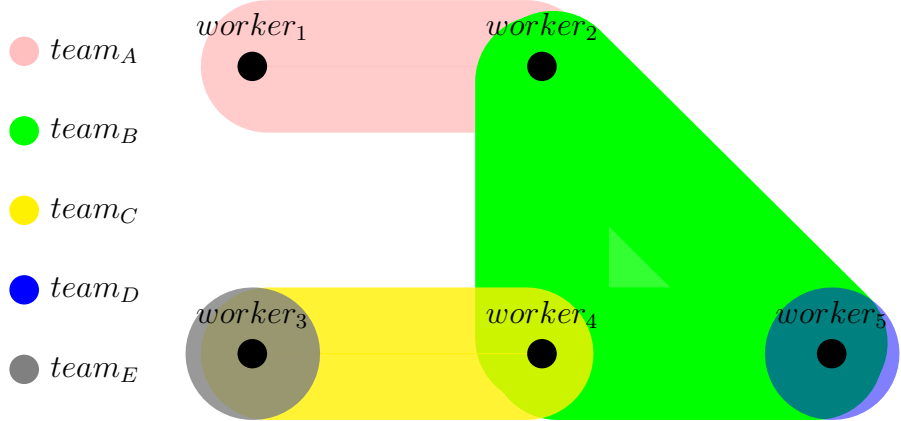
\begin{figure}
	\caption{Collaboration hypergraph with five teams and five workers\label{Fig_hyper}}
	\begin{center}
		\begin{tikzpicture}
		%\node[vertex,label=above:\(inventor_1\)] (v1) {};
		\node[vertex,label=above:\(worker_1\)] (v2) {};
		\node[vertex,right of=v2,label=above:\(worker_2\), node distance=1.5in] (v3) {};
		\node[vertex,below of=v2,label=above:\(worker_3\), node distance=1.5in] (v4) {};
		\node[vertex,right of=v4,label=above:\(worker_4\), node distance=1.5in] (v5) {};
		\node[vertex,right of=v5,label=above:\(worker_5\), node distance=1.5in] (v6) {};
		
		\begin{pgfonlayer}{background}
		\draw[edge,color=pink,opacity=.8] (v2) -- (v3);
		\begin{scope}[transparency group,opacity=.8]
		\draw[edge,opacity=1,color=green] (v3) -- (v5) -- (v6) -- (v3);
		\fill[edge,opacity=.8,color=green] (v3.center) -- (v5.center) -- (v6.center) -- (v3.center);
		\end{scope}
		\draw[edge,color=yellow,opacity=.8] (v4) -- (v5);
		\draw[edge,color=gray,opacity=.8] (v4.center) -- (v4.center);
		\draw[edge,color=blue] (v6.center) -- (v6.center);
		\end{pgfonlayer}
		
		\node[elabel,color=pink,label=right:\(team_A\)]  (e1) at (-3,0.2) {};
		\node[elabel,below of=e1,color=green,label=right:\(team_B\)]  (e2) {};
		\node[elabel,below of=e2,color=yellow,label=right:\(team_C\)]  (e3) {};
		\node[elabel,below of=e3,color=blue,label=right:\(team_D\)]  (e4) {};
		\node[elabel,below of=e4,color=gray,label=right:\(team_E\)]  (e5) {};
		\end{tikzpicture}
	\end{center}
\end{figure}

As an example, Figure \ref{Fig_hyper} represents a hypergraph of collaborations with five teams and five workers. Workers 3 and 5 produce on their own. In addition, worker 3 produces jointly with worker 4, and workers 4 and 5 produce in a team together with worker 2. Lastly, worker 1 produces with worker 2 in a team. 

Worker $i$ contributes to the team a quantity $\alpha_i$, which is unobserved to the econometrician. I assume $\alpha_i$ is constant across collaborations, and I refer to $\alpha_i$ as the \emph{type} of worker $i$.  Constancy of worker types over time is a key feature of my framework. However, assuming that individual types remain constant may be restrictive.\footnote{To relax this assumption, one could rely on hidden Markov models, which allow types to evolve as Markov processes, or one could alternatively rely on mixed membership models (e.g., Airoldi \textit{et al.}, 2008), which allow workers' types to depend on the team they participate in. I do not study these possibilities here.} For this reason, in the empirical analysis I will use observations from at most five consecutive years. Note that, while worker types are constant, type composition varies across teams. 

The output of a team $j$ with $n$ workers is denoted as $Y_{nj}$, and it is given by
\begin{equation}\label{eq_prod_fun}
Y_{nj}=\phi_n(\alpha_{i_1(j)},...,\alpha_{i_n(j)},\varepsilon_{nj}),
\end{equation}
where $\phi_n$ is the production function of an $n$-worker team, and $\varepsilon_{nj}$ represent other factors, or shocks, unobserved to the econometrician, that affect team output beyond workers' inputs $\alpha_i$.\footnote{Here and in the following, $\alpha_{i_1(j)}=\sum_{i=1}^N\boldsymbol{1}\{i_1(j)=i\}\alpha_i$, and similarly for the other workers in team $j$.}$^{,}$\footnote{Since $n=n_j$ is the size of team $j$, an alternative notation for (\ref{eq_prod_fun}) could be $$\widetilde{Y}_{j}=\phi_{n_j}(\alpha_{i_1(j)},...,\alpha_{i_n(j)},\widetilde{\varepsilon}_{j}),$$ where $\widetilde{Y}_{j}=Y_{n_j,j}$ and $\widetilde{\varepsilon}_{j}=\varepsilon_{n_j,j}$.} I assume that the positions of workers in the team do not matter, so $\phi_n$ is symmetric with respect to its first $n$ arguments. Hence, the framework abstracts from within-team hierarchy and organization.

I now state the first key assumption. Throughout, $\{A_k\,:\, k\}$ denotes the set of $A_k$ for all $k$.

\begin{assumption}{(Network exogeneity)}$\quad$\label{ass_net_ex}
	
	$\{\varepsilon_{nj}\,:\, (n,j)\}$ are independent of $\{(i_{1}(j),...,i_{n}(j))\,:\, (n,j)\}$ conditional on $\{\alpha_i\, :\, i\}$.
	\end{assumption}

Assumption \ref{ass_net_ex} restricts the team formation process. It states that team formation is independent of the team-specific shocks $\varepsilon_{nj}$, conditional on the worker types $\alpha_1,...,\alpha_N$. Hence, while the probability of joining a team may depend unrestrictedly on worker types and other factors that are independent of the $\varepsilon_{nj}$'s, it cannot depend on the $\varepsilon_{nj}$'s themselves. Assumption \ref{ass_net_ex} will thus fail if, before forming a team or joining one, workers have advance information about team-specific shocks. This assumption is important for the tractability of the approach. Yet it is restrictive, and in future work it will be important to relax it.\footnote{A possibility to relax Assumption \ref{ass_net_ex} is a joint random-effects approach, where production and team formation are estimated together (see Section \ref{Complements_sec}).}

%In future work it will be important to test it.  

I now state the second key assumption.

\begin{assumption}{(Independent shocks)}$\quad$\label{ass_indep_shocks}
	
	$\{\varepsilon_{nj}\,:\, (n,j)\}$ are mutually independent, and independent of $\{(i_{1}(j),...,i_{n}(j))\,:\, (n,j)\}$ and $\{\alpha_i\, :\, i\}$.
\end{assumption}

Assumption \ref{ass_indep_shocks} rules out dependence among team-specific shocks. As an implication, shocks to a team of workers who collaborate repeatedly over time are assumed serially independent. An alternative approach, which I do not study in this chapter, would be to model the dependence, for example using a parametric process for $\varepsilon_{nj}$. More generally, the framework does not allow for dynamics in team production, such as state dependence effects. 

Another implication of Assumption \ref{ass_indep_shocks} is the absence of ``team effects'' in the model. In the framework, a team $j$ is only a collection of workers, and there is no effect of $j$ \textit{per se} except through the independent shocks $\varepsilon_{nj}$. In particular, the model does not allow for team-specific capital (Jaravel \textit{et  al.}, 2018). In applications where firms (e.g., in R\&D) or team identities (e.g., in sports) play an important role, it will be important to extend the framework to allow for team types and other team-specific factors, in addition to worker types. 

Lastly, another assumption implicit in (\ref{eq_prod_fun}) is the absence of covariates, except for team size $n$. Covariates can be incorporated as additional inputs to production. In the empirical applications, I will account for time effects, and for age effects in robustness checks.

To specify and estimate the model, I will use two approaches. In the first one (``fixed-effects'', FE) I will treat the $\alpha_i$'s as parameters and estimate them. The FE approach has the advantage of not requiring to specify a model of team formation. Moreover, when team output is additive in worker types and Assumption \ref{ass_net_ex} holds, the FE approach is tractable. However, in the empirical collaboration networks that I will analyze, $\alpha_i$ estimates will often be very noisy. The prevalence of sample noise will require using methods for bias reduction.

In the second approach (``random-effects'', RE) I will model the joint distribution of the $\alpha_i$'s. I will use this approach to estimate nonlinear models with complementarity between worker types. The RE approach accounts naturally for sample noise. However, it requires specifying how the worker types depend on the collaborations in the network. In other words, the RE approach requires making assumptions about team formation beyond Assumption \ref{ass_net_ex}. I will compare several approaches based on simple statistical models. %Incorporating economic models of team formation into the framework is a natural next step for this research. 

\section{Additive production\label{Additive_sec}}

In this section and the next, I present methods for additive production and nonlinear production, in turn.

Suppose that the production function in (\ref{eq_prod_fun}) takes the following additive form
\begin{equation}\label{eq_prod_fun_add}Y_{nj}=\lambda_n\left(\alpha_{i_1(j)}+...+\alpha_{i_n(j)}\right)+\varepsilon_{nj},\end{equation}
where $\lambda_n$ is a team-size scaling factor. I adopt the normalization $\lambda_1=1$. In (\ref{eq_prod_fun_add}), output depends on the sum of worker types, or equivalently on the mean worker type in the team (given the presence of $\lambda_n$).\footnote{In model (\ref{eq_prod_fun_add}), types $\alpha_i$ are scalar, and their effects on output in teams of varying sizes are proportional to each other. The nonlinear model in Section \ref{Complements_sec} will allow for different patterns, possibly non-monotonic across teams of different sizes with varying worker composition.}

It is useful to write (\ref{eq_prod_fun_add}) in vector form as $Y_n=\lambda_nA_n+\varepsilon_n$; that is, equivalently, stacking all teams and team sizes together, 
\begin{equation}\label{eq_prod_fun_add_stack}Y=D_{\lambda}A\alpha+\varepsilon,\end{equation}
where $D_{\lambda}$ is a diagonal matrix, and $A$ is a matrix of zeros and ones where a one indicates that a worker (i.e., a column) participates in a team (i.e., a row). Network exogeneity then takes the form 
\begin{equation}\mathbb{E}\left(\varepsilon_{nj}\,|\, A,\alpha\right)=\mu_n,\label{eq_mean_indep}\end{equation}
where $\mu_n$ is a team-size-specific constant. In the empirical analysis, I will estimate (\ref{eq_prod_fun_add}) under the assumption that $\mu_n=0$. In addition, in a robustness check, I will report results from the following alternative specification in logarithms,
\begin{equation}\label{eq_prod_fun_add_2}\ln Y_{nj}=\mu_n+\alpha_{i_1(j)}+...+\alpha_{i_n(j)}+\varepsilon_{nj}.\end{equation}

\subsection{Identification}

To analyze identification in this setting, I treat $\alpha$, $\lambda$, $\mu$, and $A$ as non-stochastic quantities.\footnote{Hence the conditioning on $(A,\alpha)$ in (\ref{eq_mean_indep}) could be removed from the notation.} Suppose, to start with, that the $\lambda_n$'s are known, and that $\mu_n=0$ in (\ref{eq_mean_indep}). Since model (\ref{eq_prod_fun_add_stack})-(\ref{eq_mean_indep}) is a linear regression, the identification of the worker effects $\alpha_1,...,\alpha_N$ is determined by the rank properties of $A$. 

As an example, consider the matrix $A$ corresponding to Figure \ref{Fig_hyper}:
$$A=\left(\begin{array}{cccccc} 1 & 1 & 0 & 0&0\\0&1&0&1&1\\0&0&1&1&0\\0&0&0&0&1\\0&0&1&0&0\end{array}\right).$$
Since this matrix is non-singular, all $\alpha_1,...,\alpha_5$ are identified. An intuitive algorithm to show identification in this case is as follows. Let ${\cal{S}}$ denote the set of workers $i$ for which $\alpha_i$ is identified. First, focus on workers who produce on their own: since workers $3$ and $5$ work on their own, include them in ${\cal{S}}$. Then, focus on teams where the workers in ${\cal{S}}$ collaborate with others: since workers $3$ and $4$ work together, include worker $4$ in ${\cal{S}}$. Repeat this operation: since workers $2$, $4$ and $5$ work together, include worker $2$ in ${\cal{S}}$, and since workers $1$ and $2$ work together, include worker $1$ in ${\cal{S}}$. Hence ${\cal{S}}=\{1,2,3,4,5\}$.

To generalize the approach beyond this simple example, let $B=D_{\lambda}A$, continuing with the case where $D_{\lambda}$ is known. A worker type $\alpha_i$, for $i\in\{1,...,N\}$, is identified if and only if
$$\exists v_i\, :\, B'v_i=e_i,$$
where $e_i$ denotes the $i$-th canonical vector in $\mathbb{R}^N$. Letting $(B')^{\dagger}$ denote the Moore-Penrose generalized inverse of $B'$, and $I$ be a conformable identity matrix, this is equivalent to
\begin{equation}(I-B'(B')^{\dagger})e_i=0.\label{eq_ident}\end{equation}
All worker types $\alpha_i$ for which (\ref{eq_ident}) holds are identified. In general, however, not all $\alpha_i$'s are identified. For example, when two workers always work together, with or  without other co-workers, their respective $\alpha_i$'s are not separately identified.   

In practice, it is convenient to further restrict the sets of workers and teams such that the types of all workers in those teams are identified. To do this, I implement the following simple iterative algorithm.
\begin{algorithm}Iterate until convergence:
	\begin{enumerate}
	\item Select all the workers such that (\ref{eq_ident}) holds. 
	\item Select all the teams that only comprise those workers. 
	\item Let $B^{\rm sub}$ be the resulting selection of rows and columns of $B$. Rename $B^{\rm sub}$ as $B$. 
	\end{enumerate}	
\end{algorithm}

To compute Moore-Penrose generalized inverses, I rely on sparse LU matrix decompositions. With some abuse of notation, I will still denote the resulting restricted vectors and matrices as $Y$, $D_{\lambda}$, $A$, and $\varepsilon$.\footnote{Here my goal is to point-identify the $\alpha_i$ values of a set of workers, and the corresponding contributions of these workers to team output. For this reason, in the identification strategy I rely on workers who produce on their own. In other settings (e.g., team sports) this information may not be available. In such cases, it may still be possible to identify differences between $\alpha_i$'s. As an example, consider a team of five players, where player $i$ gets replaced by player $i'$. In this case the difference in the team's output identifies $\alpha_i-\alpha_{i'}$, though not $\alpha_i$ and $\alpha_{i'}$ separately. Another example is doubles tennis, as studied in Devereux (2018). In the empirical analysis, in robustness checks I will report estimates of nonlinear models that are solely based on 2-worker teams.}  

Consider now the team-size parameters $\lambda_n$, under the maintained assumption that $\mu_n=0$. Since
$$D_{\lambda}^{-1}Y=A\alpha+D_{\lambda}^{-1}\varepsilon,$$
it follows that
$$(I-AA^\dagger)D_{\lambda}^{-1}Y=(I-AA^\dagger)D_{\lambda}^{-1}\varepsilon.$$
Hence, (\ref{eq_mean_indep}) implies
\begin{equation}\mathbb{E}\left[Z_{n}'(I-AA^\dagger)D_{\lambda}^{-1}Y \right]=0,\quad \text{ for all }n,\label{eq_ident_lambda}\end{equation}
where $Z_n$ is a $J\times 1$ vector with ones where $n_j=n$ and zeros elsewhere. System (\ref{eq_ident_lambda}) is linear in the parameters $\lambda_n^{-1}$, so the conditions for identification are standard.\footnote{This quasi-differencing approach is related to Holtz-Eakin \textit{et al.} (1988) and Chamberlain (1992).} In particular, $\lambda_1,\lambda_2,\lambda_3$ are not identified in the stylized example of Figure \ref{Fig_hyper}. 

Lastly, when relaxing the assumption that $\mu_n=0$ in (\ref{eq_mean_indep}), one can use a similar strategy. To see this, let $\mu$ be the vector of $\mu_n$'s. We have
\begin{equation}\mathbb{E}\left[Z'(I-AA^\dagger)D_{\lambda}^{-1}(Y-\mu) \right]=0,\label{eq_ident_lambda2}\end{equation}
where instruments $Z$ can be constructed as functions of collaborations in the network. As an example, the $j$-th row of $Z$ may contain some functions of the team sizes of the co-workers in $j$, in addition to the size $n_j$ of team $j$. Moreover, under assumptions on the covariance matrix of $\varepsilon$, one can add covariance restrictions on $\lambda$ and $\mu$ to the mean restrictions in (\ref{eq_ident_lambda2}). I will not implement these approaches empirically in Sections \ref{Research_sec} and \ref{Inventors_sec}, and proceed under the assumption that $\mu_n=0$ for all $n$. 

\subsection{Estimation}

I will focus on the following decomposition of output variance, for every team size $n$:
\begin{align}\label{eq_vardec}
\underset{\mbox{Total variance}}{\underbrace{{\mbox{Var}}_n\left(Y_{nj}\right)}}=&\underset{\mbox{Heterogeneity}}{\underbrace{\lambda_n^2\sum_{m=1}^n\mbox{Var}_n\left(\alpha_{i_m(j)}\right)}} +\underset{\mbox{Sorting}}{\underbrace{2\lambda_n^2\sum_{m=1}^n\sum_{m'=m+1}^n\mbox{Cov}_n\left(\alpha_{i_m(j)},\alpha_{i_{m'}(j)}\right)}}+\underset{\mbox{Other factors}}{\underbrace{{\mbox{Var}}_n\left(\varepsilon_{nj}\right)}}.
\end{align}
In this decomposition, the component labeled ``heterogeneity'' reflects the variation in worker effects on output, keeping team composition constant. Equivalently, it represents the effect of worker heterogeneity if the allocation of workers to teams were random, in which case covariances would be zero. In turn, the component labeled ``sorting'' reflects the variance contribution due to team composition not being random.\footnote{It is common in the literature on matched employer-employee data to apply a related decomposition to the variance of log-wages, as opposed to (log-) output. This raises issues for the interpretation of variance components estimates in terms of heterogeneity and sorting (Eeckhout and Kircher, 2011). In contrast, here heterogeneity and sorting can be directly interpreted as contributions to production.}

To estimate the variance components in (\ref{eq_vardec}), I first estimate the team-size effects $\lambda_n$ using an empirical counterpart to the just-identified system (\ref{eq_ident_lambda}).\footnote{The additive specification (\ref{eq_prod_fun_add_2}) in logs can be estimated similarly to the one in levels, the only difference being a slightly different approach to estimate $\mu_n$.} Next, I estimate $\alpha$ using OLS as
$$\widehat{\alpha}=(B'B)^{-1}B'Y,$$
for $B=D_{\lambda}A$. Hence, for any variance component $V_Q=\alpha'Q\alpha$, where $Q$ is an $N\times N$ matrix, I can construct the fixed-effects (FE) estimator
\begin{equation}\widehat{V}_Q=\widehat{\alpha}'Q\widehat{\alpha}.\label{eq_vardec_est}\end{equation}

However, the FE estimator $\widehat{V}_Q$ is biased. To see this, note that
\begin{equation}\mathbb{E}[\widehat{V}_Q\,|\, A,\alpha]=V_Q+\underset{\mbox{Bias}}{\underbrace{\mathbb{E}[\varepsilon'(B'B)^{-1}B'QB(B'B)^{-1}\varepsilon\,|\, A,\alpha]}}.\end{equation}
Following Andrews \textit{et al.} (2008), I subtract an unbiased estimate of the bias to $\widehat{V}_Q$. To construct such an estimate, note that
$$\mbox{Bias}=\mbox{Trace}( (B'B)^{-1}B'QB(B'B)^{-1}\Omega),$$
where $\Omega$ is the covariance matrix of $\varepsilon$. To obtain an empirical counterpart of $\Omega$, I assume that all the $\varepsilon_{nj}$'s are independent (see Assumption \ref{ass_indep_shocks}), and in addition that $\mbox{Var}(\varepsilon_n|A,\alpha)=\sigma_{n}^2$ varies only with team size. I estimate $\sigma_n^2$ as
\begin{equation}\widehat{\sigma}_n^2=\frac{Y_n'(I_n-A_nA_n^{\dagger})Y_n}{\mbox{Trace}(I_n-A_nA_n^{\dagger})}.\label{eq_sigma_hat}\end{equation}
The bias correction relies on the $\varepsilon_{nj}$'s being independent, and on their variance being constant within team size. While homoskedasticity could be relaxed by following Kline \textit{et al.} (2020) and suitably restricting the sets of workers and teams, independence may be empirically strong in the applications. Under independence, Kline \textit{et al.} (2020) derive consistency results and limiting distributions for bias-corrected variance components estimates in environments where the numbers of rows (teams) and columns (workers) in $A$ both tend to infinity. 

To implement the bias correction in large data sets, I rely on a sparse LU decomposition to compute the denominator in (\ref{eq_sigma_hat}), and I approximate the trace in the bias formula using Hutchinson's approximation, as in Gaure (2014) and Kline \textit{et al.} (2020). I use 1000 draws in the approximation.

\section{Nonlinear production\label{Complements_sec}}

I now describe a nonlinear model that allows for complementarity between worker types. To keep tractability, I model types $\alpha_i$ as discrete, with $K$ points of support. I will vary $K$ in robustness checks.

I adopt a random-effects (RE) strategy that consists in modeling the distribution of types. Specifically, I suppose that $\alpha_1,...,\alpha_N$ are drawn from the joint distribution $\prod_{i=1}^N \pi(\alpha_i)$, where $\pi(\alpha_i)$ are type probabilities. I will first describe an independent RE method where $\{\alpha_i\, :\, i\}$ are independent of collaborations $\{(i_1(j),...,i_n(j))\, :\, (n,j)\}$. I will interpret $\pi$ as a prior on individual types, as in Arellano and Bonhomme (2009). While, by construction, the prior imposes that team formation is independent of types -- and thus rules out sorting -- the posterior estimates that I will report may still feature sorting (and empirically they will).\footnote{In a similar spirit, in matched employer-employee data, Woodcock (2008) points out that posterior RE estimates can still indicate the presence of sorting even if the prior on worker and firm types is independent of mobility patterns.} 

However, a natural concern is that the prior $\pi$ may be empirically informative. For example, a prior that is independent of collaborations will tend to shrink sorting patterns towards zero. Unlike the FE methods of Section \ref{Additive_sec}, RE methods rely implicitly or explicitly on a model of team formation, and can be sensitive to that model being misspecified. For this reason, I will also present estimates where I will model the dependence between the types $\alpha_i$ and the collaborations $(i_1(j),...,i_n(j))$.

With discrete types, a different approach would be to treat type membership indicators as discrete parameters, and use kmeans clustering methods for estimation (Bonhomme and Manresa, 2015, Bonhomme \textit{et al.}, 2021). Bonhomme \textit{et al.} (2019) use this grouped fixed-effects approach to incorporate discrete firm heterogeneity in a nonlinear model of wages. I will use a related approach to provide preliminary evidence of sorting and complementarity. However, accurately estimating worker types using grouped fixed-effects requires a relatively large number of observations per worker. In the data sets that I analyze in this chapter, many workers produce only a handful of times. For this reason, I instead rely on a RE approach.

\subsection{Identification}

Nonparametric identification of finite mixture models under independence assumptions akin to Assumption \ref{ass_indep_shocks} has been extensively studied. When all teams consist of a single worker, the model has independent measures and finite types, and its identification has been studied in Hall and Zhou (2003), Hu (2008), and Allman \textit{et al.} (2009), among others. When teams have multiple workers, the model has a network structure that is related to the settings studied in Allman \textit{et al.} (2011).  

Here I outline an identification argument in the case where $\{\alpha_i\, :\, i\}$ are independent of collaborations $\{(i_1(j),...,i_n(j))\, :\, (n,j)\}$, output is i.i.d. across teams, and team size is $n\in\{1,2\}$. Formally, I treat worker types $\alpha_i$ as random, and collaborations $(i_1(j),...,i_n(j))$ as non-stochastic, and assume that the type distribution $\pi$ is common across workers, independent of the collaborations. The goal is to identify the type distribution $\pi$, as well as the type-specific conditional distributions of output. 

Focusing on workers who produce at least three times on their own gives, for $\{j_1,j_2,j_3\}$ triplets such that $i_1(j_1)=i_1(j_2)=i_1(j_3)$, 
\begin{align}&\label{eq_id_n1}\Pr\left[Y_{1j_1}\leq y_1,Y_{1j_2}\leq y_2,Y_{1j_3}\leq y_3\right]\notag\\
&=\sum_{\alpha}\pi(\alpha)\Pr\left[Y_{1j_1}\leq y_1\,|\, \alpha_{i_1(j_1)}=\alpha\right]\Pr\left[Y_{1j_2}\leq y_2\,|\, \alpha_{i_1(j_2)}=\alpha\right]\Pr\left[Y_{1j_3}\leq y_3\,|\, \alpha_{i_1(j_3)}=\alpha\right],\end{align}
where I have used Assumptions \ref{ass_net_ex} and \ref{ass_indep_shocks}, and that $\pi$ does not depend on $\{(i_1(j),...,i_n(j))\, :\, (n,j)\}$. System (\ref{eq_id_n1}) identifies $\pi(\alpha)$ and $\Pr\left[Y_{1j}\leq y\,|\, \alpha_{i_1(j)}=\alpha\right]$ up to labeling of the latent types, under suitable rank conditions (Allman \textit{et al.}, 2009). 

Next, focusing now on pairs of workers who produce once on their own and once together, we similarly have, for $\{j_1,j_2,j_3\}$ triplets such that $\{i_1(j_1),i_1(j_2)\}=\{i_1(j_3),i_2(j_3)\}$, 
\begin{align}&\label{eq_id_n12}\Pr\left[Y_{1j_1}\leq y_1,Y_{1j_2}\leq y_2,Y_{2j_3}\leq y_3 \right]\notag\\
&=\sum_{\alpha,\alpha'}\pi(\alpha)\pi(\alpha')\Pr\left[Y_{1j_1}\leq y_1\,|\, \alpha_{i_1(j_1)}=\alpha\right] \Pr\left[Y_{1j_2}\leq y_2\,|\, \alpha_{i_1(j_2)}=\alpha'\right]\times\notag\\
&\quad \quad \quad \quad\quad\quad \quad \quad\quad \quad \quad\quad \quad \quad\Pr\left[Y_{2j_3}\leq y_3\,|\, \alpha_{i_1(j_3)}=\alpha,\alpha_{i_2(j_3)}=\alpha'\right].\end{align}
Given that $\pi(\alpha)$ and $\Pr\left[Y_{1j}\leq y\,|\, \alpha_{i_1(j)}=\alpha\right]$ are identified up to labeling, (\ref{eq_id_n12}) identifies $\Pr\left[Y_{2j}\leq y\,|\, \alpha_{i_1(j)}=\alpha,\alpha_{i_2(j)}=\alpha'\right]$ up to the same labeling, under a suitable rank condition.\footnote{Specifically, it suffices that there exist a set $\{y_r\}$ of values in the support of $Y_{1j}$ such that the matrix with $\left((y_r,y_{r'}),(\alpha,\alpha')\right)$-element $\Pr\left[Y_{1j}\leq y_r\,|\, \alpha_{i_1(j)}=\alpha\right]\Pr\left[Y_{1j}\leq y_{r'}\,|\, \alpha_{i_1(j)}=\alpha'\right]$ has full column rank.} The argument requires no monotonicity assumptions about how output depends on worker types. In addition, identification can be shown using a related argument when $\pi$ depends on collaborations and worker types are not independent of each other within teams, provided none of the joint probabilities $\pi(\alpha,\alpha')$, for pairs of workers who produce once on their own and once together, are positive.
 
While the previous argument relies on workers producing on their own, identification can also be shown in settings without 1-worker teams; see Allman \textit{et al.} (2011) for identification results in certain random graph models.\footnote{As an example, for two workers who collaborate together at least three times, we have, for $\{j_1,j_2,j_3\}$ triplets such that $\{i_1(j_1),i_2(j_1)\}=\{i_1(j_2),i_2(j_2)\}=\{i_1(j_3),i_2(j_3)\}$,    
\begin{align*}&\Pr\left[Y_{2j_1}\leq y_1,Y_{2j_2}\leq y_2,Y_{2j_3}\leq y_3\right]=\sum_{\alpha,\alpha'}\pi(\alpha)\pi(\alpha')\Pr\left[Y_{2j_1}\leq y_1\,|\, \alpha_{i_1(j_1)}=\alpha,\alpha_{i_2(j_1)}=\alpha'\right]\times\notag\\
&\quad\quad\quad\quad\quad\quad\quad\quad\quad\Pr\left[Y_{2j_2}\leq y_2\,|\, \alpha_{i_1(j_2)}=\alpha,\alpha_{i_2(j_2)}=\alpha'\right]\Pr\left[Y_{2j_3}\leq y_3\,|\, \alpha_{i_1(j_3)}=\alpha,\alpha_{i_2(j_3)}=\alpha'\right],\end{align*}
which identifies $\pi(\alpha)$ and $\Pr\left[Y_{2j}\leq y\,|\, \alpha_{i_1(j)}=\alpha,\alpha_{i_2(j)}=\alpha'\right]$ up to labeling of the components, under suitable rank conditions.} As robustness checks, I will report empirical results based on 2-worker teams only.

\subsection{Estimation using a variational approach}

Estimating the random-effects model of team production is challenging. To see this, consider a setup where the density $f^n_{\alpha_{i_1(j)},...,\alpha_{i_n(j)}}(y)$ of $Y_{nj}$ conditional on $\alpha_{i_1(j)},...,\alpha_{i_n(j)}$ is parametric, indexed by a finite-dimensional vector $\theta$. The RE likelihood 
$${\cal{L}}(\theta,\pi)=\sum_{\alpha_1}...\sum_{\alpha_N}\prod_i \pi(\alpha_i)\prod_n \prod_j f^n_{\alpha_{i_1(j)},...,\alpha_{i_n(j)}}\left(Y_{nj};\theta\right)$$
involves an intractable $N$-dimensional sum over all possible worker type realizations. The likelihood does not factor in simple ways, except in the special case where all teams consist of one worker.\footnote{In 1-worker teams, the identification strategy leads to practical nonparametric estimators; see Bonhomme \textit{et al.} (2016) for example. This could be used to nonparametrically estimate models of team production, using (\ref{eq_id_n12}) and its extensions for $n>2$.}

To reduce computational complexity, I follow a mean-field variational approach that is increasingly popular in related settings in machine learning and statistics. The idea is to introduce an auxiliary distribution $\prod_{i=1}^N q_i(\alpha_i)$, and set it to be as close as possible to the posterior density of $\alpha_1,...,\alpha_N$. Unlike the posterior density, its variational approximation factors across $i$. This makes estimation feasible even in large data sets.

Formally, the variational objective function is
\begin{equation}\mbox{ELBO}(\theta,\pi)=\underset{q_1,...,q_N}{\mbox{max}}\, \ln\left({\cal{L}}(\theta,\pi)\right)-\underset{\mbox{KL divergence}}{\underbrace{\mathbb{E}_{q_1...q_N} \ln \frac{\prod_iq_i(\alpha_i)}{p(\alpha_1,...,\alpha_N;\theta,\pi)}}},\label{eq_ELBO2}\end{equation}
where $p(\alpha_1,...,\alpha_N;\theta,\pi)$ is the (computationally intractable) posterior density of worker types given $\theta$ and $\pi$ values. The second term on the right-hand side of (\ref{eq_ELBO2}) is the Kullback-Leibler (KL) divergence between the posterior density and its variational approximation. Hence, the variational objective and the RE likelihood are closer when the variational approximation is more accurate. The variational objective is a lower bound on the RE likelihood, and it is often referred to as the ``evidence lower bound'' (ELBO).

To see why adding the KL term can improve tractability, note that we equivalently have 
\begin{align}\mbox{ELBO}(\theta,\pi)&=\underset{q_1,...,q_N}{\mbox{max}}\, \sum_{n}\sum_j\mathbb{E}_{q_{i1(j)}...q_{i_n(j)}} \ln f_{\alpha_{i_1(j)},...,\alpha_{i_n(j)}}(Y_{nj};\theta)+\sum_{i}\mathbb{E}_{q_i} \ln \pi-\sum_i\mathbb{E}_{q_i} \ln q_i\notag\\
&=\underset{q_1,...,q_N}{\mbox{max}}\, \sum_{n}\sum_j\sum_{\alpha_1}...\sum_{\alpha_n}q_{i_1(j)}(\alpha_1)...q_{i_n(j)}(\alpha_n)\ln f_{\alpha_{1},...,\alpha_{n}}(Y_{nj};\theta)\notag\\
&\quad\quad \quad \quad \quad  \quad \quad \quad \quad \quad  +\sum_{i}\sum_{\alpha}q_i(\alpha) \ln \pi(\alpha)-\sum_i\sum_{\alpha}q_i(\alpha) \ln q_i(\alpha).\label{eq_ELBO}\end{align}
Since this expression no longer involves an $N$-dimensional sum, it can be evaluated easily, at least when team size $n$ is relatively low. To estimate the nonlinear models, I will maximize the evidence lower bound using the EM algorithm; see Bishop (2006) and Mariadassou \textit{et al.} (2010). Specifically, the algorithm alternates between updates of the $q_i(\alpha)$'s given $(\theta,\pi)$ via Newton steps, and updates of $(\theta,\pi)$ given the $q_i(\alpha)$'s via weighted maximum likelihood estimation.

Computation of the evidence lower bound in (\ref{eq_ELBO}) becomes more complex as team size increases. In the applications, I will restrict the analysis to 1-worker and 2-worker teams. It will be important to devise computational strategies to efficiently handle larger teams.\footnote{Moreover, to estimate models with larger teams, one will need to impose additional structure on how production depends on types. A possibility, in the spirit of Ahmadpoor and Jones (2019), would be to model type-specific mean output using a CES function.} 

\paragraph{Statistical properties of variational estimates.}

An obvious issue with replacing the RE likelihood by the evidence lower bound is that this leads to optimizing a different objective function. Indeed, in the present setting as well as in most network applications, the posterior density $p$ does not factor across workers, whereas the variational density $q$ does. This discrepancy can make the variational estimates inconsistent even when the assumptions of the RE model hold.

The analysis in Bickel \textit{et al.} (2013) suggests that variational estimates can be theoretically justified in network models with discrete types (see also Celisse \textit{et al.}, 2012). Bickel \textit{et al.} (2013) focus on a standard stochastic blockmodel (e.g., Snijders and Nowicki, 1997) with binary outcomes, where choice probabilities are functions of both agents' types. In their asymptotic environment, expected degree in the network grows at a rate faster than $\ln N$. They first show that the RE maximum likelihood estimator (MLE) is asymptotically equivalent to an infeasible MLE where the agents' types are observed.\footnote{Related perfect classification results have been derived in panel data models under discrete heterogeneity (e.g., Hahn and Moon, 2010, Bonhomme and Manresa, 2015).} As a result, the RE estimator is consistent and asymptotically normal. They then show that the mean-field variational estimator (Daudin \textit{et al.}, 2008) is asymptotically equivalent to both the infeasible MLE and the RE MLE, hence that it is consistent with the same asymptotically normal distribution.

Stochastic blockmodels with binary outcomes are closely related to the model I focus on. However, there are also important differences, since in my model workers both form teams and produce, and output is continuous. In addition, the conditions in Bickel \textit{et al.} (2013) rely on the network becoming sufficiently dense in large samples. It is unclear whether conditions akin to the ones they assume provide a good approximation to the settings I study. For this reason, in the appendix I probe the accuracy of the variational estimates using Monte Carlo simulations. In simulated samples that mimic the data on economic researchers, I find that variational estimates recover the true parameter values well when the numbers of productions and collaborations are sufficiently large.

\subsection{Accounting for the link between worker types and collaborations}

The assumption that the distribution of worker types $\alpha_i$ does not depend on collaborations $\{(i_1(j),...,i_n(j))\, :\, (n,j)\}$ may be restrictive. To give intuition, consider a simple assignment model of team formation, along the lines of the roommate matching problem studied by Chiappori \textit{et al.} (2019). In the model, workers are assigned to teams of size $n\in\{1,2\}$ in a way that maximizes expected output. Let $\mu_1(\alpha)=\mathbb{E}(Y_{1j}\,|\, \alpha_{i_1(j)}=\alpha)$ and $\mu_2(\alpha,\alpha')=\mathbb{E}(Y_{2j}\,|\, \alpha_{i_1(j)}=\alpha,\alpha_{i_2(j)}=\alpha')$. Let $T_{1\alpha},T_{2\alpha}\in\mathbb{N}$ be exogenous ``budgets'' for type-$\alpha$ workers, indicating the maximum numbers of teams of size $1$ and $2$ in which they can participate. Consider an allocation solving
\begin{align}\label{max_surplus}
&\mbox{max}\, \sum_{\alpha} \tau_{\alpha} \mu_1(\alpha)+\sum_{\alpha}\tau_{\alpha\alpha}\mu_2(\alpha,\alpha)+\frac{1}{2}\sum_{\alpha}\sum_{ \alpha'\neq \alpha} \tau_{\alpha\alpha'} \mu_2(\alpha,\alpha')\\
&\mbox{s.t.}\quad \tau_\alpha,\tau_{\alpha\alpha'}\in \mathbb{N},\, \tau_{\alpha\alpha'}=\tau_{\alpha'\alpha},\, \tau_\alpha\leq T_{1\alpha},\,2\tau_{\alpha\alpha}+\sum_{\alpha'\neq \alpha} \tau_{\alpha\alpha'}\leq T_{2\alpha}, \notag
\end{align}   
where $\tau_{\alpha}$ denotes the number of teams with one worker of type $\alpha$, and $\tau_{\alpha\alpha'}$ denotes the number of teams with one worker of type $\alpha$ and another one of type $\alpha'$.

The optimization in (\ref{max_surplus}) is related to the planner's objective in the Becker (1973) two-sided marriage model.\footnote{In the Becker model, the optimal allocation can be represented as a stable equilibrium in an economy with transfers. In one-sided problems, existence of a decentralized solution is not guaranteed in general (e.g., Talman and Yang, 2011). Chiappori \textit{et al.} (2019) show existence when types are discrete and budget sizes are even. Unlike in their setup, here there are two budgets, for 1-worker and 2-worker teams, respectively. Determining optimal allocations of workers across teams of different sizes would require measuring the costs of teamwork -- e.g., effort costs -- which I do not consider here.} As in the Becker model, in (\ref{max_surplus}) the assignment of individuals to teams will be driven by complementarity patterns in the expected payoff function $\mu_2(\alpha,\alpha')$. Intuitively, the higher the degree of complementarity between workers, the higher the assortativeness of the optimal allocation. I will illustrate the link between complementarity and assortativeness by computing optimal allocations of economists and inventors in Sections \ref{Research_sec} and \ref{Inventors_sec}. In particular, this simple team assignment model implies that types will generally \emph{not} be uniformly distributed across teams.

To relax the independence assumption between types and collaborations, I will rely on two approaches. In the first approach, I will maintain prior independence between the $\alpha_i$'s, however I will let $\pi_i(\alpha_i)$ depend on worker $i$'s characteristics. This correlated random-effects approach (Chamberlain, 1984) is widely used in traditional panel data applications. However, in a network of teams, it is not \textit{a priori} obvious which characteristics $\alpha_i$ should depend on. I will experiment with features of the degree distributions in the collaboration hypergraph. In addition, note that this approach imposes that worker types be conditionally independent of one another given collaborations, which is unlikely when the assignment of workers to teams is given by a team formation model such as (\ref{max_surplus}).

In the second approach, I will jointly model output production and team formation. This adds another estimation step, since one needs to estimate the team formation process as well. Identification can be analyzed using related techniques,\footnote{In this case there are two kinds of dependent variables: the output variables and the team membership indicators. See Bonhomme (2017) for an example.} and the mean-field approximation can be implemented similarly. A specific feature of the joint model of team formation and production is that worker types influence both the quality and the quantity of output. For example, in the case of researchers in economics, the type of a worker will influence both how many articles she publishes (i.e., quantity) and in which journals they appear (i.e., quality). To implement this approach, I will specify a simple stochastic blockmodel of team formation. Though of great interest, using an economic team formation model instead would raise several challenges, and I do not consider this possibility in this chapter.

%\section{Researchers in economics\label{Research_sec}}
%
%Illustrate the method. 

\section{Academic production in economics\label{Research_sec}}

In this section and the next I present two applications. I start by applying the method to study economists and their academic production.

\subsection{Data and sample selection}

\begin{table}[h!]
	\caption{Descriptive statistics on economic researchers\label{Tab_desc_1}}
	\begin{center}
		\begin{tabular}{||l||c|c|c|c||c|c|c||}\hline\hline
			&\multicolumn{4}{c}{(a) Sample}& \multicolumn{3}{c||}{(b) Subsample}\\
			&All & $n=1$ & $n=2$ & $n=3$	&All & $n=1$ & $n=2$\\\hline\hline
			\# Authors & 6509&6194 & 4937&1643  & 4825&4724 & 3148\\
			\# Articles & 41150& 31085&8987 &1078 & 31782& 26558&5224\\
			Mean output & 5.55& 4.44& 7.12&7.56& 5.37& 4.55& 7.45\\
			Std output &11.18 & 9.80& 12.74&12.95&10.85 & 9.89& 12.75\\
			p10\% output & 0.52& 0.52& 0.52&0.54 & 0.52& 0.52& 0.52\\
			p50\% output &0.76 &0.66 & 1.81&2.14&0.67&0.66 & 2.06\\
			p90\% output & 14.88
			&12.09 & 19.65 &19.72& 14.86
			&12.42 & 20.68 \\
			p95\% output &24.85 &22.73 & 35.13&37.06&24.85 &23.02 & 35.13\\
			p99\% output &58.90 & 57.66& 62.29&64.92&58.58 & 56.67& 63.65\\
			p10\% \#articles &5 &3 & 2&1&5 &4 & 2\\
			p50\% \#articles & 8& 6& 5&2& 8& 6& 4\\
			p90\% \#articles & 18&12 & 11&7& 16&13 & 10\\
			p95\% \#articles & 23 & 16& 14&8 & 21 & 17& 13\\
			p99\% \#articles &39 & 31& 20&13&35 & 35& 18\\\hline\hline		
		\end{tabular}
	\end{center}
	{\footnotesize \textit{Notes: Data from Ductor \textit{et al.} (2014), 1995-1999. Output is a measure of journal quality, net of multiplicative year effects (reference year is 1999). Percentiles of the distribution of number of articles per author are indicated in the bottom five rows. In panel (a) the sample is restricted to at least 5 publications per author during the period. In panel (b) the subsample is restricted to collaborations between at most two authors, where all authors have at least 5 publications during 1995-1999. I use the subsample in panel (b) to estimate nonlinear models.}}
\end{table}

I use data from Ductor \textit{et al.} (2014). These data, initially constructed by Goyal \textit{et al.} (2006), are drawn from the EconLit database, a bibliography of journals compiled
by the Journal of Economic Literature. I restrict the sample to articles published between 1995 and 1999, with at most three co-authors.

I only include authors who produced at least five articles during the period. While a natural selection when aiming to infer individual contributions, restricting the number of publications per author in this way is restrictive. Indeed, starting from 62615 authors and 89834 articles in the original 1995-1999 data, the selection restricts the sample to 6509 authors and 41150 articles. To assess how representative the results on this sample are, I will run checks using larger samples. 

As measure of academic output, I follow Ductor \textit{et al.} (2014) and use the journal quality variable from Kodrzycki and Yu (2006), as extended by Ductor and co-authors. From the output measure I net out multiplicative year fixed-effects, using 1999 as reference year. I show descriptive statistics in panel (a) of Table \ref{Tab_desc_1} for the main sample. Research output, as measured by the journal quality variable, is right-skewed. The number of collaborations per author is also skewed, with 50\% of authors producing at most 8 articles, and 1\% producing at least 39. The subsample in panel (b), which I will use to estimate nonlinear models of 1-author and 2-author teams, shows similar patterns.

\subsection{Additive production}

I start by estimating the additive model of academic production (\ref{eq_prod_fun_add}). I first restrict the set of authors and articles in order to ensure identification. This gives 6479 authors and 41049 articles. Hence, relative to the sample in panel (a) of Table \ref{Tab_desc_1}, only 30 authors and 101 articles drop out.

Using this sample, I first estimate the team-size effects $\lambda_n$. I find $\lambda_2=0.67$ and $\lambda_3=0.48$. This suggests that, keeping author type(s) constant, collaborations between two co-authors increase output by $2\times 0.67-1=34\%$, and collaborations between three co-authors increase output by $3\times 0.48-1=44\%$. Next, I estimate the variance components in the decomposition (\ref{eq_vardec}). I report uncorrected estimates, which are obtained using fixed-effects, and bias-corrected estimates, which I obtain using the method described in Section \ref{Additive_sec}.

\begin{table}[h!]
	\caption{Additive production, economic researchers\label{Tab_AKM_RE}}
	\begin{center}
		\begin{tabular}{||l||c|c|c||}\hline\hline
			& $n=1$ & $n=2$ & $n=3$\\\hline\hline
			Total variance & 95.97&163.21& 161.38\\
			Heterogeneity & 32.53& 31.71&22.58 \\
			\color{gray}{Heterogeneity (uncorrected)} & 	\color{gray}{42.80}&	\color{gray}{53.26}&	\color{gray}{42.70}\\
			Sorting & -&  28.41& 32.73\\
			\color{gray}{Sorting (uncorrected) }& 	\color{gray}{-}&	\color{gray}{20.67}& 	\color{gray}{23.34}\\
			Other factors &63.86 & 82.82& 86.81\\
			\color{gray}{Other factors (uncorrected)} &	\color{gray}{53.17} & 	\color{gray}{92.26}& 	\color{gray}{99.41}\\	Team scale $\lambda_n$ & 1.00& 0.67& 0.48\\\hline\hline		
		\end{tabular}
	\end{center}
	{\footnotesize \textit{Notes: Estimates of variance components in equation (\ref{eq_vardec}), for different team sizes $n$. Descriptive statistics on the sample are given in panel (a) of Table \ref{Tab_desc_1}.}}
\end{table}

In Table \ref{Tab_AKM_RE} I show the results of the variance decomposition (\ref{eq_vardec}), for the three team sizes: $n=1,2,3$. The results suggest that author heterogeneity matters substantially, while playing a smaller role in larger teams. Indeed, the variance share explained by author heterogeneity is 33\% for sole-authored articles, 19\% in articles with two co-authors, and 14\% in articles with three co-authors. Sorting also contributes a substantial share: 17\% in articles with two co-authors, and 20\% in articles with three co-authors. Hence, according to this analysis, both heterogeneity and sorting contribute to the variation in academic output.

Another finding in Table \ref{Tab_AKM_RE} is the magnitude of the component due to other factors, $\sigma_n^2$. This component explains 67\% of the output variance in articles with one author, 57\% in articles with two co-authors, and 62\% in articles with three co-authors. The large residual variance contributes to the fixed-effects estimates of variance components being substantially biased. For example, the fixed-effects estimate of the variance contribution of heterogeneity is biased upward by 33\% with one author, 69\% with two co-authors, and 91\% with three co-authors. This suggests that, as has been documented in matched employer-employee settings (e.g., Bonhomme \textit{et al.}, 2020), bias correction is needed in order to obtain reliable estimates of variance components in collaboration networks. 

\paragraph{Robustness analysis.}

The adequacy of the additive model (\ref{eq_prod_fun_add}) depends on how one measures the output variable. To explore the sensitivity to this choice, I use two alternative measures of output. First, I estimate the model in logs, accounting for additive team-size effects as in (\ref{eq_prod_fun_add_2}). The results are shown in Appendix Table \ref{Tab_AKM_rob}, panel (a). Second, I use year-specific ranks of journal quality as dependent variables, instead of journal quality itself. The results are shown in Appendix Table \ref{Tab_AKM_rob}, panel (b). While it is reassuring that the variance shares in the two specifications are broadly comparable to the baseline, this exercise does not fully address the issue of the sensitivity to the choice of output measure. The nonlinear model that I estimate in the next section will be less sensitive to this issue.\footnote{In this application, it would be interesting to link the journal quality variable to economic payoffs and career outcomes.}

The estimation sample represents a relatively small share of authors and articles during the period. To probe the robustness of the findings to sample definition, I consider two less restrictive rules for inclusion in the sample, requiring every author to produce at least one or two articles, respectively, as opposed to at least five in the baseline sample. The results are shown in Appendix Table \ref{Tab_AKM_rob}, panels (c) and (d). When requiring at least two articles per author, the sample contains $24555$ authors and $70550$ articles, and $23498$ authors and $69331$ articles in the sample where all author types are identified. Panel (c) of the table shows that bias-corrected variance shares are broadly similar to the baseline, and that biases are larger, so bias correction matters more. When removing all restrictions and including all authors with at least one article, the sample where all author types are identified contains $44852$ authors and $79659$ articles. Panel (d) of Appendix Table \ref{Tab_AKM_rob} shows larger differences compared to the baseline subsample, and a substantially higher amount of bias.

\subsection{Nonlinear model}

Before presenting the results of the nonlinear model, I provide some preliminary evidence of complementarity, as well as sorting and heterogeneity, in the data. To do so, I select authors who produce at least five articles \emph{on their own}. This gives a set of 3263 authors. Then, for every author I compute the average output (i.e., average journal quality) among their sole-authored articles, and divide this measure into four quartiles. I call the resulting quartile the ``type proxy'' of the author. The reason for focusing on individuals with at least five sole-authored articles is to reduce noise in the proxy. At the same time, this does not fully remove the noise and requires a selected sample.

\begin{figure}[h!]
	\caption{Preliminary evidence based on type proxies, economic researchers \label{fig_prelim}}
	\begin{center}
		\begin{tabular}{cc}
			(a) Sorting & (b) Heterogeneity \& complementarity \\
			\includegraphics[width=60mm, height=60mm]{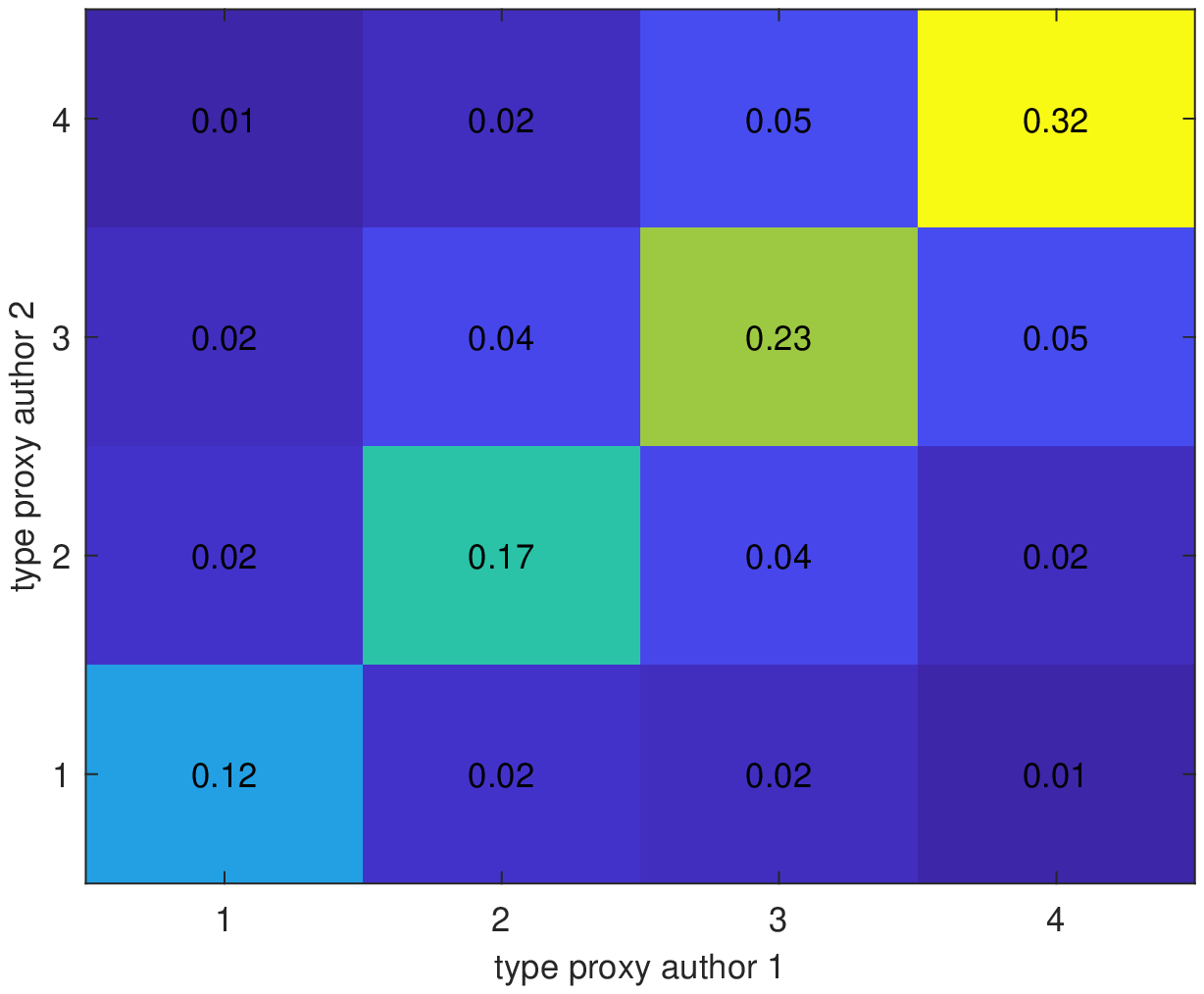}&\includegraphics[width=60mm, height=60mm]{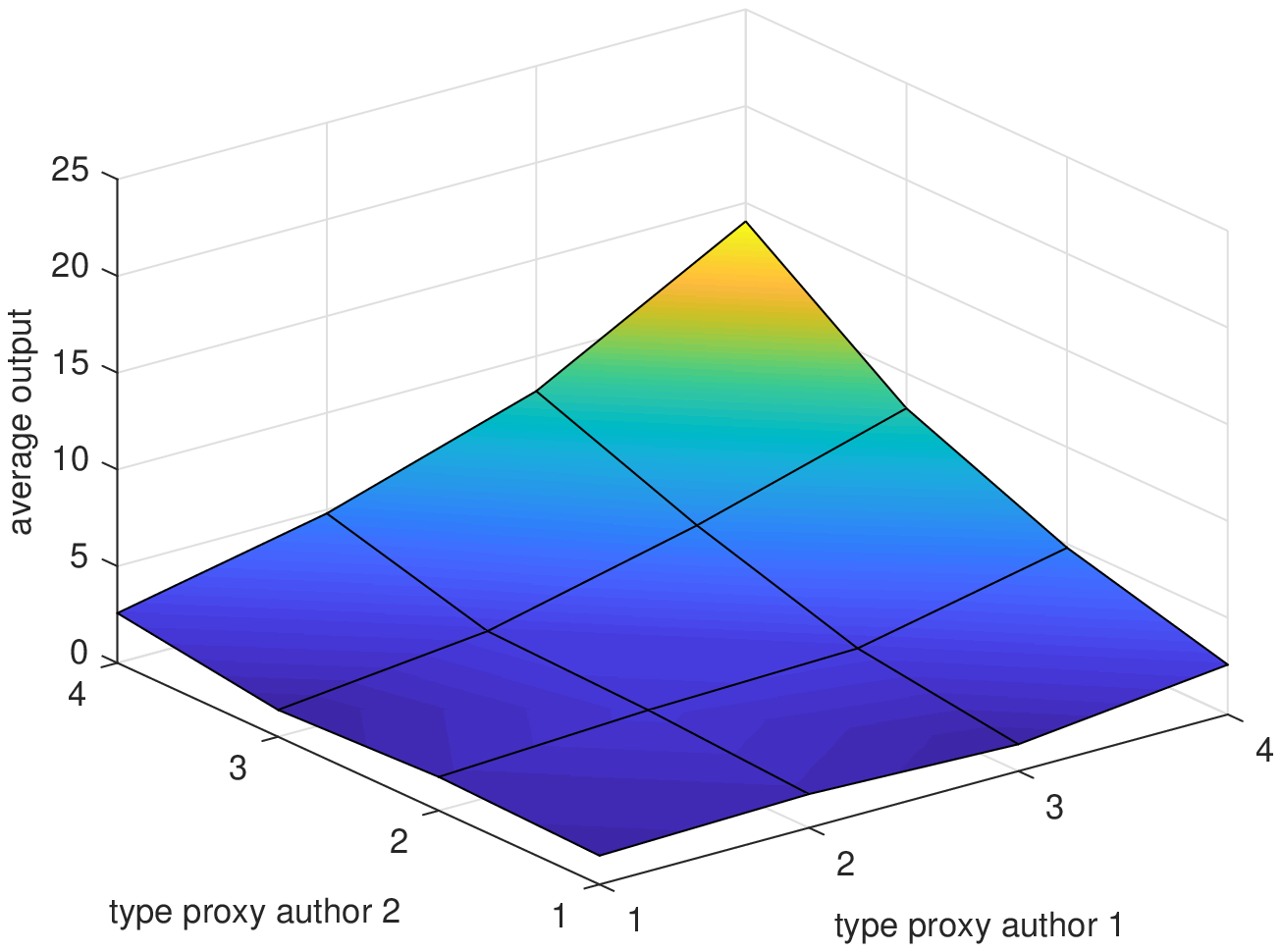}. \\\end{tabular}
	\end{center}
	\par
	\textit{{\footnotesize Notes: Subsample of authors who produce at least 5 articles on their own. I compute type proxies as quartiles of average sole-authored output. In panel (a) I show the proportions of type proxies for authors producing together in a 2-author team. In panel (b) I show average output (i.e., average journal quality) for different combinations of the type proxies.}}
\end{figure}

Given this proxy of type, I focus on 2-author teams and report two sets of results, which I interpret as reflecting sorting and complementarity, respectively. In panel (a) of Figure \ref{fig_prelim}, I show the proportions of every type proxy pairs in the sample. The percentages suggest that higher-type authors, who produce higher-quality articles on their own, tend to work together, implying the presence of sorting between authors of similar productivity. In panel (b) of Figure \ref{fig_prelim}, I compute the average output for every type proxy pairs. Authors who produce higher-quality articles on their own tend to produce higher-quality output as well when they work in teams, which is consistent with the presence of author heterogeneity. In addition, joint output is highest when authors with the highest type proxies (i.e., in the top quartile of sole-authored production) work together, and the figure is suggestive of the presence of complementarity. Since the additive model (\ref{eq_prod_fun_add}) rules those out, this motivates estimating a nonlinear model of academic production.

\begin{figure}[tbp]
	\caption{Nonlinear model estimates, economic researchers ($K=4$) \label{fig_RE}}
	\begin{center}
		\begin{tabular}{cc}
			(a) Sorting & (b) Heterogeneity \& complementarity \\
			\includegraphics[width=60mm, height=60mm]{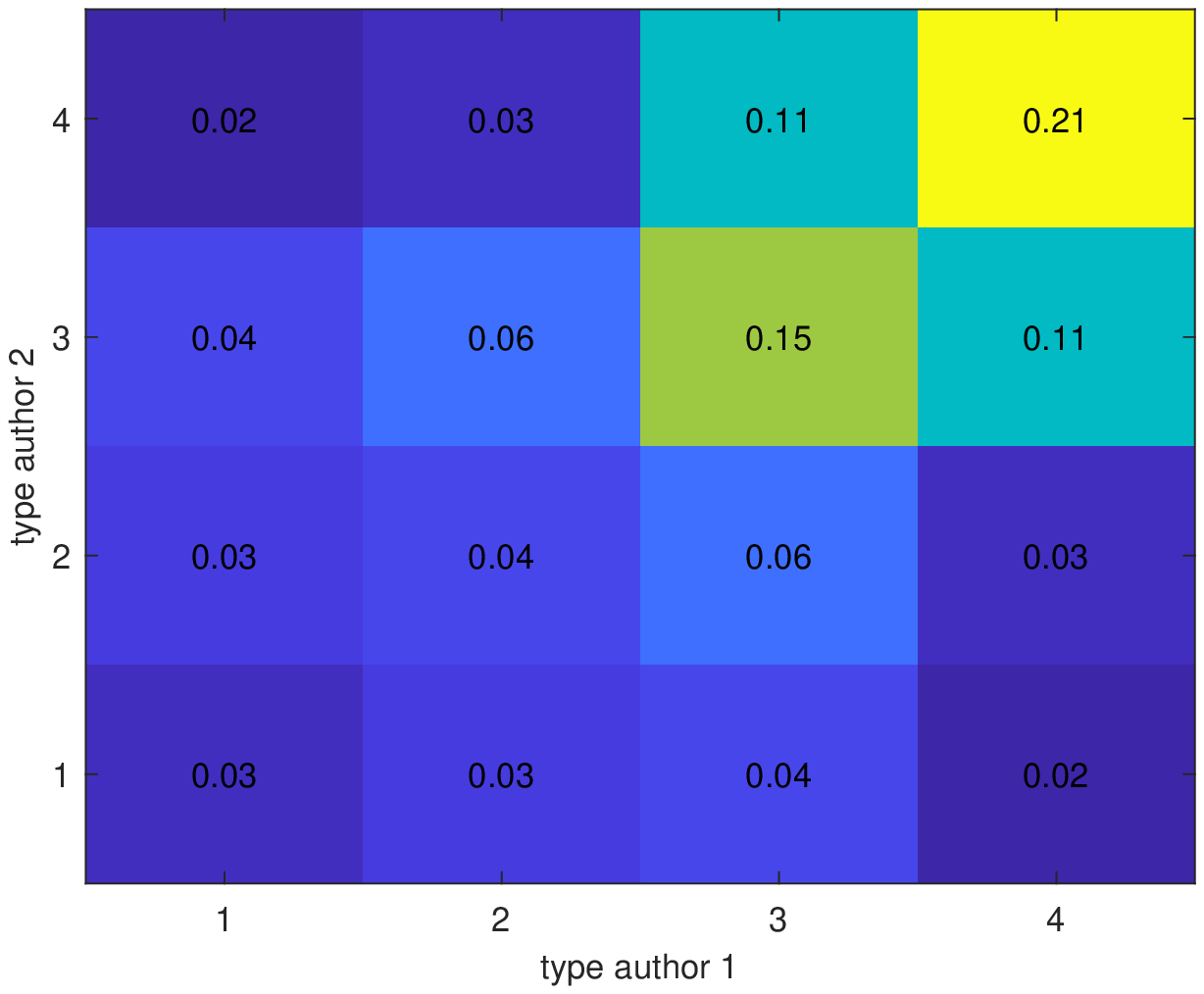}&\includegraphics[width=60mm, height=60mm]{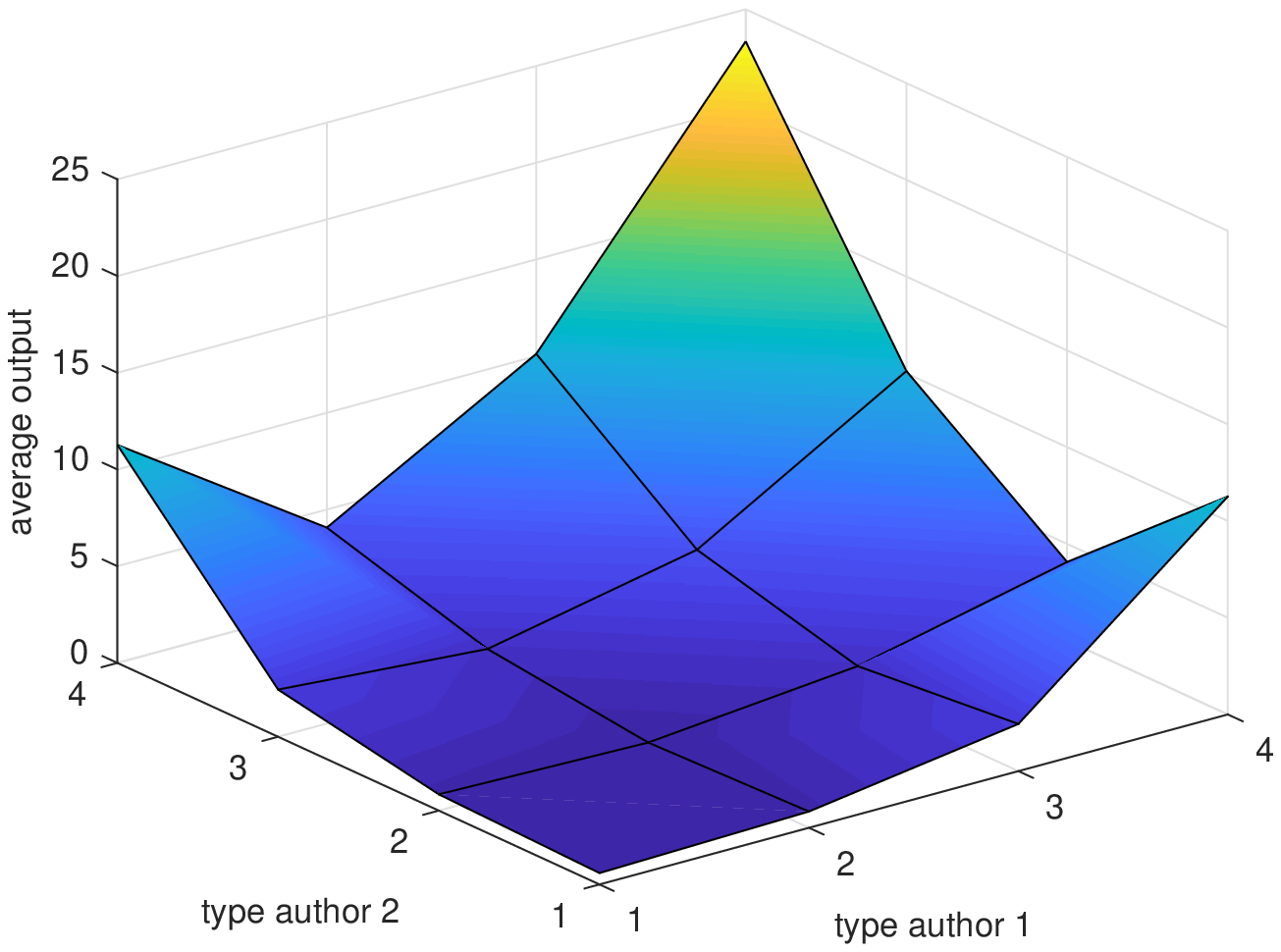}. \\\end{tabular}
	\end{center}
	\par
	\textit{{\footnotesize Notes: Random-effects estimates of a finite mixture model with $K=4$ types. In panel (a) I show the proportions of types for authors producing together in a 2-author team. In panel (b) I show average output (i.e., average journal quality) for different combinations of the types. Descriptive statistics on the sample are given in panel (b) of Table \ref{Tab_desc_1}.}}
\end{figure}

\paragraph{Estimates of the nonlinear model, baseline specification.}

I now report estimates of a finite mixture model with $K$ types, varying $K$ between $2$ and $6$. In this analysis I only consider 1-author and 2-author productions; see panel (b) of Table \ref{Tab_desc_1} for summary statistics on the sample. I model the output distribution as a log-normal, with mean and variance that depend on the types of authors in the team: for 1-author teams mean and variance are functions of the type, and for 2-author teams mean and variance are symmetric functions of the two types. The log-normal specification is restrictive, and in future work it will be important to check how robust the results are to removing this functional form assumption. I use mean-field variational EM for estimation.\footnote{I declare convergence when the increment in evidence lower bound is less than $10^{-3}$. To alleviate local maxima issues, I started the variational EM algorithm from different sets of parameters. The values that I obtained upon reaching the tolerance threshold differed somewhat, however they had similar implications for heterogeneity, sorting and complementarity. }     

For conciseness, I only comment in detail on the results for $K=4$. In this case, the type proportions are $15\%$, $20\%$, $36\%$, and $29\%$. The means of the log-normal output distribution are, in 1-author teams:
$$\left(\begin{array}{c}0.59\\
0.80\\
2.56\\
14.94\end{array}\right),$$
and in 2-author teams:
$$\left(\begin{array}{cccc}0.59  &  0.84  &  2.44 &   11.27\\
0.84 &   0.59 &   1.61&   4.07\\
2.44  &  1.61   & 3.81  & 10.12\\
11.27   & 4.07 & 10.12 &  23.34\\
\end{array}\right).$$
This suggests that authors have heterogeneous productivity levels, and these differences affect both 1-author and 2-author productions. 

To visualize the implications of the estimated model for heterogeneity, sorting and complementarity, in Figure \ref{fig_RE} I report similar quantities as in Figure \ref{fig_prelim}, except that I now use the types estimated under the model instead of the type proxies. Panel (a) of Figure \ref{fig_RE} shows that high-type authors have a stronger propensity to work together, although this evidence of sorting is less pronounced than in Figure \ref{fig_prelim}.

Panel (b) of Figure \ref{fig_RE} suggests the presence of complementarity, since the return of two high types working together is 1.5 standard deviations above the return of two low or middle types working together. This suggests gains from complementarity in economic research, which are reflected in stronger sorting at the top. In addition, the results show that sole-authored productivity and productivity in teams are not one-to-one: while the lowest type produces slightly lower quality output than the second-lower type on her own, she benefits substantially more from working with others, in particular with a high-type co-author.  

In addition to documenting the patterns of heterogeneity, sorting and complementarity, the nonlinear model can be used to refine the variance decomposition in (\ref{eq_vardec}). Indeed, the nonlinear model accounts for interaction effects between co-authors, over and beyond the additive effects of author types. This gives a fourth variance component, which I will refer to as ``nonlinearities'', to report in the decomposition, in addition to the contributions of heterogeneity, sorting, and other factors. This component can be computed as the difference between the variance of the mean of $Y_{nj}$ given the worker types -- which depends on both additive and interactive terms in general -- and the variance of its best linear approximation.  

\begin{table}[tbp]
	\caption{Nonlinear production, economic researchers\label{Tab_nonlin_RE}}
	\begin{center}
		\begin{tabular}{||l||cc|cc|cc|cc||}\hline\hline
			& \multicolumn{2}{c}{$K=2$} & \multicolumn{2}{c}{$K=4$} &\multicolumn{2}{c}{$K=5$} &\multicolumn{2}{c||}{$K=6$} \\
			& $n=1$ & $n=2$ & $n=1$ & $n=2$ & $n=1$ & $n=2$ & $n=1$ & $n=2$\\\hline\hline
			Total variance & 97.81&162.64& 97.81&162.64& 97.81&162.64& 97.81&162.64\\
			Heterogeneity & 2.18& 13.84&35.71 &46.85 & 42.77 & 51.80 & 35.52 & 49.86\\
			Sorting & -&  1.65& -&16.87& -&19.66& -&17.09\\
			Nonlinearities  & -& 1.36& -&3.92& -&2.80& -&3.48\\
			Other factors &95.63 & 145.79&62.11 & 90.00 & 55.05 & 88.39& 62.29 & 92.22\\\hline\hline		
		\end{tabular}
	\end{center}
	{\footnotesize \textit{Notes: Estimates of variance components for different team sizes $n$ in the nonlinear model with $K$ types. Descriptive statistics on the sample are given in panel (b) of Table \ref{Tab_desc_1}.}}
\end{table}

I report the results of the variance decomposition in Table \ref{Tab_nonlin_RE}, for $K=2,4,5,6$. While $K=2$ seems to allow for too little heterogeneity, the results for $K=4,5,6$ are broadly similar to one another. They show that author heterogeneity explains approximately 30\% of output variance, and sorting explains between 10\% and 12\%. Despite different type modeling and form of the production function, these variance shares are not too different from those based on the additive model (see Table \ref{Tab_AKM_RE}). In addition, nonlinearities explain between 1.5\% and 2.5\% of the variance. While this share may seem small, the presence of complementarity at the top of the productivity distribution, as shown by panel (b) of Figure \ref{fig_RE}, is likely to be important to study policies that re-allocate researchers across teams.

\begin{figure}[tbp]
	\caption{Optimal allocation, economic researchers ($K=4$) \label{fig_RE_alloc}}
	\begin{center}
		\begin{tabular}{cccc}
			\includegraphics[width=60mm, height=60mm]{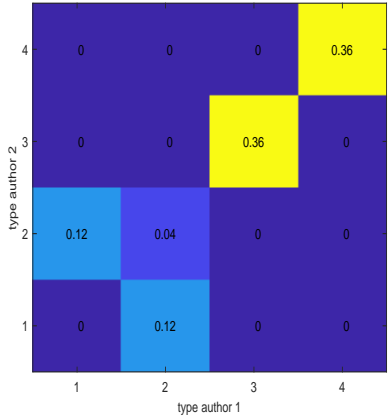} \\\end{tabular}
	\end{center}
	\par
	\textit{{\footnotesize Notes: Proportions of types for authors producing together in a 2-author team, in the allocation that solves (\ref{max_surplus}). Descriptive statistics on the sample are given in panel (b) of Table \ref{Tab_desc_1}.}}
\end{figure}

To assess how complementarities might affect worker allocations, I compute an allocation that maximizes total output by solving (\ref{max_surplus}). I compute expected payoffs and ``budgets'' based on the estimates of the nonlinear model with $K=4$. In this case, when budget sizes are even, an optimal integer-valued allocation can be computed by solving a linear program that does not impose the integral constraints. The resulting allocation in Figure \ref{fig_RE_alloc} shows sorting at the top of the type distribution, but not at the bottom.

\begin{figure}[tbp]
	\caption{Nonlinear model estimates and optimal allocation, economic researchers, other specifications ($K=4$) \label{fig_RE_spec}}
	\begin{center}
		\begin{tabular}{ccc}
			\multicolumn{3}{c}{1. Correlated RE}\\
			(1a) Sorting & (1b) Heterogeneity \& complementarity & (1c) Optimal allocation  \\
			\includegraphics[width=40mm, height=40mm]{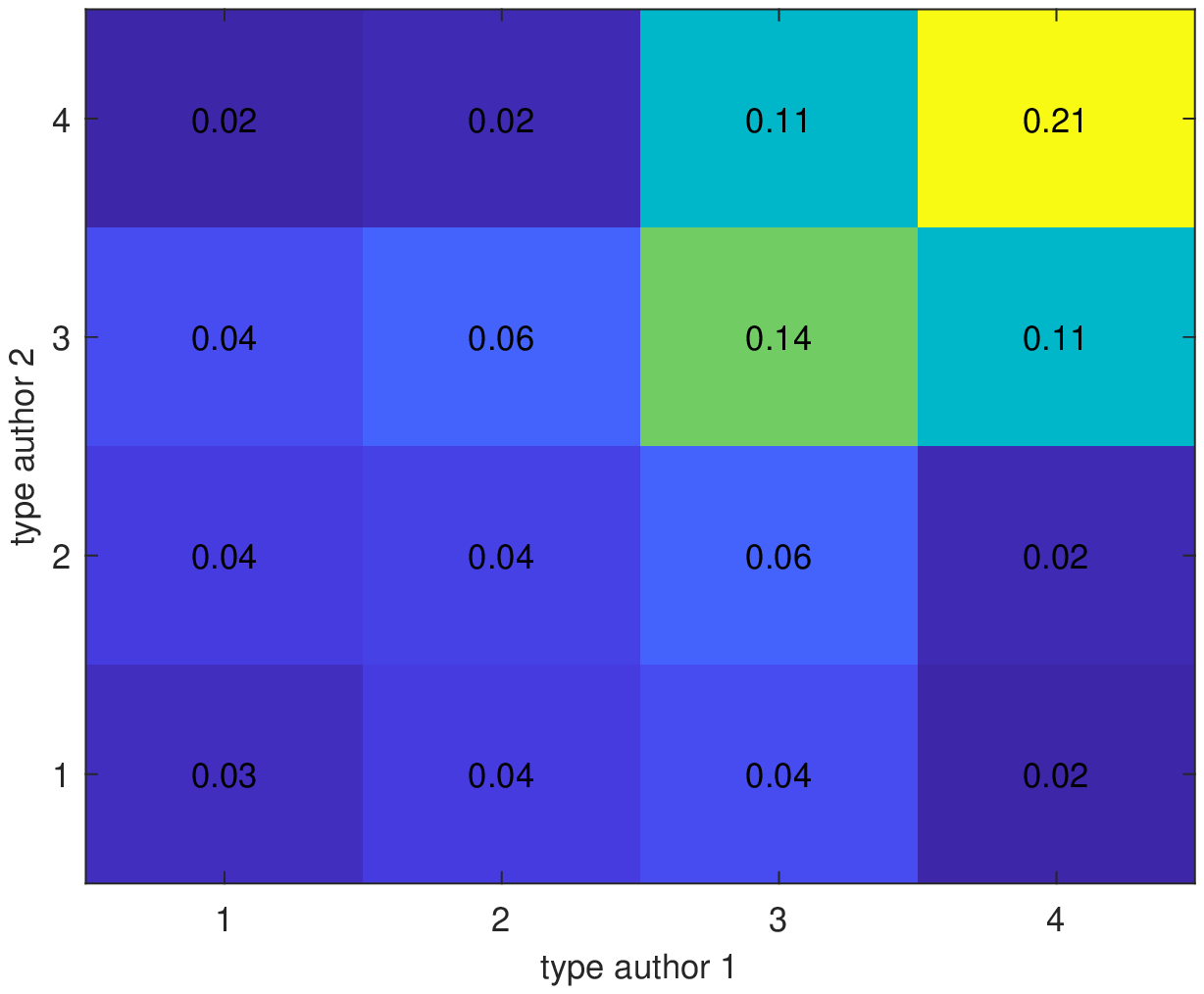}&\includegraphics[width=40mm, height=40mm]{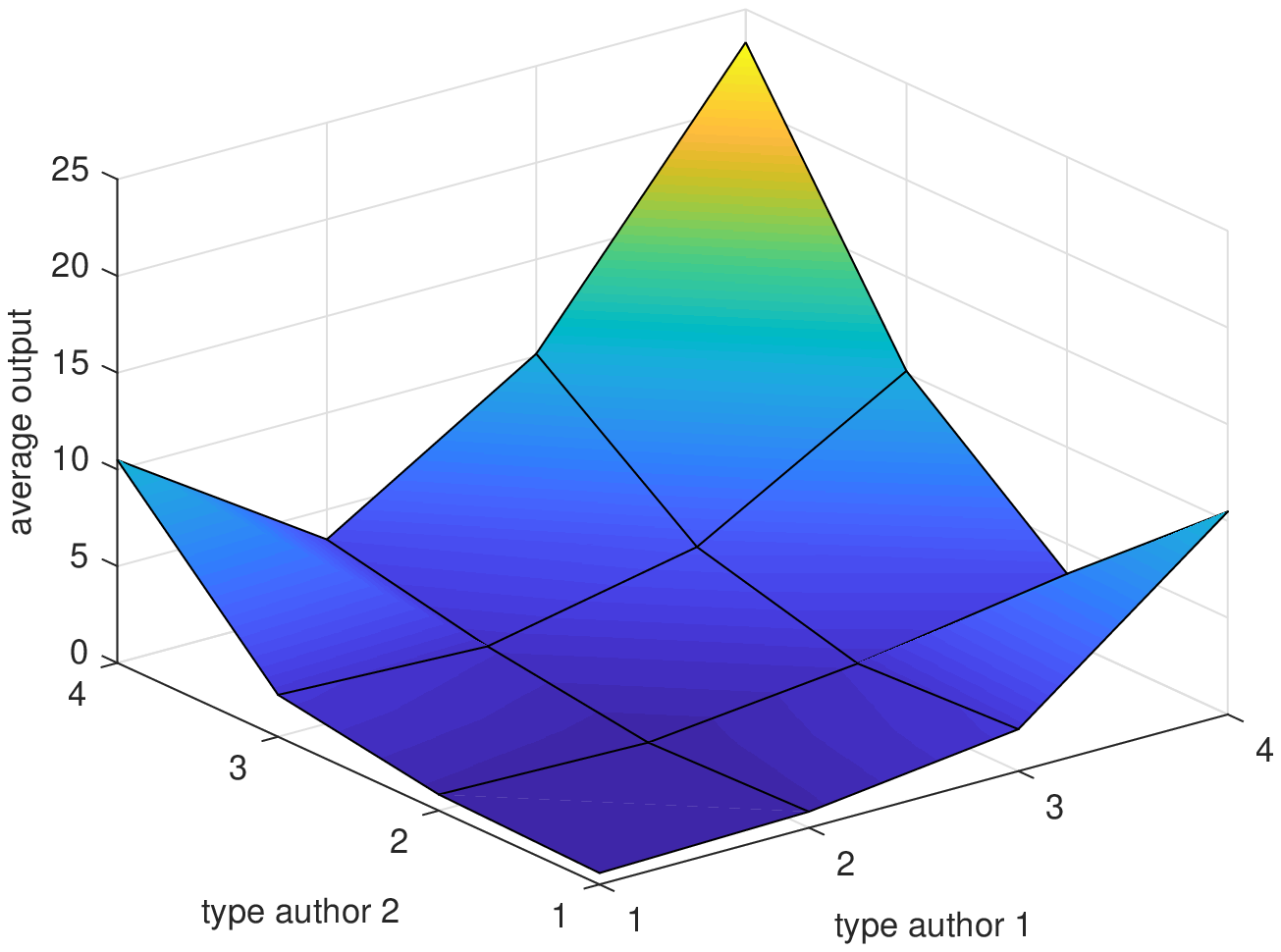} & \includegraphics[width=40mm, height=40mm]{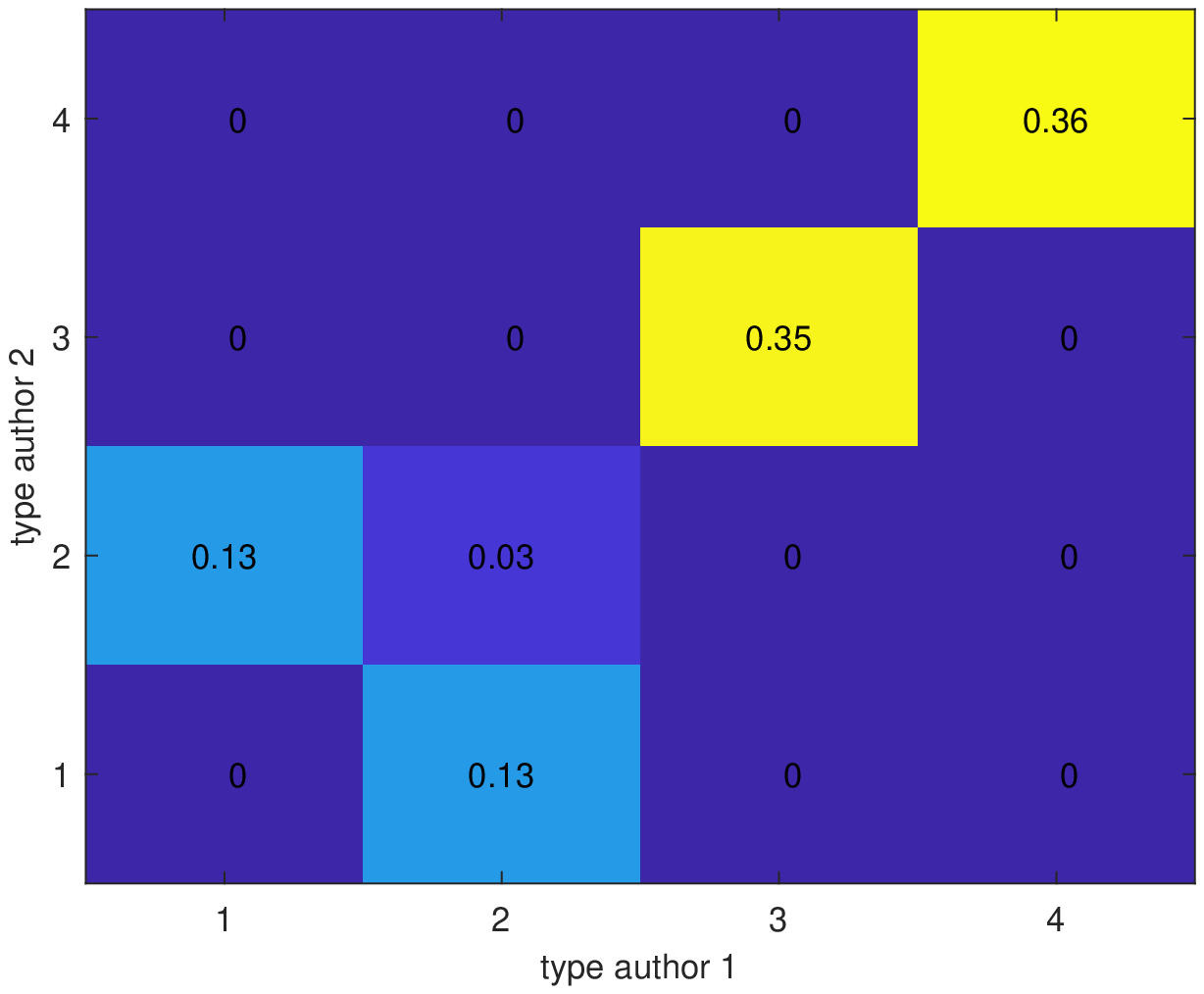}\\
				\multicolumn{3}{c}{2. Joint RE}\\
				(2a) Sorting & (2b) Heterogeneity \& complementarity & (2c) Optimal allocation \\
				\includegraphics[width=40mm, height=40mm]{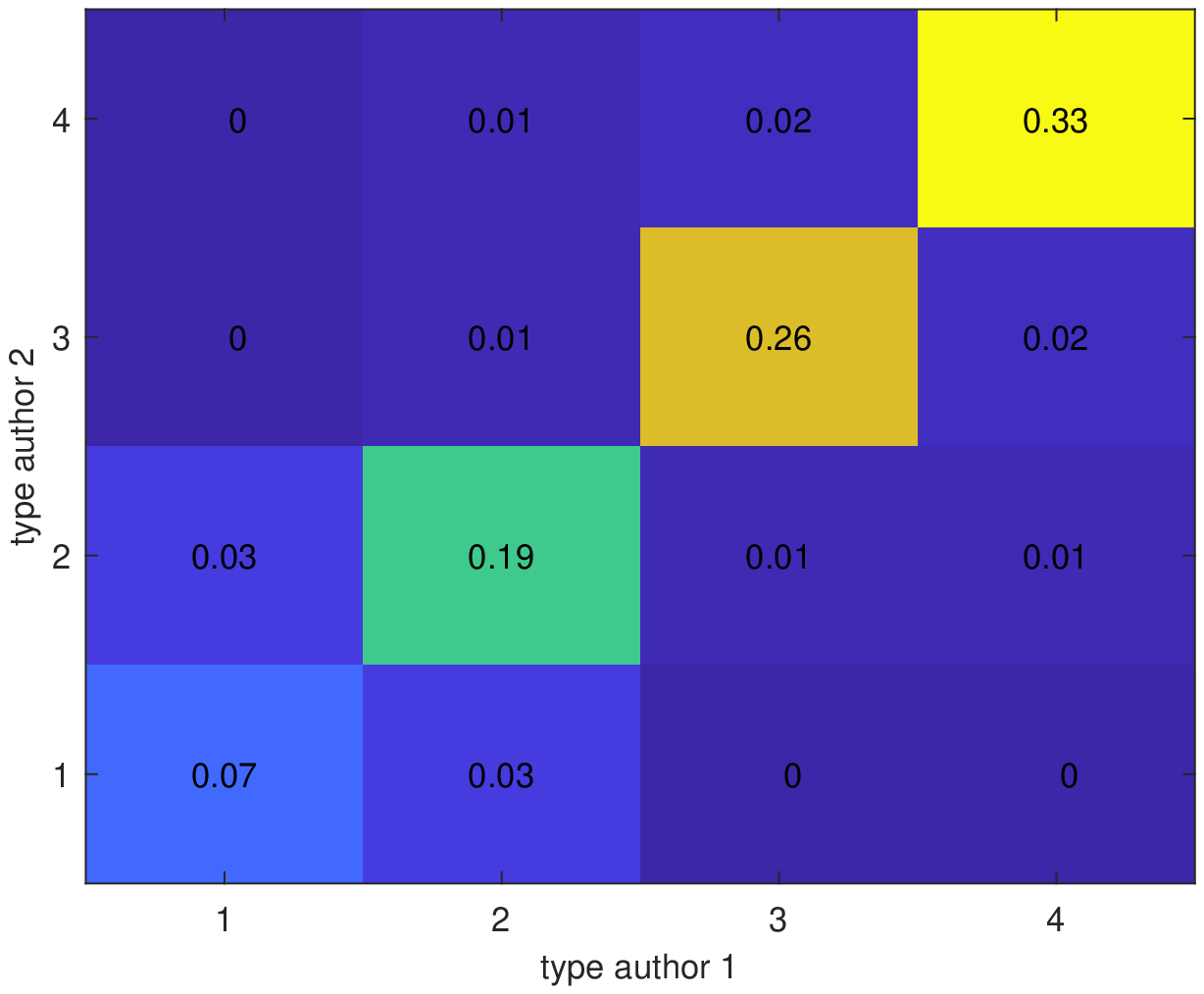}&\includegraphics[width=40mm, height=40mm]{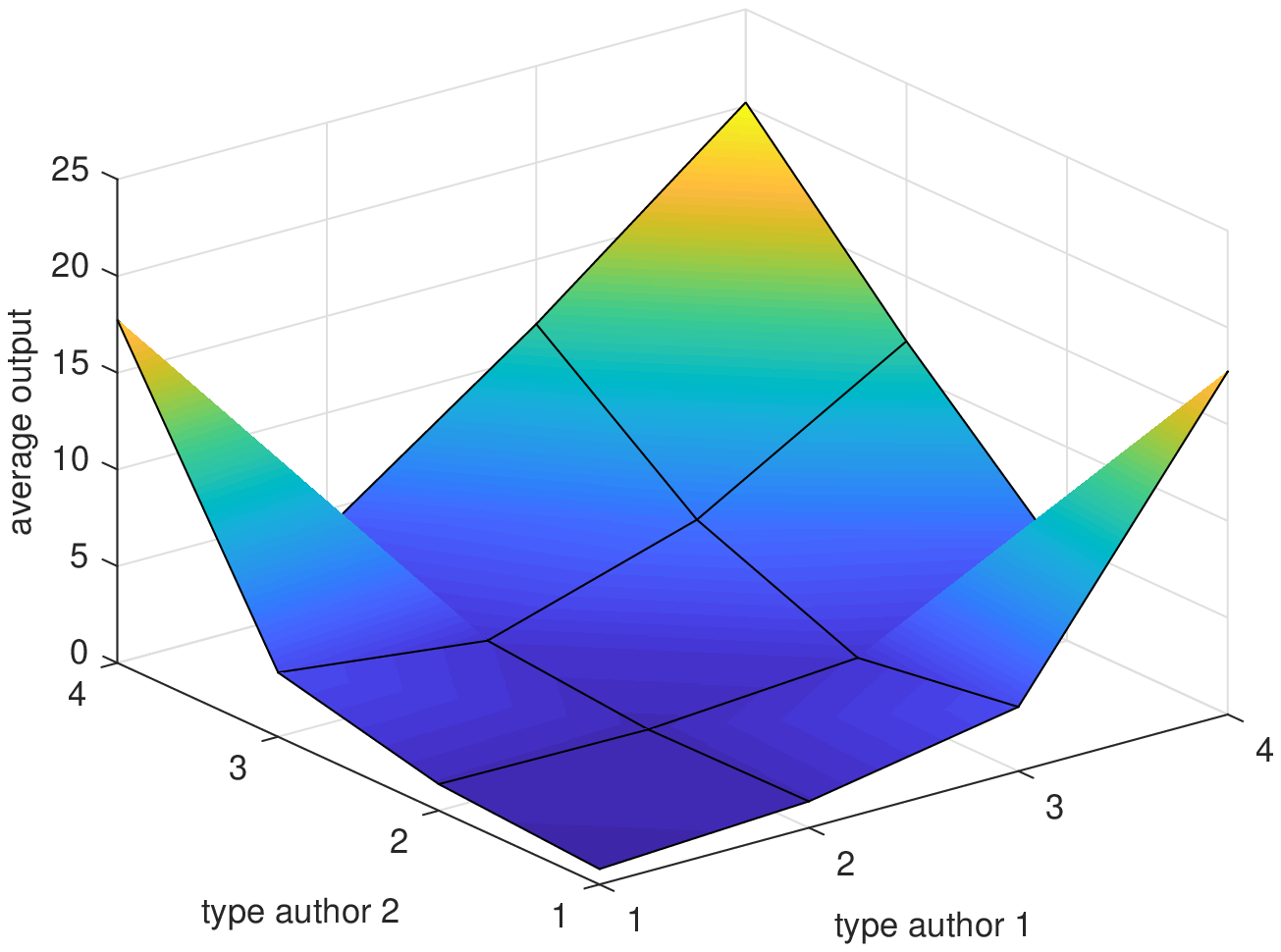} & \includegraphics[width=40mm, height=40mm]{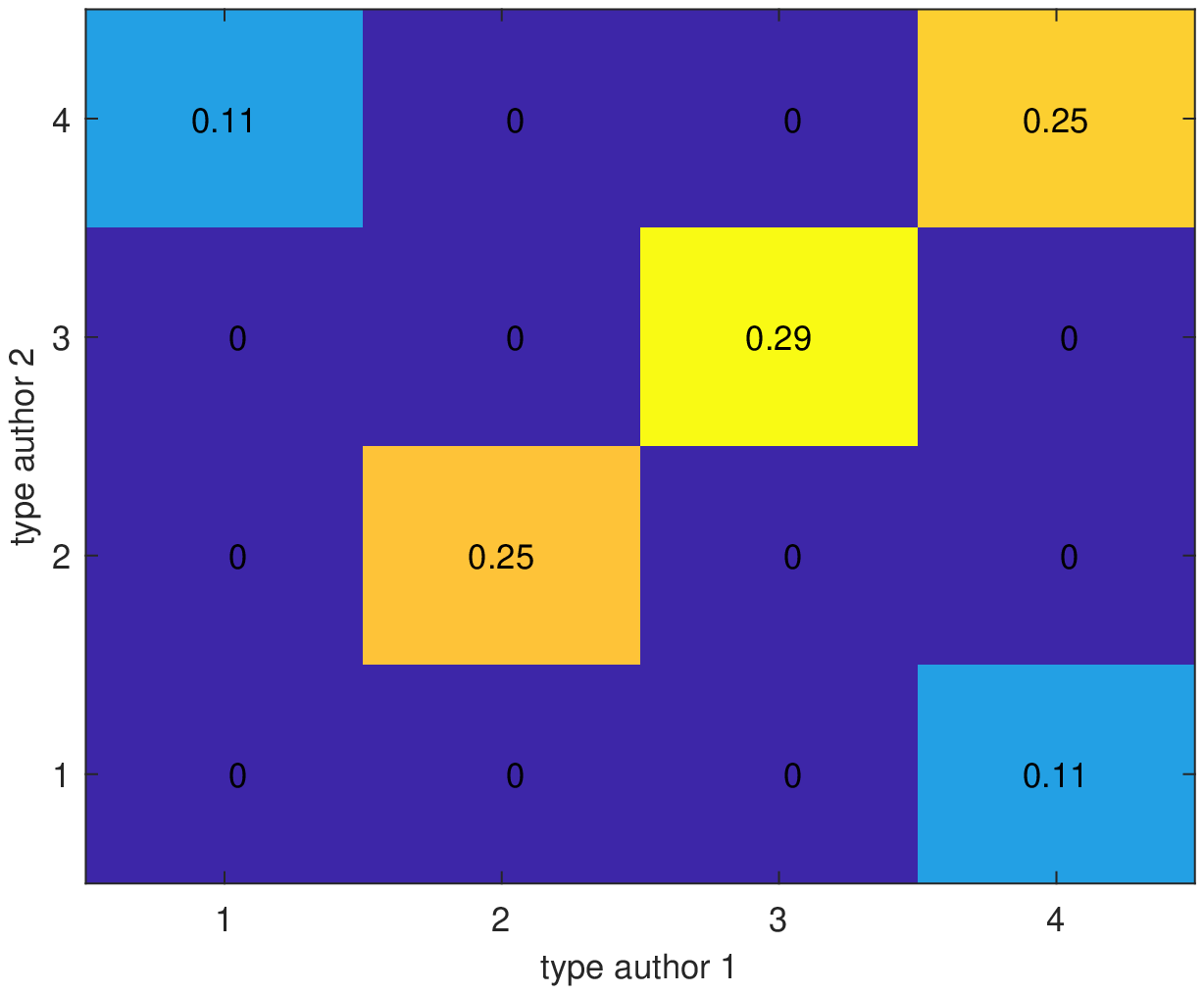}\\
				\multicolumn{3}{c}{3. 2-author only}\\
				(3a) Sorting & (3b) Heterogeneity \& complementarity & (3c) Optimal allocation \\
				\includegraphics[width=40mm, height=40mm]{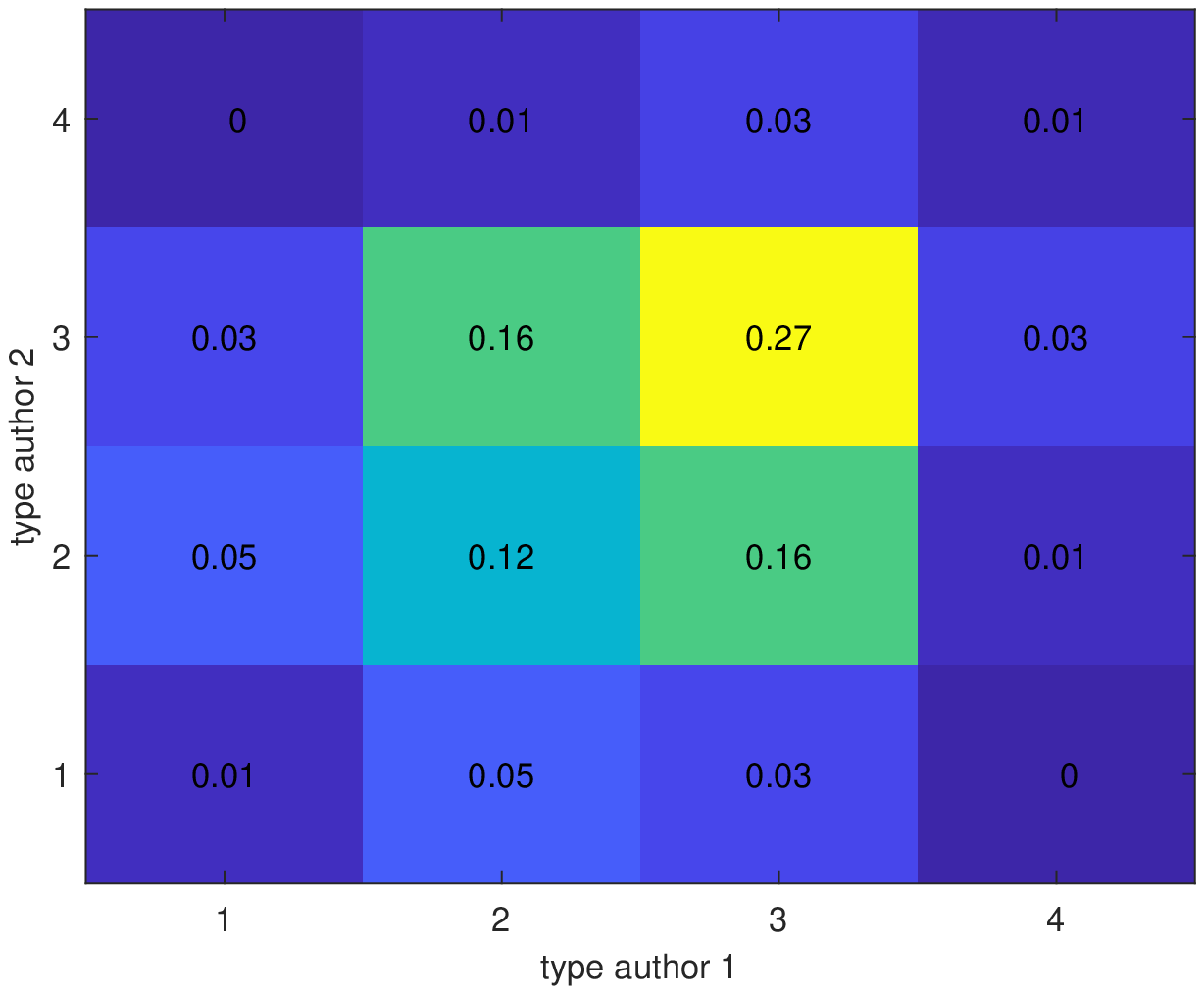}&\includegraphics[width=40mm, height=40mm]{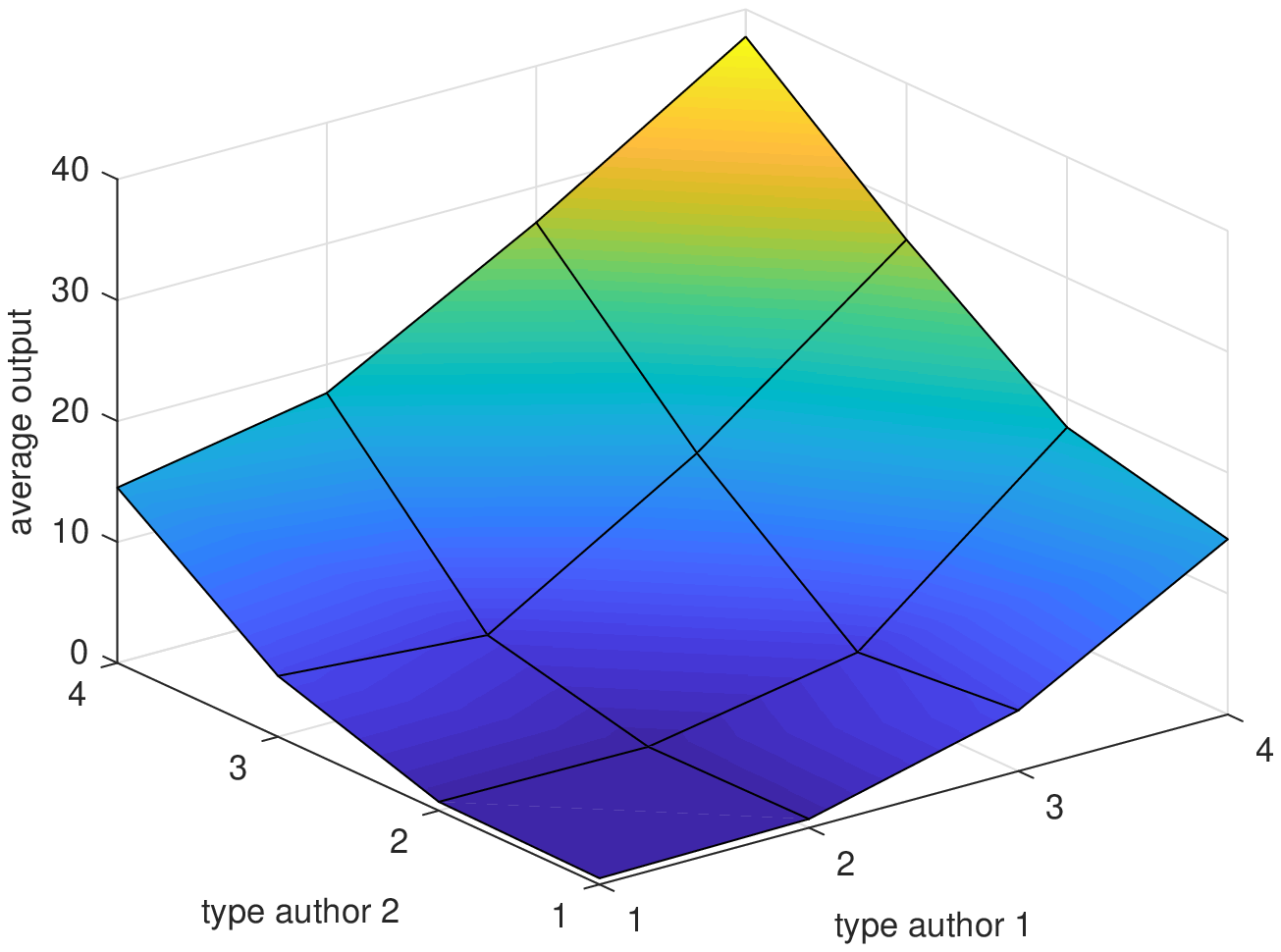}& \includegraphics[width=40mm, height=40mm]{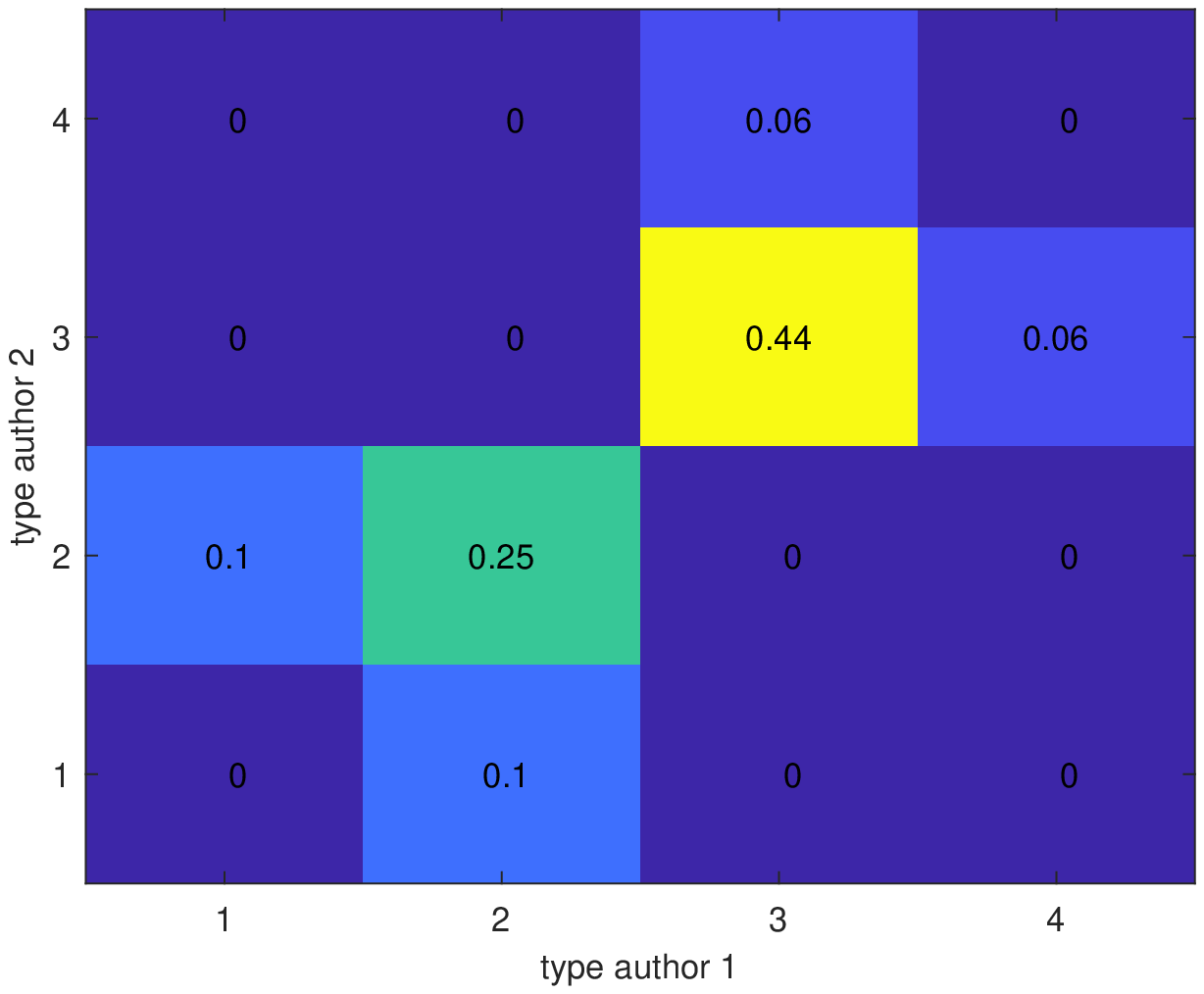} \\\end{tabular}
	\end{center}
	\par
	\textit{{\footnotesize Notes: Random-effects estimates of a finite mixture model with $K=4$ types. In panels (a) I show the proportions of types for authors producing together in a 2-author team. In panels (b) I show average output (i.e., average journal quality) for different combinations of the types. In panels (c) I show the proportions of types for authors producing together in a 2-author team, in the allocation that solves (\ref{max_surplus}). ``Correlated RE'' comes from a model where the types follow a multinomial logit distribution given the numbers of 1-author and 2-author collaborations at the extensive and intensive margins. ``Joint RE'' comes from a model where collaborations follow independent type-specific Poisson distributions. ``2-author only'' only uses information from 2-author teams. Descriptive statistics on the sample are given in panel (b) of Table \ref{Tab_desc_1}.}}
\end{figure}

 \paragraph{Other specifications.}
 
 I now present the results of three exercises based on other specifications of the nonlinear model. In the baseline specification, the distribution of author types does not depend on collaborations. To assess the impact of this assumption, I estimate two alternative specifications. In the first specification, I model the type distribution as a function of author characteristics, using a multinomial logit specification. As characteristics, I use the numbers of teams in which author $i$ participates, separately for 1-author and 2-author teams, as well as the indicators that these numbers are zero. In panels (1a) and (1b) of Figure \ref{fig_RE_spec}, I show type-pair proportions and mean output by pairs of types, and in panel (1c) I show the optimal allocation obtained by solving (\ref{max_surplus}). In Appendix Table \ref{Tab_nonlin_RE_spec} I show the variance decomposition estimates. The results are similar to the ones from the baseline specification.

   In the second specification, I augment the team production model with a model of team formation, and I jointly estimate the parameters using a variational EM algorithm. I specify team formation using a stochastic blockmodel for hypergraphs, where 1-author and 2-author collaborations follow independent Poisson distributions, with a parameter that is type-specific in the first case and pair-of-types-specific in the second case. In this model, as in most models of team formation, the conditional distribution of $\{\alpha_i\, :\, i\}$ given $\{(i_1(j),...,i_n(j))\, :\, (n,j)\}$ does not factor across $i$. In addition, in such a model, author types do not only influence heterogeneity in research quality, but also in quantity. The results show some differences with the baseline. In particular, panel (2a) of Figure \ref{fig_RE_spec} shows stronger evidence of sorting. In panel (2b), I show mean output in 2-author teams, by type pairs. The estimates, and the implied optimal allocation in panel (2c), are again consistent with complementarity between high-type authors.\footnote{However, note that the high degree of sorting prevents one from assessing with confidence the gains from collaboration between highest and lowest types (i.e., between type-1 and type-4 authors): while the point-estimates in panel (2b) are large, panel (2a) shows that such collaborations are virtually non-existent. This is likely to affect the optimal allocation numbers shown in panel (2c) of Figure \ref{fig_RE_spec}.} In Appendix Table \ref{Tab_nonlin_RE_spec} I report variance decomposition estimates. In 2-author teams, heterogeneity accounts for a lower share of variance compared to the baseline (22\% versus 29\%), sorting accounts for a larger share (19\% versus 10\%), and the variance share accounted for by nonlinearities is smaller than in the baseline.

   In a last exercise, I report estimates based on 2-author teams only. Hence, sole-authored productions are discarded. Panel (3a) of Figure \ref{fig_RE_spec} shows a higher degree of sorting in the middle of the type distribution, rather than at the top. Panel (3b) again shows evidence of complementarity, although the patterns differ somewhat from the baseline, as shown by the implied optimal allocation in panel (3c). Lastly, the variance decomposition results in Appendix Table \ref{Tab_nonlin_RE_spec} show that, in this case, heterogeneity accounts for 32\% of variance, and sorting accounts for 5\% (as opposed to 29\% and 10\% in the baseline).

\section{Inventors and innovation\label{Inventors_sec}}

In this section I apply the method to study patents and inventors.

\subsection{Data}

I use data from Akcigit \textit{et al.} (2016). Their main source is the disambiguated inventor data of Li \textit{et al.} (2014), which identifies unique inventors in the USPTO data. I restrict the analysis to US patents, which were granted between 1995 and 1999. Throughout, I focus on patents in the technology class ``Computers and Communications''.\footnote{Among all inventors who patent in this class, 80\% patent only in this class during the period.} For the output measure, I follow Akcigit \textit{et al.} (2016) and use Hall \textit{et al.}'s (2001) measure of patent quality, which is a truncation-adjusted measure of forward citations of the patent. From this measure I net out multiplicative year fixed-effects, using 1999 as the reference year. In this illustration, I do not use information about the firms where inventors work, although it would be interesting to incorporate this information in future research.

\begin{table}[h!]
	\caption{Descriptive statistics on patents and inventors\label{Tab_desc_pat}}
	\begin{center}
		\begin{tabular}{||l||c|c|c|c||c|c|c||}\hline\hline
				&\multicolumn{4}{c}{(a) Sample}& \multicolumn{3}{c||}{(b) Subsample}\\
			&All & $n=1$ & $n=2$ & $n\geq 3$&All & $n=1$ & $n=2$\\\hline\hline
			\# Inventors & 5896&4328 & 4267&3814& 1871&1719 & 1011\\
			\# Patents & 30068& 17261&7740 &5067& 12793& 10520&2273\\
			Mean output & 33.96& 31.36&33.03&37.06& 29.67& 29.26&30.69\\
			Std output &40.34 & 38.98& 38.97&42.40&35.41 & 35.55& 35.09\\
			p10\% output & 4.43& 4.23& 4.43&5.46& 3.96& 3.96& 4.23\\
			p50\% output &21.72 &19.02 & 21.13&24.23 &19.02 &18.46 & 19.34\\
			p90\% output & 76.08
			&69.74 & 73.97 &81.76&65.52
			&65.47& 65.94\\
			p95\% output &105.67 &100.05& 103.56&111.13&92.99 &92.25& 95.10\\
			p99\% output &192.83&189.46& 196.47&192.83&175.42&175.41& 175.41\\
			p10\% \#patents &5 &2 & 2&3 &5 &4 & 2\\
			p50\% \#patents & 9& 5& 5&6& 8& 7& 6\\
			p90\% \#patents & 25&13 & 11&16& 21&16 & 15\\
			p95\% \#patents & 32 & 19& 16&22& 29 & 26& 17\\
			p99\% \#patents &68 & 37& 26&42&62 & 44& 27\\\hline\hline		
		\end{tabular}
	\end{center}
	{\footnotesize \textit{Notes: Sample from Akcigit \textit{et al.} (2016), where the patent is from the US and granted between 1995 and 1999, and belongs to the class ``Computers and Communications''. Output is the truncation-adjusted measure of forward citations of Hall \textit{et al.} (2001), net of multiplicative year effects (reference year is 1999). Percentiles of the distribution of number of patents per inventor are indicated in the bottom five rows. In panel (a) the sample is restricted to at least 5 patents per inventor during the period. In panel (b) the subsample is restricted to collaborations between at most two inventors, where all inventors have at least 5 patents during 1995-1999. I use the subsample in panel (b) to estimate nonlinear models.}}
\end{table}

%\begin{table}[h!]
%	\caption{Descriptive statistics on patents and inventors, subsample used for nonlinear models\label{Tab_desc_pat_2}}
%	\begin{center}
%		\begin{tabular}{||l||c|c|c||}\hline\hline
%			&All & $n=1$ & $n=2$ \\\hline\hline
%			\# Inventors & 1871&1719 & 1011\\
%			\# Patents & 12793& 10520&2273 \\
%			Mean output & 29.67& 29.26&30.69\\
%			Std output &35.41 & 35.55& 35.09\\
%			p10\% output & 3.96& 3.96& 4.23\\
%			p50\% output &19.02 &18.46 & 19.34\\
%			p90\% output &65.52
%			&65.47& 65.94 \\
%			p95\% output &92.99 &92.25& 95.10\\
%			p99\% output &175.42&175.41& 175.41\\
%			p10\% \#patents &5 &4 & 2\\
%			p50\% \#patents & 8& 7& 6\\
%			p90\% \#patents & 21&16 & 15\\
%			p95\% \#patents & 29 & 26& 17\\
%			p99\% \#patents &62 & 44& 27\\\hline\hline		
%		\end{tabular}
%	\end{center}
%	{\footnotesize \textit{Notes: See notes to Table \ref{Tab_desc_pat}. The sample is restricted to collaborations between at most two inventors, where all inventors have at least 5 patents during 1995-1999.}}
%\end{table}

 In panel (a) of Table \ref{Tab_desc_pat}, I provide summary statistics about the sample, where I restrict inventors to participate in at least five patents during the period. In this setting also, output and the number of collaborations are right-skewed. The original sample of US patents in the class ``Computers and Communications'' that were granted between 1995 and 1999 contains 65848 inventors and  62927 patents. Hence, imposing that all inventors be on at least 5 patents restricts the sample size substantially. In the sample, team size exhibits a range of variation but teams tend to be small: 57\% of teams are 1-inventor teams, 26\% of teams have 2 members, and less than 1\% of teams have more than 6 members. In panel (b) of Table \ref{Tab_desc_pat}, I show summary statistics for the subsample of 1-inventor and 2-inventor teams that I will use to estimate the nonlinear models.

 \subsection{Additive model}
 
  I first estimate the additive model (\ref{eq_prod_fun_add}). Restricting the set of inventors and patents to ensure identification gives 5547 inventors and 29101 patents. I estimate the model by allowing for four team sizes $n$ in $\lambda_n$, where $n=4$ corresponds to all teams with at least four inventors. I find $\lambda_2=0.54$, $\lambda_3=0.39$, and $\lambda_{4}=0.29$. Hence, keeping inventor type(s) constant, patents with two inventors are cited 8\% more than patents with one inventor, 3-inventor patents are cited 17\% more, and patents with at least 4 inventors are cited 16\% more.

\begin{table}[tbp]
	\caption{Additive production, patents and inventors\label{Tab_AKM_PAT}}
	\begin{center}
		\begin{tabular}{||l||c|c|c||}\hline\hline
			& $n=1$ & $n=2$ & $n=3$ \\\hline\hline
			Total variance & 1519.3&1450.0& 1388.5\\
			Heterogeneity &578.2& 467.6&261.3 \\
			\color{gray}{Heterogeneity (uncorrected)} & 	\color{gray}{782.1}&	\color{gray}{710.7}&	\color{gray}{559.3}\\
			Sorting & -&  90.2& 169.6\\
			\color{gray}{Sorting (uncorrected)} & 	\color{gray}{-}&	\color{gray}{-25.9}& 	\color{gray}{-12.0}\\
			Other factors &941.1 & 733.1& 712.8\\
			\color{gray}{Other factors (uncorrected)} &	\color{gray}{737.2} & 	\color{gray}{727.3}& 	\color{gray}{792.0}\\	Team scale $\lambda_n$ & 1.00&  0.54& 0.39\\\hline\hline		
		\end{tabular}
	\end{center}
	{\footnotesize \textit{Notes: Estimates of variance components in equation (\ref{eq_vardec}), for different team sizes $n$. Descriptive statistics on the sample are given in panel (a) of Table \ref{Tab_desc_pat}.}}
\end{table}

In Table \ref{Tab_AKM_PAT}, I show the estimates of the variance components in (\ref{eq_vardec}), for $n=1,2,3$. The variance share explained by inventor heterogeneity is 38\% in patents with one inventor, 32\% in 2-inventor patents, and 19\% in 3-inventor patents. Sorting contributes 6\% of variance in 2-inventor patents, and 12\% in 3-inventor patents. Here also, the other factors $\varepsilon_{nj}$ account for the main share of the output variance. Overall, the variance decomposition estimates in the patent sample are thus not very different from those in the sample of economists, with a somewhat larger contribution of worker heterogeneity and a smaller contribution of sorting.

 In Appendix Table \ref{Tab_AKM_PAT_rob}, I augment the sample and require every inventor to produce at least one or two patents, as opposed to at least five in the baseline sample. In addition, as in all the other robustness checks in this application, using a Poisson regression I net out from the output the effects of year and ``inventor age'', as measured by the difference between the year of observation and the first year where the inventor produced a patent. Compared to the baseline estimates, the results in Appendix Table \ref{Tab_AKM_PAT_rob} show that, in these larger samples, inventor fixed-effects are estimated with more noise, and uncorrected variance components are very large in magnitude. The bias-corrected estimates indicate a larger role of inventor heterogeneity than in the baseline, and a negative sorting contribution, although, given the amount of noise, these findings should be interpreted with caution.

   \subsection{Nonlinear model}

%Map types to characteristics?

 \begin{figure}[tbp]
 	\caption{Nonlinear model estimates, patents and inventors ($K=4$) \label{fig_PAT}}
 	\begin{center}
 		\begin{tabular}{cc}
 			(a) Sorting & (b) Heterogeneity \& complementarity \\
 			\includegraphics[width=60mm, height=60mm]{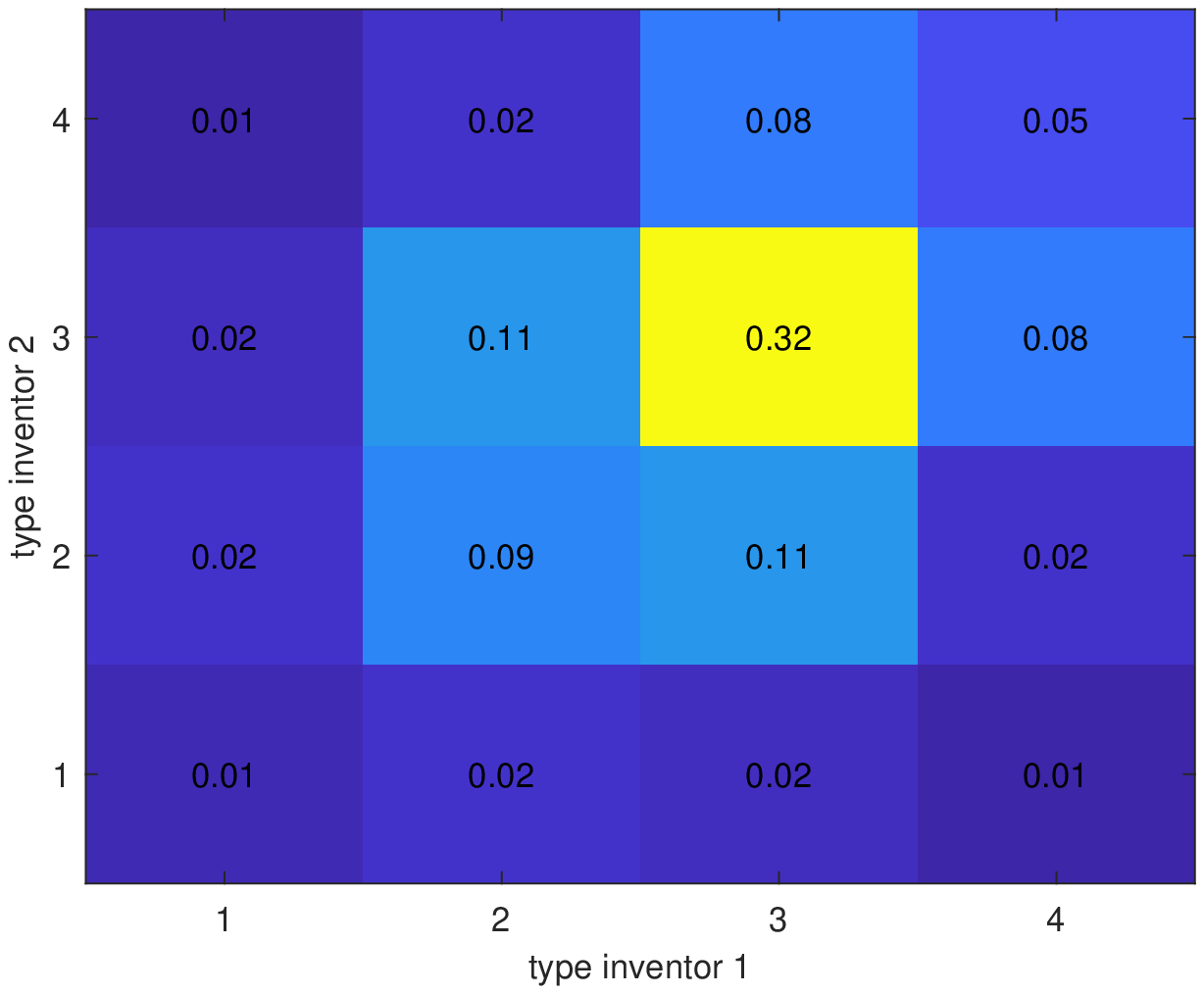}&\includegraphics[width=60mm, height=60mm]{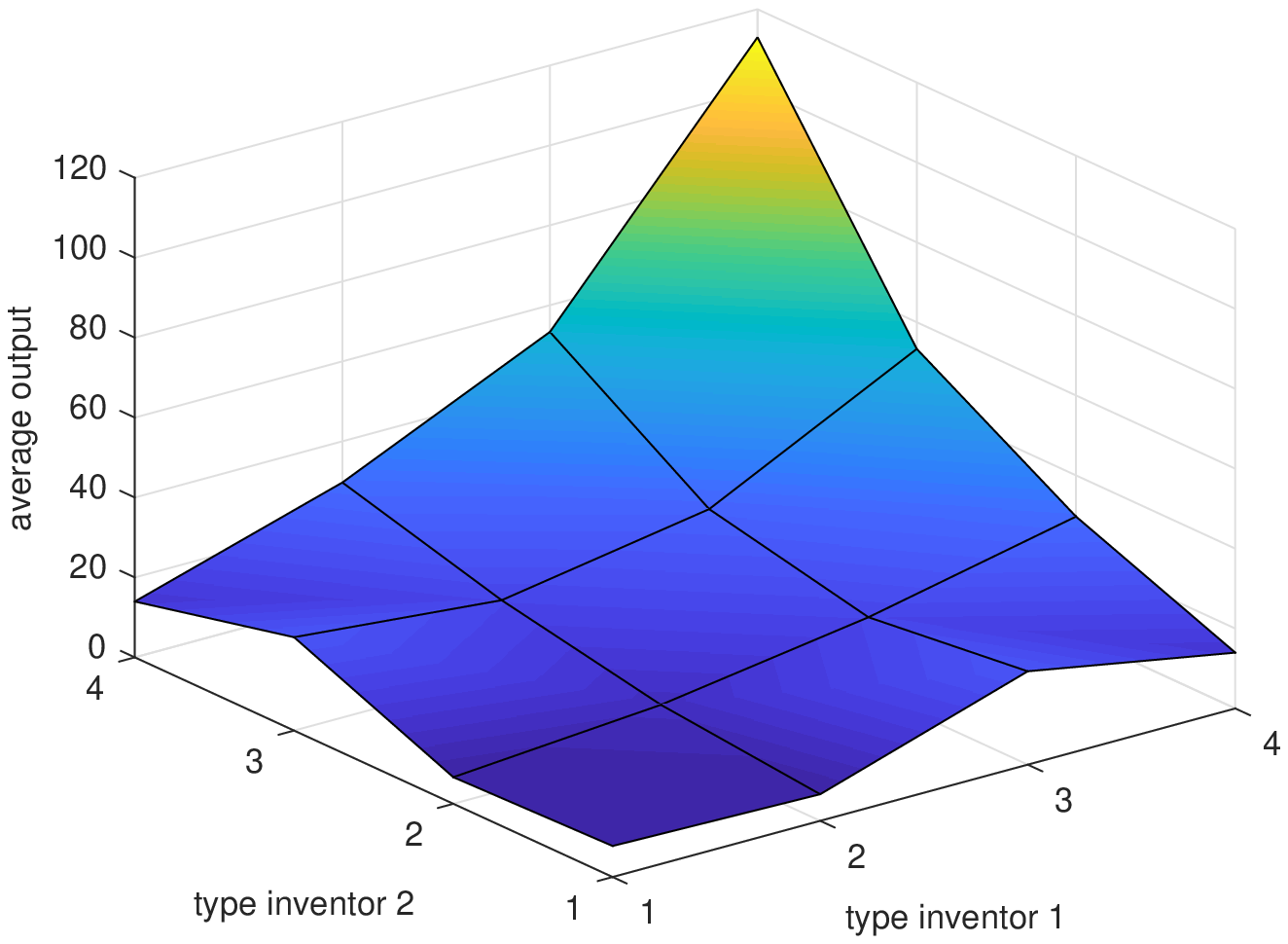}. \\\end{tabular}
 	\end{center}
 	\par
 	\textit{{\footnotesize Notes: Random-effects estimates of a finite mixture model with $K=4$ types. In panel (a) I show the proportions of types for inventors producing together in a 2-inventor team. In panel (b) I show average output (i.e., average truncation-adjusted forward citations) for different combinations of the types. Descriptive statistics on the sample are given in  panel (b) of Table \ref{Tab_desc_pat}.}}
 \end{figure}

To specify the nonlinear model,\footnote{In Appendix Figure \ref{fig_prelim_pat}, I construct type proxies as in Figure \ref{fig_prelim}. I group inventors according to the quartiles of the citations of their sole-authored patents, and restrict the sample to inventors with at least 5 patents on their own, so the sample is small, with 1554 inventors.  The figure suggests the presence of heterogeneity and sorting, although complementarity is less salient than for economists, and sorting seems more evenly spread out along the diagonal (compare with Figure \ref{fig_prelim}).} I model the output distribution as a negative binomial with mean and variance that depend on the types of inventors in the team. This parametric form allows for a convenient treatment of the zeros in the dependent variable. I now comment in detail on the results for $K=4$. In this case, the type proportions are $6\%$, $25\%$, $52\%$, and $17\%$. The means of the negative binomial output distribution are, in 1-inventor teams:
$$\left(\begin{array}{c}9.80\\
12.31\\
27.03\\
65.08\end{array}\right),$$
and in 2-inventor teams:
$$\left(\begin{array}{cccc}7.76 &   6.65  & 23.36 &   13.97\\
6.65 &  10.71 &   18.56 &   29.70\\
23.36 &  18.56 &  27.30 &  53.32\\
13.97  & 29.70 &  53.32 & 112.99\end{array}\right).$$

In Figure \ref{fig_PAT}, I report similar quantities as in Figure \ref{fig_RE} to illustrate sorting, heterogeneity, and complementarity in the sample of inventors. Panel (a) shows some evidence of sorting towards the top, however the pattern is less concentrated than in the sample of economists (compare with panel (a) of Figure \ref{fig_RE}). Panel (b) of Figure \ref{fig_PAT} suggests the presence of complementarity, since the return to two high-type inventors working together is 2 standard deviations above the return to two low or middle types working together. This suggests gains from complementarity in patent production, with somewhat less sorting than in the case of economists.

In Table \ref{Tab_vardec_PAT}, I show variance decomposition results based on the nonlinear model, for $K=2,4,5,6$. Focusing on the results for $K=4,5,6$, which are broadly comparable, inventor heterogeneity explains approximately 30\% of output variance, and sorting explains between 6\% and 9\%. Nonlinearities explain a larger share than in the economists' sample, between 5\% and 8\% of the variance.

 \begin{table}[tbp]
 	\caption{Nonlinear production, patents and inventors\label{Tab_vardec_PAT}}
 	\begin{center}
 		\begin{tabular}{||l||cc|cc|cc|cc||}\hline\hline
 			& \multicolumn{2}{c}{$K=2$} & \multicolumn{2}{c}{$K=4$} &\multicolumn{2}{c}{$K=5$} &\multicolumn{2}{c||}{$K=6$} \\
 			& $n=1$ & $n=2$ & $n=1$ & $n=2$ & $n=1$ & $n=2$ & $n=1$ & $n=2$\\\hline\hline
 			Total variance & 1263.5&1231.7& 1263.5&1231.7& 1263.5&1231.7& 1263.5&1231.7\\
 			Heterogeneity & 237.2& 302.9&324.6 & 366.9&375.2 & 412.8 & 390.1& 384.4\\
 			Sorting & -& 46.6& -&80.1& -&72.7& -&109.9\\
 			Nonlinearities  & -& 42.2& -&60.9& -&96.7& -&70.2\\
 			Other factors &1026.4 & 840.0& 938.9 & 723.7 & 888.3 & 649.5& 873.4 & 667.1\\\hline\hline		
 		\end{tabular}
 	\end{center}
 	{\footnotesize \textit{Notes: Estimates of variance components for different team sizes $n$ in the nonlinear model with $K$ types. Descriptive statistics on the sample are given in panel (b) of Table \ref{Tab_desc_pat}.}}
 \end{table}
 
\begin{figure}[tbp]
	\caption{Optimal allocation, patents and inventors ($K=4$) \label{fig_PAT_alloc}}
	\begin{center}
		\begin{tabular}{c}
			\includegraphics[width=60mm, height=60mm]{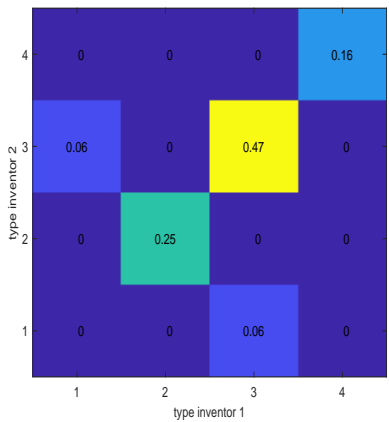} \\\end{tabular}
	\end{center}
	\par
	\textit{{\footnotesize Notes: Proportions of types for inventors producing together in a 2-inventor team, in the allocation that solves (\ref{max_surplus}). Descriptive statistics on the sample are given in panel (b) of Table \ref{Tab_desc_pat}.}}
\end{figure}

In addition, in Figure \ref{fig_PAT_alloc} I report the allocation that maximizes total output according to (\ref{max_surplus}). As in the case of economists, the allocation has perfect assortative matching for the top type but not for the bottom types. This optimal allocation differs quite substantially from the estimated allocation shown in panel (a) of Figure \ref{fig_PAT}.

\paragraph{Other specifications.}

\begin{figure}[tbp]
	\caption{Nonlinear model estimates and optimal allocation, patents and inventors, other specifications ($K=4$) \label{fig_PAT_spec}}
	\begin{center}
		\begin{tabular}{ccc}
			\multicolumn{3}{c}{1. Correlated RE}\\
			(1a) Sorting & (1b) Heterogeneity \& complementarity & (1c) Optimal allocation \\
			\includegraphics[width=40mm, height=40mm]{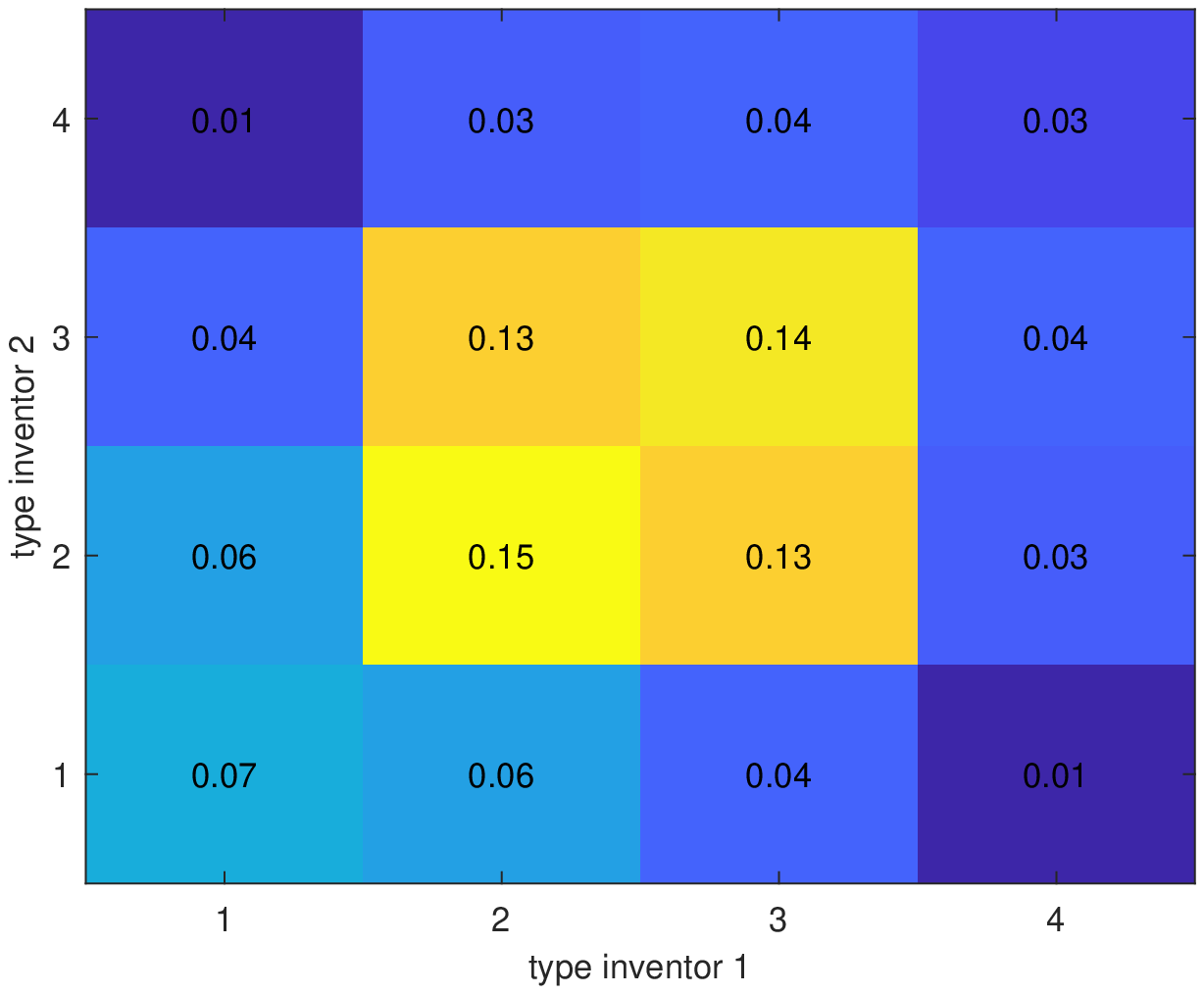}&\includegraphics[width=40mm, height=40mm]{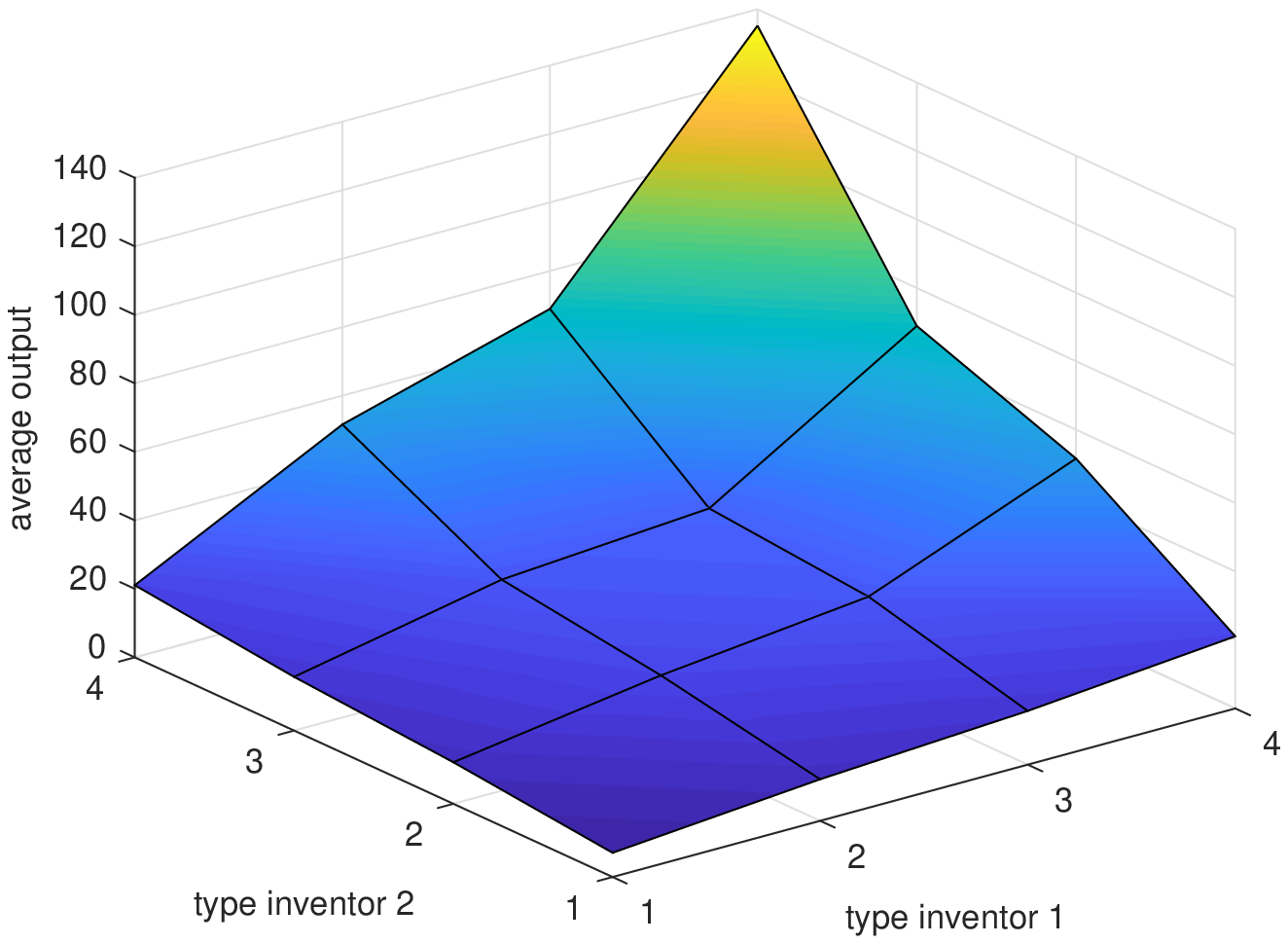}&\includegraphics[width=40mm, height=40mm]{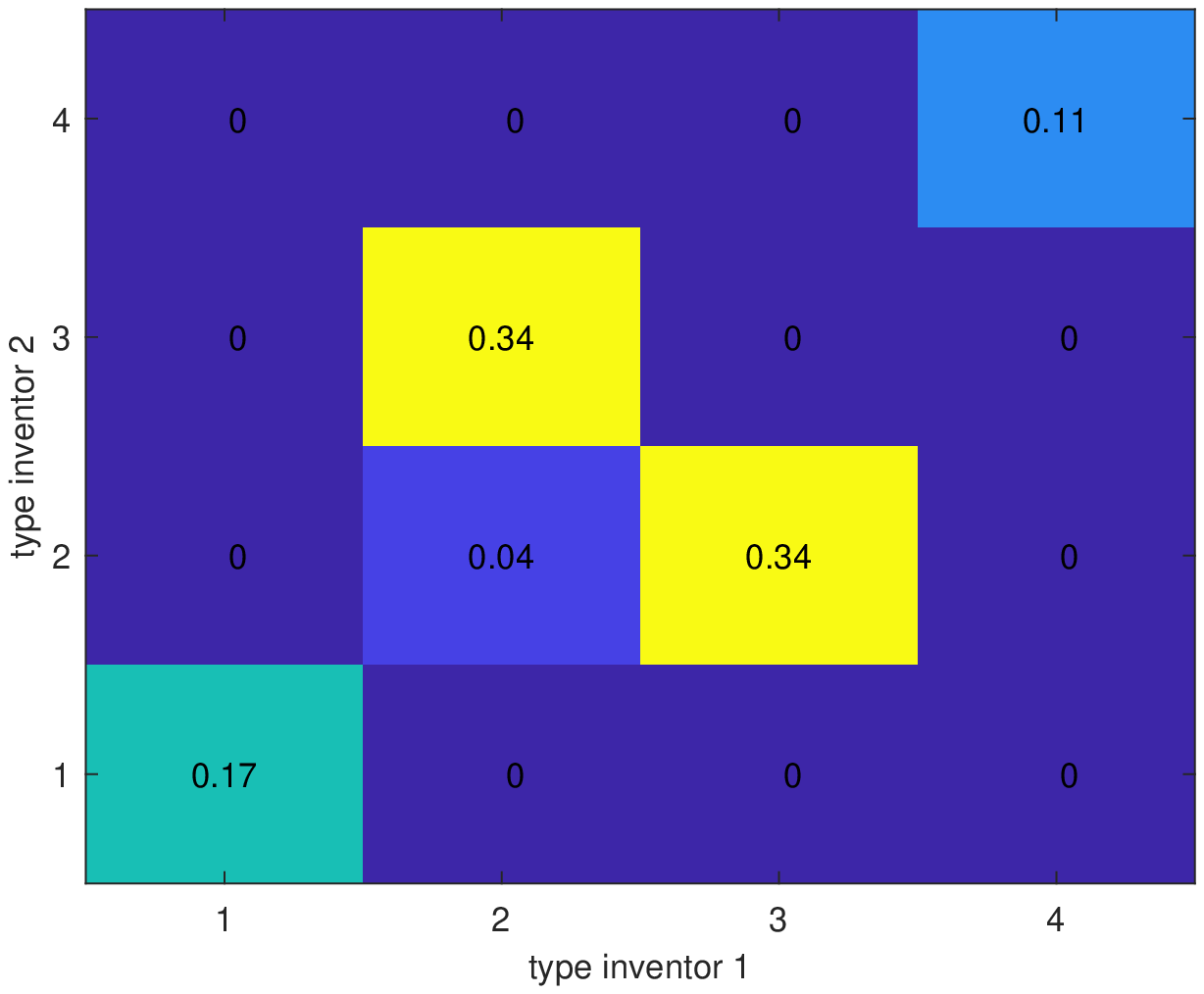} \\
			\multicolumn{3}{c}{2. Joint RE}\\
			(2a) Sorting & (2b) Heterogeneity \& complementarity & (2c) Optimal allocation \\
			\includegraphics[width=40mm, height=40mm]{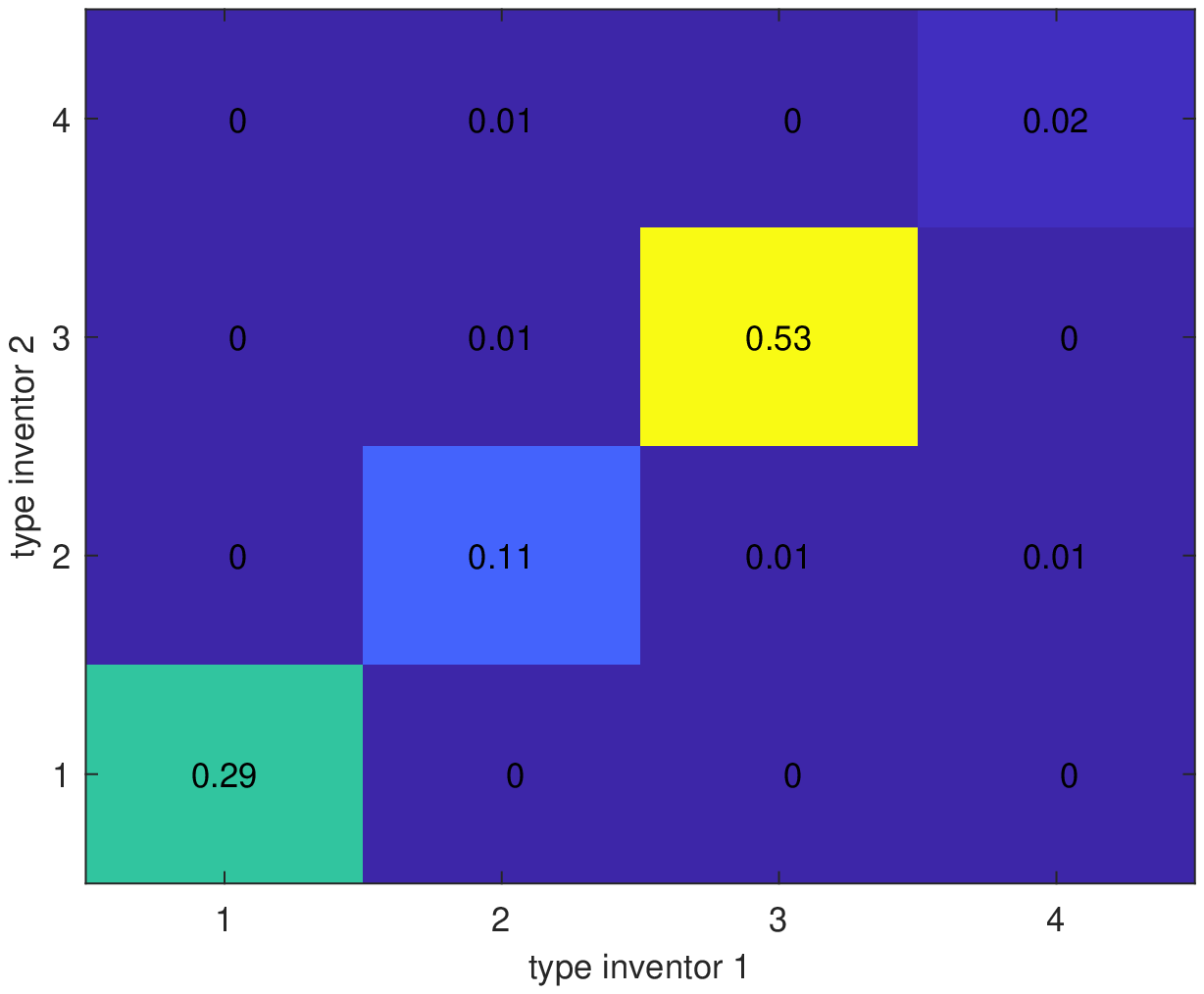}&\includegraphics[width=40mm, height=40mm]{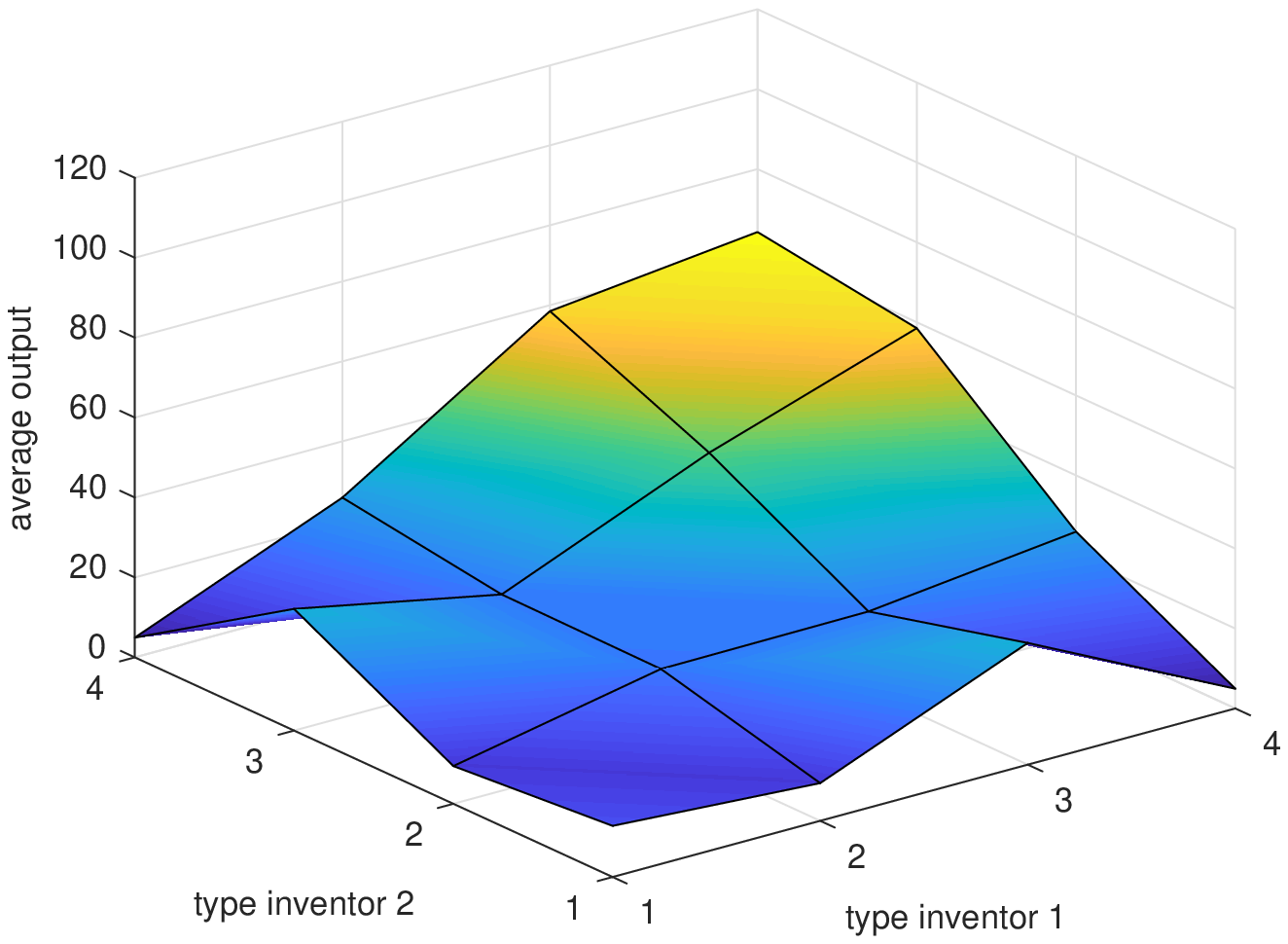}&\includegraphics[width=40mm, height=40mm]{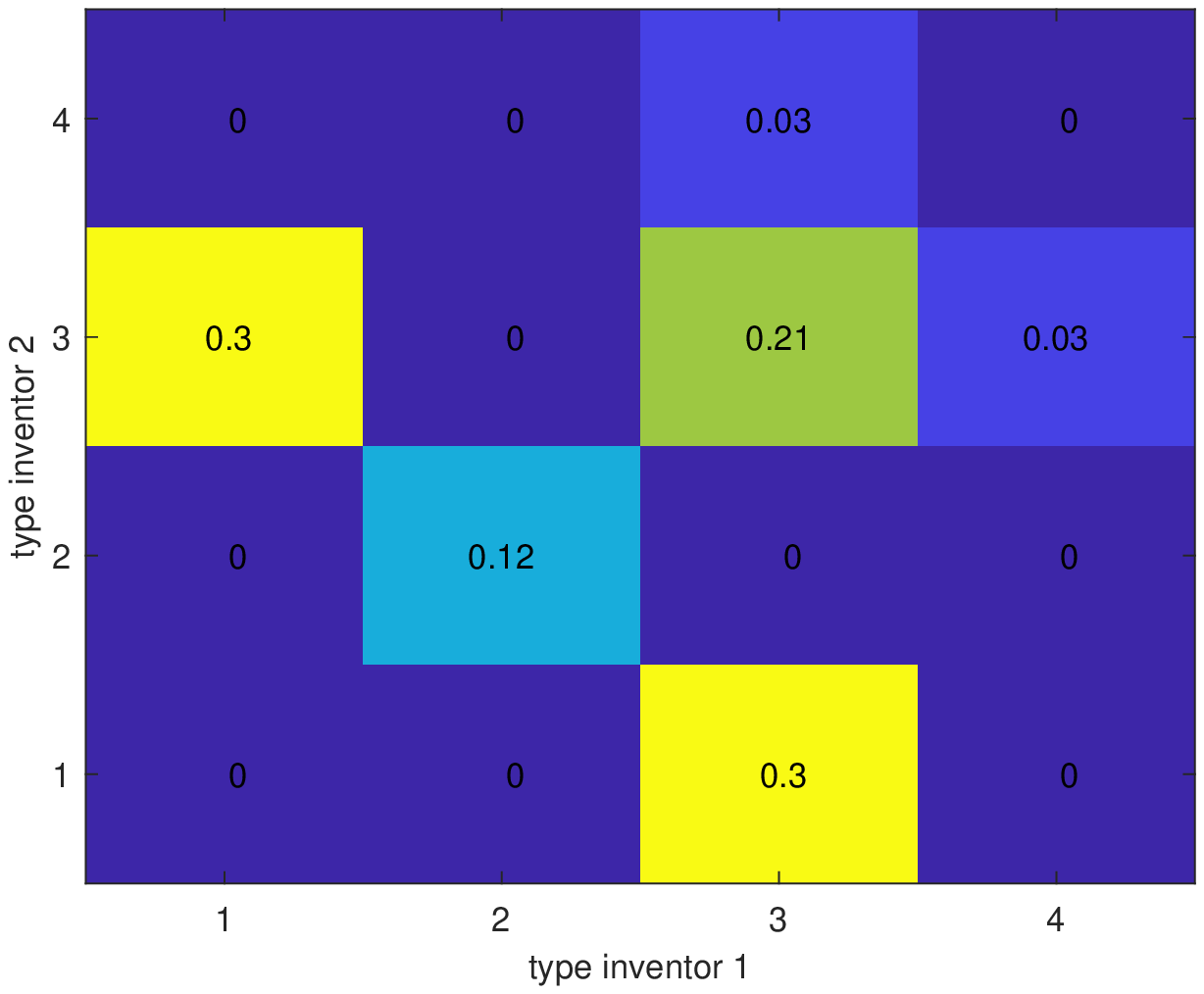}\\
			\multicolumn{3}{c}{3. 2-inventor only}\\
			(3a) Sorting & (3b) Heterogeneity \& complementarity & (3c) Optimal allocation \\
			\includegraphics[width=40mm, height=40mm]{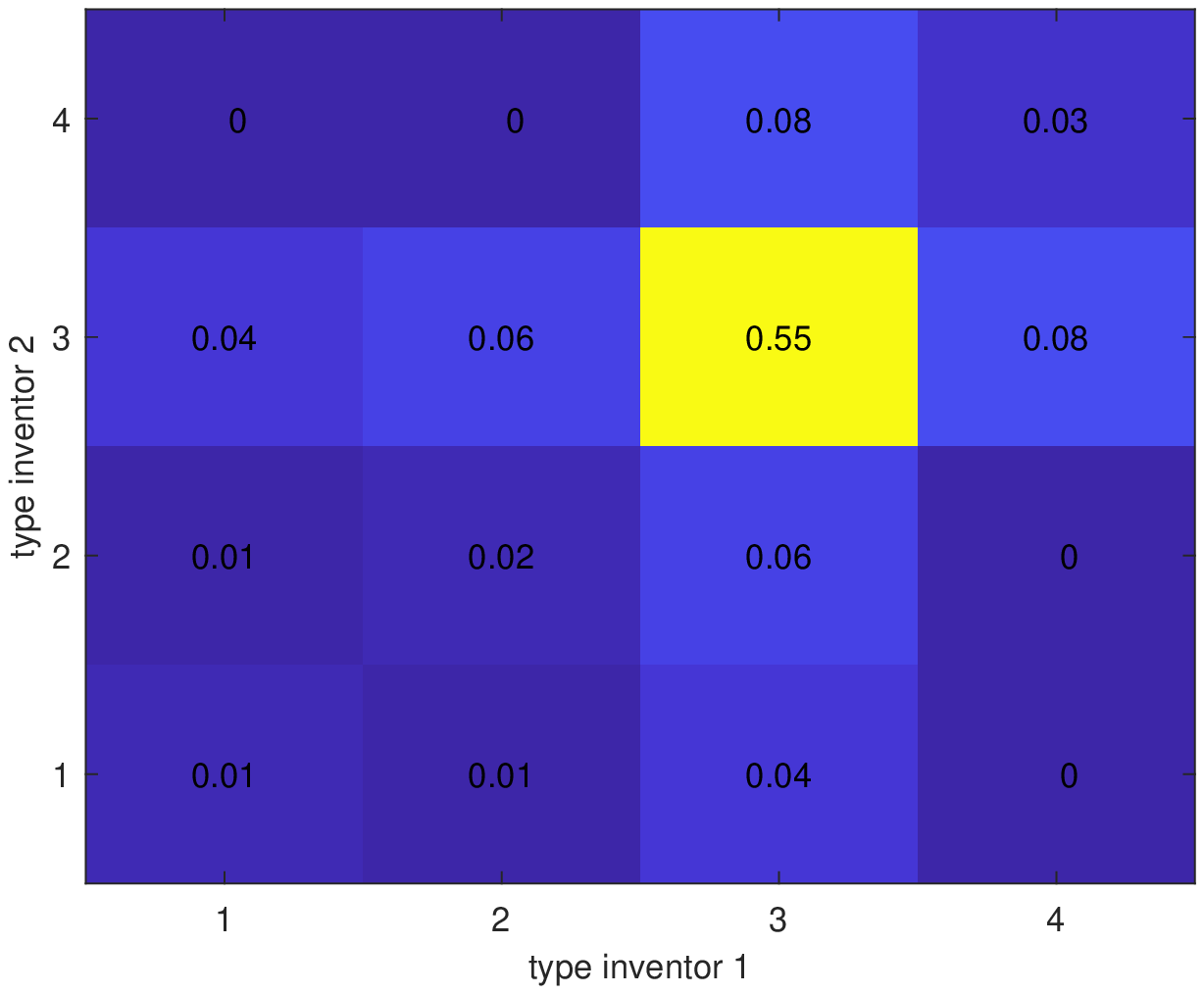}&\includegraphics[width=40mm, height=40mm]{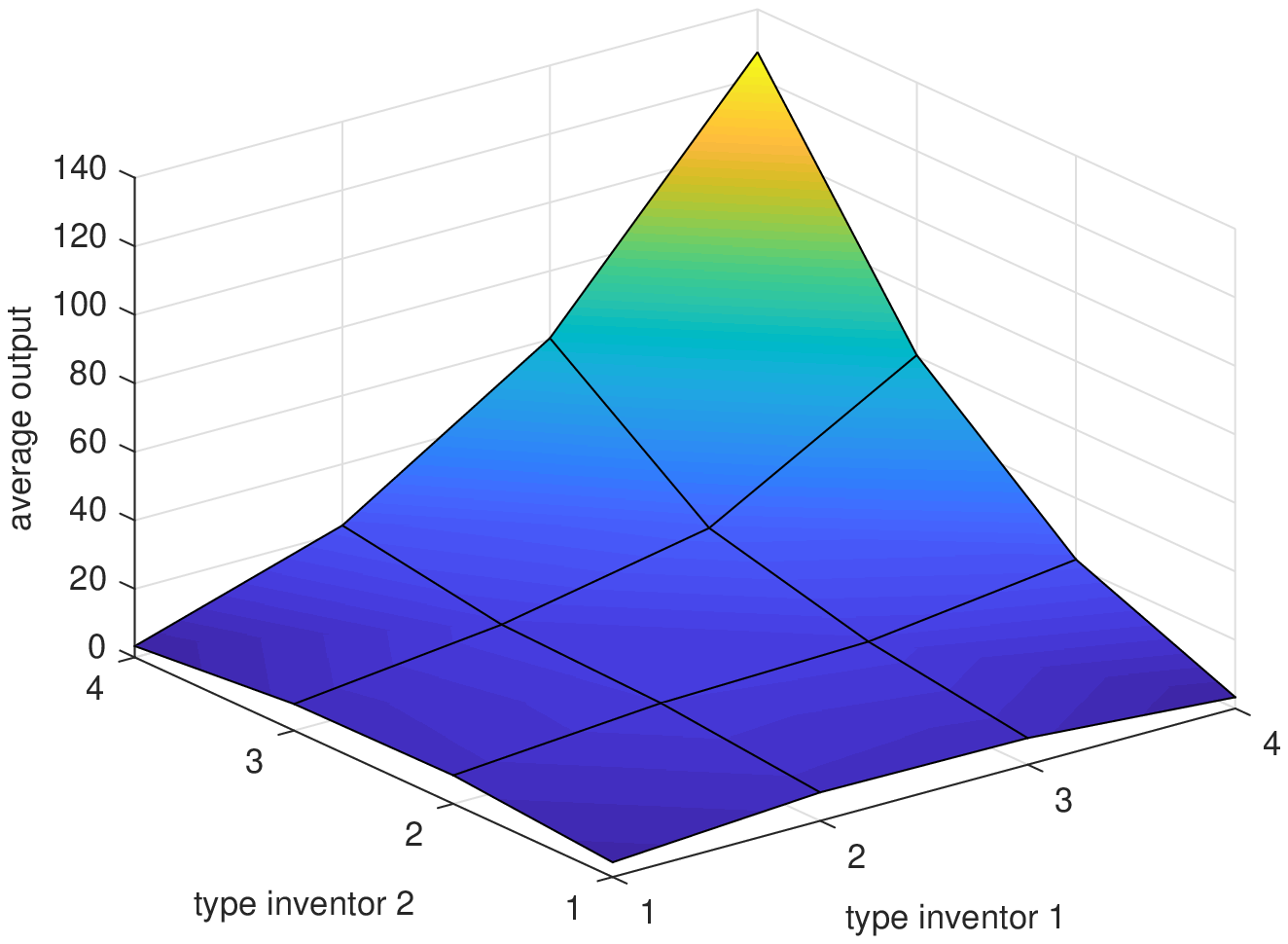}&\includegraphics[width=40mm, height=40mm]{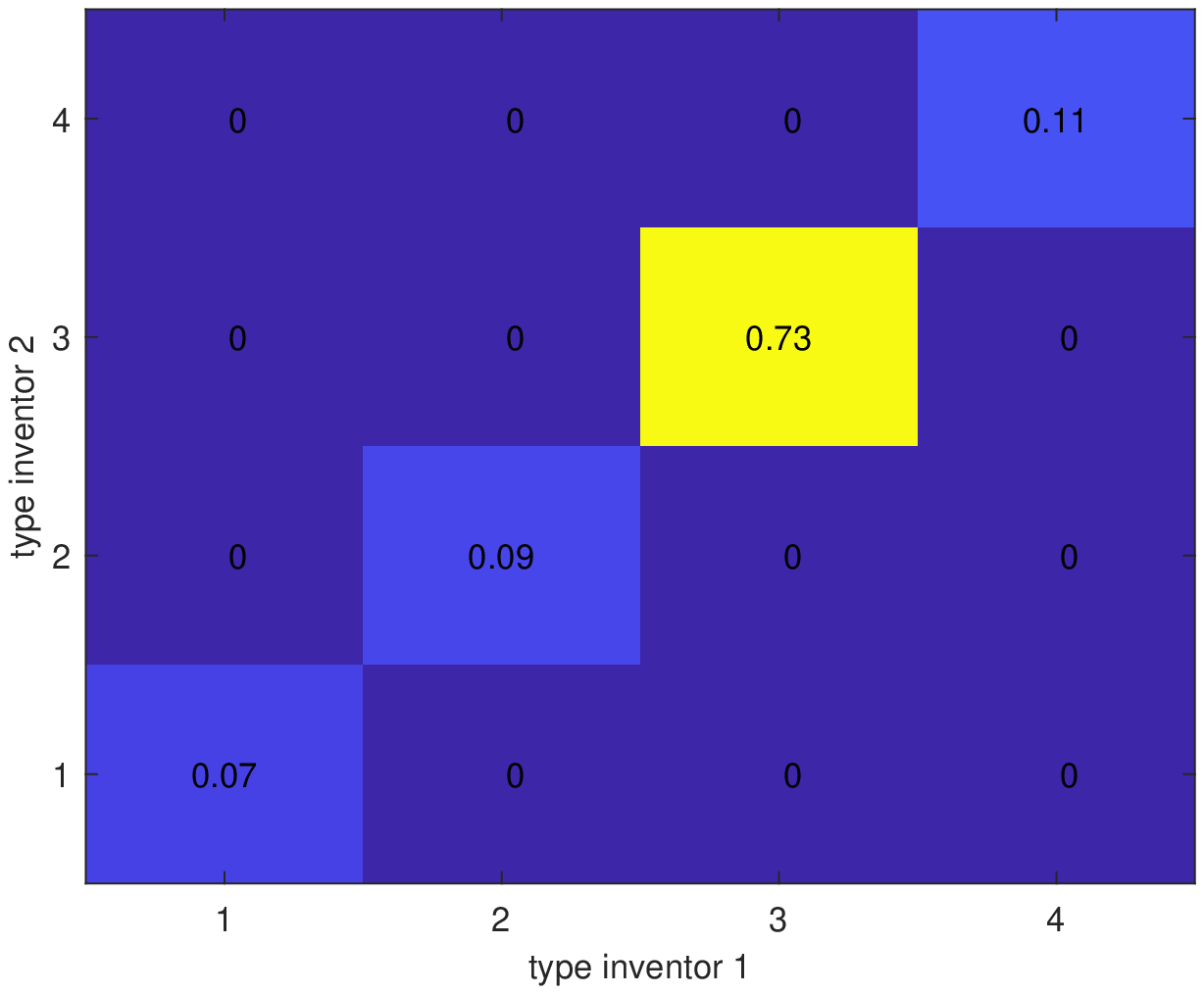} \\\end{tabular}
	\end{center}
	\par
	\textit{{\footnotesize Notes: Random-effects estimates of a finite mixture model with $K=4$ types. In panels (a) I show the proportions of types for inventors producing together in a 2-inventor team. In panels (b) I show average output (i.e., average truncation-adjusted forward citations) for different combinations of the types. In panels (c) I show the proportions of types for inventors producing together in a 2-inventor team, in the allocation that solves (\ref{max_surplus}). ``Correlated RE'' comes from a model where the types follow a multinomial logit distribution given the numbers of 1-inventor and 2-inventor collaborations at the extensive and intensive margins. ``Joint RE'' comes from a model where collaborations follow independent type-specific Poisson distributions. ``2-inventor only'' only uses information from 2-inventor teams. Descriptive statistics on the sample are given in panel (b) of Table \ref{Tab_desc_pat}. For these results I net out multiplicative year and inventor-age effects from the output.}}
\end{figure}

In Figure \ref{fig_PAT_spec}, I show estimates based on three alternative specifications: a correlated random-effects estimator, a joint random-effects estimator where 1-inventor and 2-inventor collaborations follow independent Poisson distributions with type-specific parameters, and an estimator that is solely based on 2-inventor collaborations. In these specifications I net out multiplicative inventor-age effects from the output, in addition to year effects. I show the  variance decomposition results in Appendix Table \ref{Tab_nonlin_PAT_spec}. As in the case of economists, I find that the correlated RE estimates do not substantially differ from the baseline. The optimal allocation in panel (1c) of Figure \ref{fig_PAT_spec} does differ somewhat from the baseline (compare with Figure \ref{fig_PAT_alloc}), yet both allocations exhibit perfect assortative matching at the top. 

In addition, the estimates based on 2-inventor teams are also broadly similar to the ones based on both 1-inventor and 2-inventor teams in this sample. One difference is that the group shares are more unbalanced when using only 2-inventor collaborations. Another difference is that the implied optimal allocation in panel (3c) of Figure \ref{fig_PAT_spec} exhibits perfect assortative matching along the entire distribution, and not only at the top.

\begin{figure}[tbp]
	\caption{Heterogeneity, sorting and complementarity out of sample (2000 to 2005, $K=4$) \label{fig_PAT_OOS}}
	\begin{center}
		\begin{tabular}{cc}
			(a) Sorting & (b) Heterogeneity \& complementarity \\
			\includegraphics[width=60mm, height=60mm]{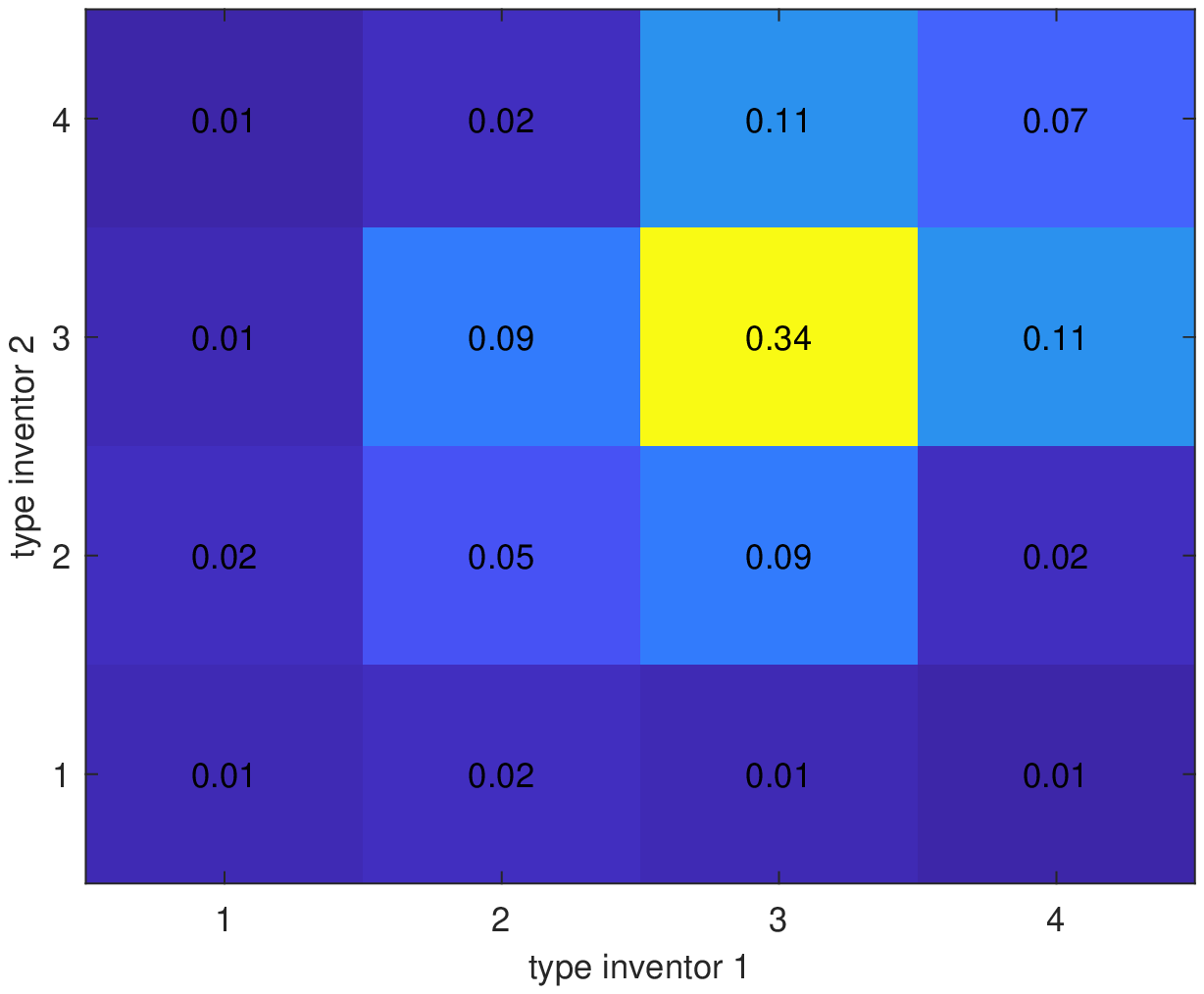}&\includegraphics[width=60mm, height=60mm]{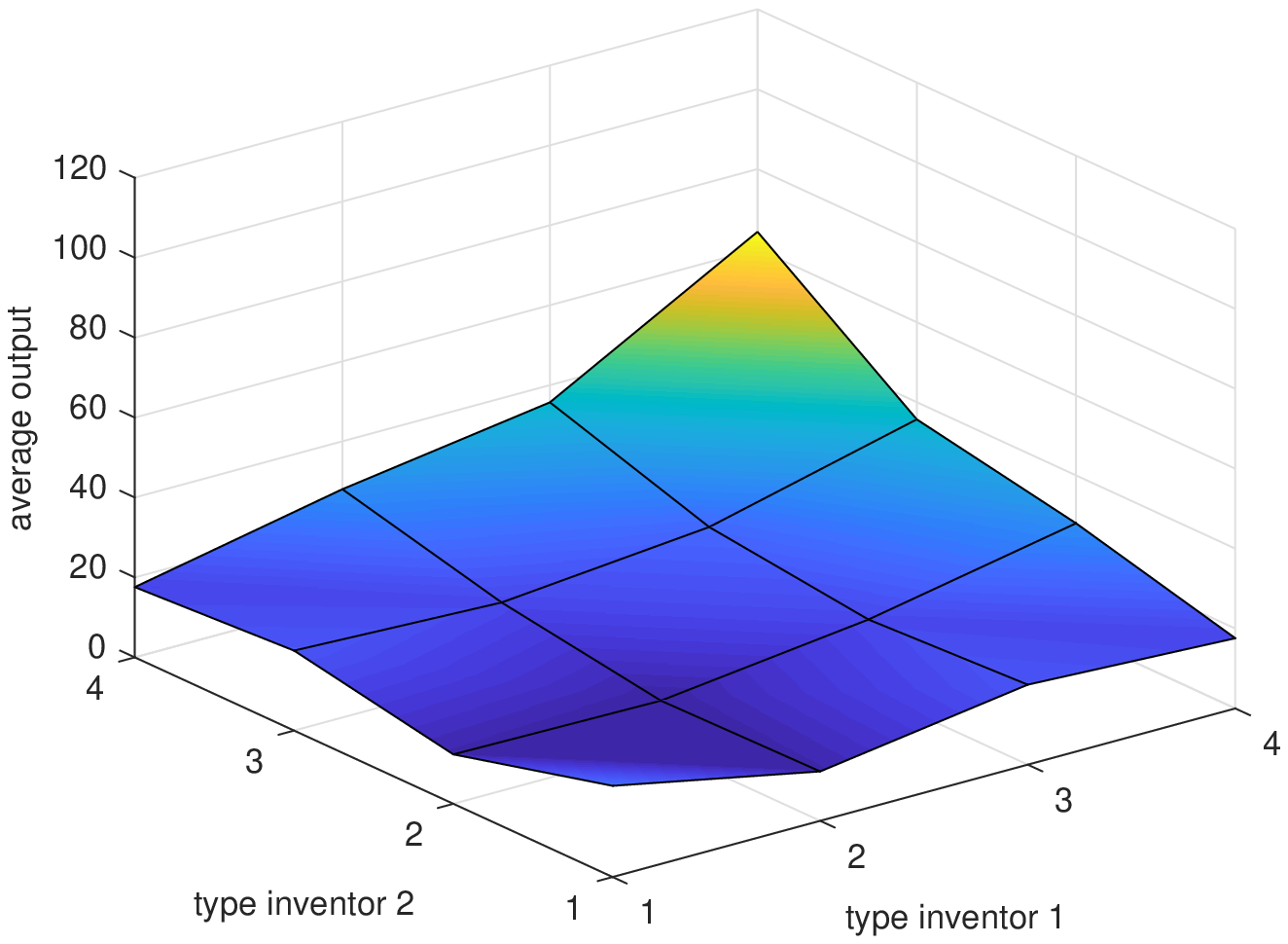}. \\\end{tabular}
	\end{center}
	\par
	\textit{{\footnotesize Notes: The random-effects model with $K=4$ inventor types is estimated on the 1995-1999 period. In panel (a) I show the proportions of types, and in panel (b) I show average output (i.e., average truncation-adjusted forward citations) for different combinations of the types, for inventors producing together in 2-worker teams between 2000 and 2005.}}
\end{figure}

However, the joint RE estimates show important differences compared to the baseline. Note that, as in the corresponding specification in the economists' sample, here inventor types do not only reflect heterogeneity in the quality of innovation through patent citations, but also in the quantity of patents produced by an inventor. In 2-inventor teams, heterogeneity, sorting and nonlinearity account for 9\%, 8\%, and less than 1\% of the variance of output, respectively, compared to 30\%, 7\%, and 5\% in the baseline (see Appendix Table \ref{Tab_nonlin_PAT_spec}). While the estimates in panels (2a) and (2b) of Figure \ref{fig_PAT_spec} show a high degree of sorting, they exhibit less evidence of complementarity compared to the other specifications. The implied optimal allocation in panel (2c) is also quite different in this case. This suggests that the modeling of inventor types and how types affect team formation is important to accurately assess the contributions of heterogeneity, sorting and complementarity in this sample.

\paragraph{How do inventor types perform out of sample?} In a last exercise, I study the predictive performance of inventor types out of sample. To do so, I use the parameters of the baseline random-effects model with $K=4$ types, estimated on the 1995-1999 period. Using the estimated variational posterior type probabilities, I then calculate average patent output produced by inventors of particular types between 2000 and 2005, and the type proportions in those collaborations. Figure \ref{fig_PAT_OOS} shows that sorting patterns and productivity differences between types persist out of sample, and that the higher returns to collaborations between high inventor types relative to other type combinations persist as well. However, the out-of-sample estimates show less inventor heterogeneity than in the baseline (compare with panel (b) of Figure \ref{fig_PAT}). There may be several explanations for this: posterior type probabilities are only based on five years of observations, truncation issues with patent citations may make the outcome less informative towards the end of the sample (Akcigit \textit{et al.}, 2016), and over a 10-year period inventor quality may change.

\section{Conclusion\label{Conclu_sec}}

In this chapter I have outlined a measurement framework to assess the contributions of individuals to team output. While an additive  production specification leads to a tractable estimator, it rules out complementarity that appears to be a feature of the samples of economists and inventors that I study. In the applications, a natural next step will be to relate the latent types to characteristics of authors and inventors, such as their national origin or education background. 

The discrete-type approach that I have implemented is promising, yet it needs to be studied more. In particular, it will be important to provide formal consistency arguments and to derive asymptotic distributions for variational estimators in this setting. Spectral clustering methods (e.g., Lei and Rinaldo, 2015) could be possible alternatives to the nonlinear random-effects methods that I have described. It will also be important to augment the framework to account for the productive effects of team-specific factors, both observed and latent. 

The literature on matching and sorting has made substantial progress in one-to-one settings (Chiappori and Salani\'e, 2016, Chade \textit{et al.}, 2017). However, less is known about many-to-one and many-to-many sorting environments. Eeckhout and Kircher (2018) model sorting in large firms, while abstracting from complementarities between workers. Chade and Eeckhout (2018) propose a model of information aggregation in teams that has tight implications for sorting.\footnote{A related strand of the literature proposes models of co-authorship networks, see among others Goyal \textit{et al.} (2004) and Gans and Murray (2014), and the co-author model in Jackson and Wolinsky (1996).} Moreover, it is well-known that models with general complementarities may not have equilibria (Kelso and Crawford, 1982). Addressing these challenges, and combining the quantitative framework of team production that I have introduced, with economic models of team formation and effort allocation (such as those recently proposed by Hsieh \textit{et al.}, 2018, and Anderson and Richards-Shubik, 2019), is an interesting avenue for future work.

\clearpage

\appendix

\renewcommand{\theequation}{\thesection \arabic{equation}}

\renewcommand{\thelemma}{\thesection \arabic{lemma}}

\renewcommand{\theproposition}{\thesection \arabic{proposition}}

\renewcommand{\thecorollary}{\thesection \arabic{corollary}}

\renewcommand{\thetheorem}{\thesection \arabic{theorem}}

\renewcommand{\theassumption}{\thesection \arabic{assumption}}

\renewcommand{\thefigure}{\thesection \arabic{figure}}

\renewcommand{\thetable}{\thesection \arabic{table}}

\setcounter{equation}{0}
\setcounter{table}{0}
\setcounter{figure}{0}
\setcounter{assumption}{0}
\setcounter{proposition}{0}
\setcounter{lemma}{0}
\setcounter{corollary}{0}
\setcounter{theorem}{0}

\begin{center}
	{ {\LARGE APPENDIX} }
\end{center}

\small

\section{Monte Carlo simulations\label{AppA}}

To probe the accuracy of the variational estimator in finite sample, I run Monte Carlo simulations. I generate models based on two sets of collaborations, both drawn from the data on economic researchers who work either on their own or with one co-author (see Section \ref{Research_sec}). For the smaller sample I select articles that were published in 1999. Since I impose that all authors have at least 5 articles, this gives a small number of authors and teams (about 150 and 900, respectively). For the larger sample I select articles that were published between 1998 and 1999. This gives a sample of about 900 authors and 5000 articles.

\begin{table}[tbp]
	\caption{Monte Carlo simulation\label{Tab_MC_K4}}
	\begin{center}
		\footnotesize
		
		\begin{tabular}{||l||c|ccc||ccc||}\hline\hline
			&&\multicolumn{3}{c||}{Smaller sample}&\multicolumn{3}{c||}{Larger sample} \\
			& True & Mean &p2.5\% &p97.5\%  & Mean &p2.5\%&p97.5\% \\\hline\hline
			\multicolumn{8}{||c||}{(a) $K_0=2,K=2$}\\\hline\hline
			Mean type 1 &   0.00 &  0.00 &  -0.05  &  0.06&     0.00 &  -0.03 &   0.02\\
			Mean type 2 &  2.00  &  2.00&   1.92 &   2.07  &    2.00 &    1.97  &  2.03
			\\
			Var. type 1 &  0.50&    0.50&  0.43&   0.56&    0.50&    0.47&    0.53\\
			Var. type 2 &    0.50&    0.50&    0.42&  0.57  &  0.50&    0.47&    0.53\\
			Mean type (1,1) &   0.00  &  0.05&  -0.49&    0.74&   0.01&   -0.10&   0.12\\
			Mean type (1,2) &1.00&  1.01&  0.59&   1.42 &    1.01 &    0.92&    1.10\\
			Mean type (2,2) & 4.00 &  3.71&   0.78&    4.72&    4.00&    3.85&    4.16	\\
			Var. type (1,1) &0.50&    0.46   & 0.11&    0.93&	     0.50&    0.40 &   0.62\\	
			Var. type (1,2) & 0.50&    0.46&   0.16&    0.84&   0.50&    0.42&    0.59\\
			Var. type (2,2) & 0.50&   0.38  &  0.00  &  1.24 &    0.49&    0.35  &  0.69	\\
			Prop. type 1 &  0.60&   0.60&    0.53&    0.67 &   0.60 &   0.57   & 0.63	\\\hline\hline
			\multicolumn{8}{||c||}{(b) $K_0=4,K=4$}\\\hline\hline
			Mean type 1 &    -0.52&   -0.52&   -0.54&   -0.50&   -0.52&   -0.53&  -0.52\\
			Mean type 2& -0.47&   -0.42&   -0.56&   -0.22&   -0.47&   -0.52&   -0.41\\
			Mean type 3&  0.07&    0.21&   -0.09&    0.71&    0.07&   -0.03&    0.18\\
			Mean type 4& 1.62&    1.68&    1.39&    2.01  &    1.64&    1.53  &  1.75\\
			Var. type 1 &    0.01&    0.02&    0.01&    0.08&    0.01&    0.01&    0.01\\
			Var. type 2 &    0.48&    0.64&    0.38&    1.15&    0.48&    0.41&    0.55\\
			Var. type 3 &     1.74&    1.88&    1.42&    2.44&    1.74&    1.59&    1.89\\
			Var. type 4 &     2.16&    2.10&    1.65&    2.59&    2.15&    1.96&    2.31\\   
			Mean type (1,1) &       -0.53&   -0.03&  -1.48&  2.90&   -0.52& -0.60&  -0.34\\
			Mean type (1,2) &       -0.39&   -0.28&   -1.83&   1.58&  -0.38&   -0.73&    0.06\\
			Mean type (2,2) &     -0.53&   -0.01&  -1.73&  2.17&   -0.50&   -0.63&   -0.08\\
			Mean type (1,3) &      0.01&  0.26&   -2.13&   2.94&   0.02&   -0.43&    0.52\\
			Mean type (2,3) &      -0.11&    0.25&   -1.50&    2.25&  -0.07&   -0.42&    0.26\\
			Mean type (3,3) &      0.35&   0.83&   -1.28&    3.23&    0.42&   -0.13&    0.95\\
			Mean type (1,4) &     1.38&   1.07&   -0.68&    3.34&  1.37&    0.75&    1.95\\
			Mean type (2,4) &       0.66&   0.94&   -1.01&    3.01&    0.67&    0.12&   1.17\\
			Mean type (3,4) &     1.36&   1.46&   -0.59&    3.53&    1.35&  0.96&    1.71\\
			Mean type (4,4) &        2.35&    2.02&  -0.47&    4.19&    2.24&    1.71&    2.75\\
			Var. type (1,1) &   0.01&   0.58&   0.00&    2.99&    0.02&    0.00&    0.24\\
			Var. type (1,2) &      0.43&   0.55&    0.00 &   2.89&    0.47&    0.11&    1.18\\
			Var. type (2,2) &      0.01&    0.61&    0.00&    2.89&    0.13&    0.00&    0.98\\
			Var. type (1,3) &      1.76&   0.79 &   0.00&    4.03&    1.68&    0.89&    2.80\\
			Var. type (2,3) &      1.18&    0.83&    0.00&    3.42&    1.25&    0.76&    1.85\\
			Var. type (3,3) &      1.98&    1.06&    0.00&    4.84&    1.92&    1.02&    2.94\\
			Var. type (1,4) &      2.09&    1.02&    0.00&    4.31&    1.98&    1.13&    3.11\\
			Var. type (2,4) &      1.48&    0.88&    0.00&    3.27&    1.43&    0.70&    2.49\\
			Var. type (3,4) &      1.91&    1.02&   0.00&    3.60&    1.87&    1.24&    2.60\\
			Var. type (4,4) &       1.61&    0.94&    0.00&    3.80&    1.65&    0.77&    2.58\\
			Prop. type 1 &       0.15&    0.15&    0.10&    0.22&    0.14&    0.12&    0.17\\
			Prop. type 2 &       0.20&   0.26&    0.15&   0.42&    0.20&    0.17&    0.24\\
			Prop. type 3 &       0.36&   0.31&    0.13&    0.45&    0.36&    0.31&    0.41\\
			\hline\hline		
		\end{tabular}
	\end{center}
	{\footnotesize \textit{Notes: Estimates of the random-effects model with $K$ groups, in data generated according to that model with $K_0$ groups. 500 simulations. The smaller sample has 156 workers and 896 teams, the larger sample has 921 workers and 5447 teams. In panel (a) $K_0=K=2$, in panel (b) $K_0=K=4$.}}
\end{table}

I first consider a log-normal model with $K_0=2$ types, and use $K=2$ types in estimation. The true parameters, as well as means and 2.5\% and 97.5\% quantiles of the variational estimator across 500 simulations, are given in panel (a) of Table \ref{Tab_MC_K4}. The results suggest that both the parameters of sole-authored productions and the type proportions are well recovered in both samples. However, the parameters of 2-author teams are not as well recovered in the smaller sample. In particular, while the true mean output for two high-type workers is 4, the mean Monte Carlo estimate is 3.71. This suggests that the variational approximation, which only matters for 2-author collaborations, is imperfectly accurate in this case. In contrast, estimates in the larger sample are close to unbiased. For example, the mean output for two high-type workers is indistinguishable from the truth. In addition, Monte Carlo estimates are quite tightly concentrated around true values. This suggests that the variational approximation is accurate in the larger sample.

I next consider a log-normal model with $K_0=4$ types, and use $K=4$ types in estimation. In this case, I set the true parameter values to the parameters of the model estimated on the 1995-2000 sample; see the column labeled ``True'' in  panel (b) of Table \ref{Tab_MC_K4}. While this second simulation design is closer to the data, it is also more challenging for the variational estimator. Panel (b) of Table \ref{Tab_MC_K4} shows that, in the smaller sample, both the parameters of sole-authored productions and those of 2-author teams are biased and imprecise. In the larger sample, biases are quite small in general, and estimates are more precise. In the empirical analysis in Section \ref{Research_sec}, I focus on a much larger sample of articles produced between 1995 and 1999. Since I also impose that every author produces at least 5 articles, I expect the variational method to be accurate in this case. Developing methods to assess parameter uncertainty in the variational approach is an important avenue for future work.

\section{Additional tables and figures\label{AppB}}

{\small
\begin{table}[h!]
	\caption{Additive production, economic researchers, robustness\label{Tab_AKM_rob}}
	\begin{center}\small
		\begin{tabular}{||l||c|c|c||}\hline\hline
			& $n=1$ & $n=2$ & $n=3$\\\hline\hline
			\multicolumn{4}{||c||}{(a) Log-journal quality as dependent variable}\\\hline\hline
			Total variance &2.00&2.41&2.30\\
			Heterogeneity & 0.64&0.85&1.02\\
			\color{gray}{Heterogeneity (uncorrected)} & 	\color{gray}{0.80}&	\color{gray}{1.33}&	\color{gray}{1.76}\\
			Sorting & -&  0.59& 0.93\\
			\color{gray}{Sorting (uncorrected)} & 	\color{gray}{-}&	\color{gray}{0.38}& 	\color{gray}{0.49}\\
			Other factors &1.15&1.29&1.19\\
			\color{gray}{Other factors (uncorrected)} &	\color{gray}{1.20} &	\color{gray}{1.12}&	\color{gray}{1.20}\\	Team shift $\mu_n$ & 0.00& -0.20& -0.61\\\hline\hline		
			\multicolumn{4}{||c||}{(b) Ranks as dependent variables}\\\hline\hline
			Total variance &0.089&0.090& 0.086\\
			Heterogeneity & 0.034&0.019&0.013\\
			\color{gray}{Heterogeneity (uncorrected)} & 	\color{gray}{0.043}&	\color{gray}{0.029}&	\color{gray}{0.023}\\
			Sorting & -&  0.015& 0.016\\
			\color{gray}{Sorting (uncorrected)} & 	\color{gray}{-}&	\color{gray}{0.012}& 	\color{gray}{0.012}\\
			Other factors &0.055&0.055& 0.054\\
			\color{gray}{Other factors (uncorrected)} &	\color{gray}{0.046} & 	\color{gray}{0.049}& 	\color{gray}{0.052}\\	Team scale $\lambda_n$ & 1.00& 0.55& 0.38\\\hline\hline
			\multicolumn{4}{||c||}{(c) $\geq 2$ articles per author}\\\hline\hline
			Total variance & 65.05&127.78& 155.92\\
			Heterogeneity & 21.76& 23.53&29.29\\
			\color{gray}{Heterogeneity (uncorrected)} & 	\color{gray}{36.63}&	\color{gray}{69.89}&	\color{gray}{78.47}\\
			Sorting & -&  24.04& 35.41\\
			\color{gray}{Sorting (uncorrected)} & 	\color{gray}{-}&	\color{gray}{2.81}& 	\color{gray}{8.65}\\
			Other factors &41.11&73.13& 65.96\\
			\color{gray}{Other factors (uncorrected)} &	\color{gray}{28.42} & 	\color{gray}{49.85}& 	\color{gray}{63.53}\\	Team scale $\lambda_n$ & 1.00& 0.67& 0.49\\\hline\hline	
			\multicolumn{4}{||c||}{(d) $\geq 1$ articles per author}\\\hline\hline
			Total variance &53.68&115.17& 145.22\\
			Heterogeneity & 6.36&0.40&29.92\\
			\color{gray}{Heterogeneity (uncorrected)} & 	\color{gray}{32.72}&	\color{gray}{95.30}&	\color{gray}{139.92}\\
			Sorting & -&  30.58& 31.67\\
			\color{gray}{Sorting (uncorrected)} &	\color{gray}{ -}&	\color{gray}{-18.58}& 	\color{gray}{-34.85}\\
			Other factors &47.89&34.63& 70.54\\
			\color{gray}{Other factors (uncorrected)} &	\color{gray}{20.96} & 	\color{gray}{37.97}& 	\color{gray}{63.53}\\	Team scale $\lambda_n$ & 1.00& 0.68& 0.50\\\hline\hline	
		\end{tabular}
	\end{center}
	{\footnotesize \textit{Notes: Estimates of variance components for different team sizes $n$. In panels (a) and (b) I use log-journal quality and year-specific ranks of journal quality, respectively, as dependent variables in the baseline sample. In panels (c) and (d) I enlarge the sample to include at most 2 and 1 articles per author, compared to at most 5 in the baseline sample, using journal quality as the dependent variable. }}
\end{table}
}

\begin{table}[tbp]
	\caption{Nonlinear production, economic researchers, other specifications ($K=4$)\label{Tab_nonlin_RE_spec}}
	\begin{center}
		\begin{tabular}{||l||cc|cc|c||}\hline\hline
			& \multicolumn{2}{c}{Correlated RE} & \multicolumn{2}{c}{Joint RE} &\multicolumn{1}{c||}{2-author only} \\
			& $n=1$ & $n=2$ & $n=1$ & $n=2$ & $n=2$\\\hline\hline
			Total variance & 97.81&162.64& 97.81&162.64 & 162.64\\
			Heterogeneity &36.49& 46.71&32.86 & 35.03 & 51.97\\
			Sorting & -& 17.04& -&30.90 &7.84 \\
			Nonlinearities  & -&  4.29& -&1.15&2.82\\
			Other factors &61.32 & 94.61&64.96& 95.56& 100.01\\\hline\hline		
		\end{tabular}
	\end{center}
	{\footnotesize \textit{Notes: Estimates of variance components for different team sizes $n$ in the nonlinear model with $K$ types, for three specifications. ``Correlated RE'' comes from a model where the types follow a multinomial logit distribution given the numbers of 1-author and 2-author collaborations at the extensive and intensive margins. ``Joint RE'' comes from a model where collaborations follow independent type-specific Poisson distributions. ``2-author only'' only uses information from 2-author teams. Descriptive statistics on the sample are given in panel (b) of Table \ref{Tab_desc_1}.}}
\end{table}

\begin{table}[tbp]
	\caption{Additive production, patents and inventors, robustness\label{Tab_AKM_PAT_rob}}
	\begin{center}\small
		\begin{tabular}{||l||c|c|c||}\hline\hline
			& $n=1$ & $n=2$ & $n=3$\\\hline\hline
			\multicolumn{4}{||c||}{(a) $\geq 2$ patents per inventor}\\\hline\hline
			Total variance & 1443.5&1277.1&1597.1\\
			Heterogeneity &591.5& 699.7&886.7\\
			\color{gray}{Heterogeneity (uncorrected)} & 	\color{gray}{958.9}&	\color{gray}{1368.6}&	\color{gray}{2029.5}\\
			Sorting & -& -97.4& -179.5\\
			\color{gray}{Sorting (uncorrected)} & 	\color{gray}{-}&	\color{gray}{-515.3}& 	\color{gray}{-1000.4}\\
			Other factors &872.6&596.6& 756.0\\
			\color{gray}{Other factors (uncorrected)} &	\color{gray}{484.6} & 	\color{gray}{428.4}& 	\color{gray}{497.2}\\	Team scale $\lambda_n$ & 1.00& 0.54& 0.40\\\hline\hline	
			\multicolumn{4}{||c||}{(b) $\geq 1$ patents per inventor}\\\hline\hline
			Total variance & 1446.9&1265.8& 1546.5\\
			Heterogeneity &  538.7&1079.3&698.1\\
			\color{gray}{Heterogeneity (uncorrected)} & 	\color{gray}{1098.6}&	\color{gray}{1989.1}&	\color{gray}{2241.2}\\
			Sorting & -&  -479.8& -72.9\\
			\color{gray}{Sorting (uncorrected)} &	\color{gray}{ -}&	\color{gray}{-1021.9}& 	\color{gray}{-1135.2}\\
			Other factors &918.1&637.8& 854.0\\
			\color{gray}{Other factors (uncorrected)} &	\color{gray}{348.3} & 	\color{gray}{311.8}& 	\color{gray}{381.4}\\	Team scale $\lambda_n$ & 1.00& 0.54& 0.39\\\hline\hline	
		\end{tabular}
	\end{center}
	{\footnotesize \textit{Notes: Estimates of variance components for different team sizes $n$. In panels (a) and (b) I enlarge the sample to include at most 2 and 1 patents per inventor, compared to at most 5 in the baseline sample. For these results I net out multiplicative year and inventor-age effects from the output.}}
\end{table}

 \begin{figure}[h!]
 	\caption{Preliminary evidence based on type proxy, patents and inventors \label{fig_prelim_pat}}
 	\begin{center}
 		\begin{tabular}{cc}
 			(a) Sorting & (b) Heterogeneity \& complementarity \\
 			\includegraphics[width=60mm, height=60mm]{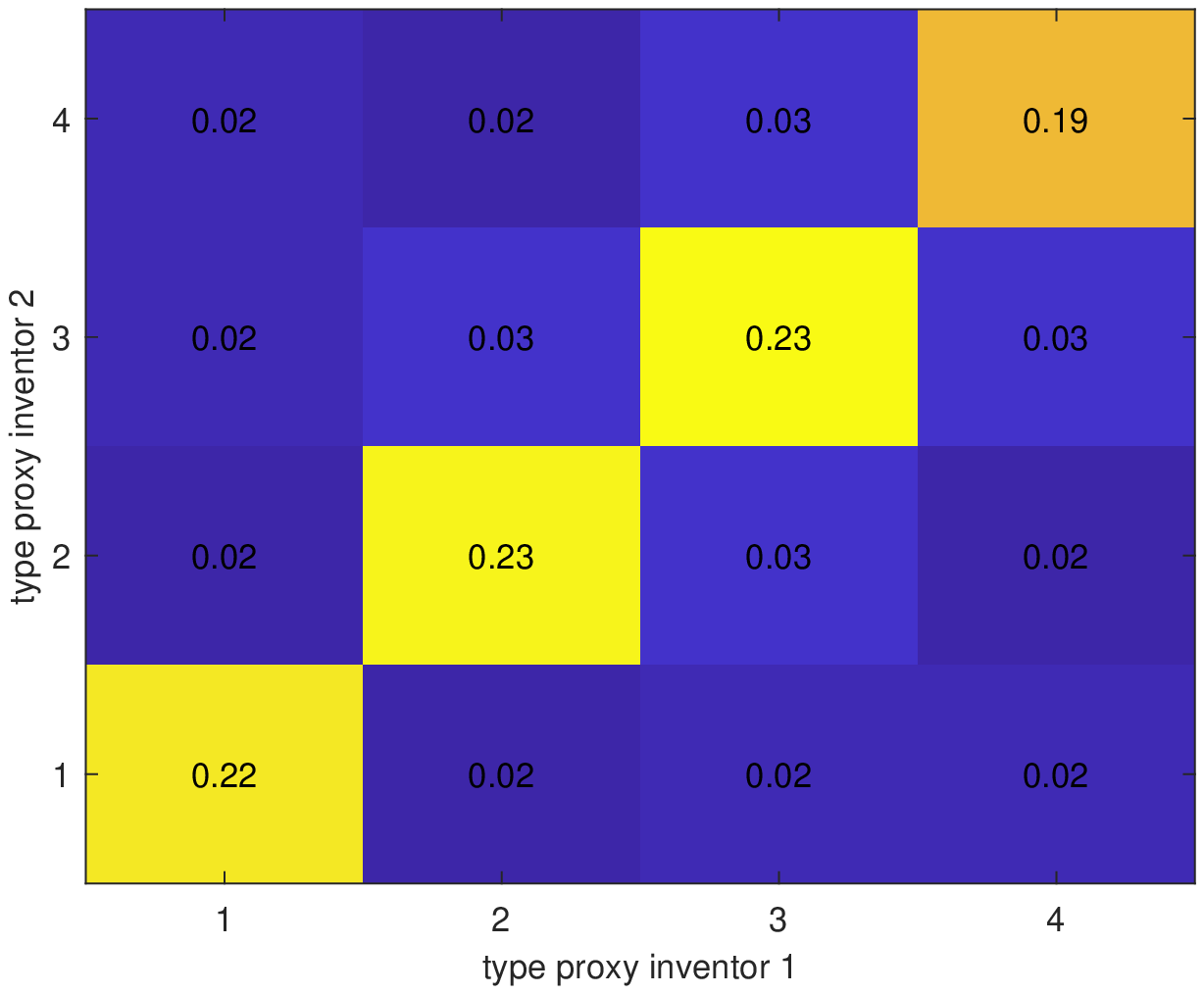}&\includegraphics[width=60mm, height=60mm]{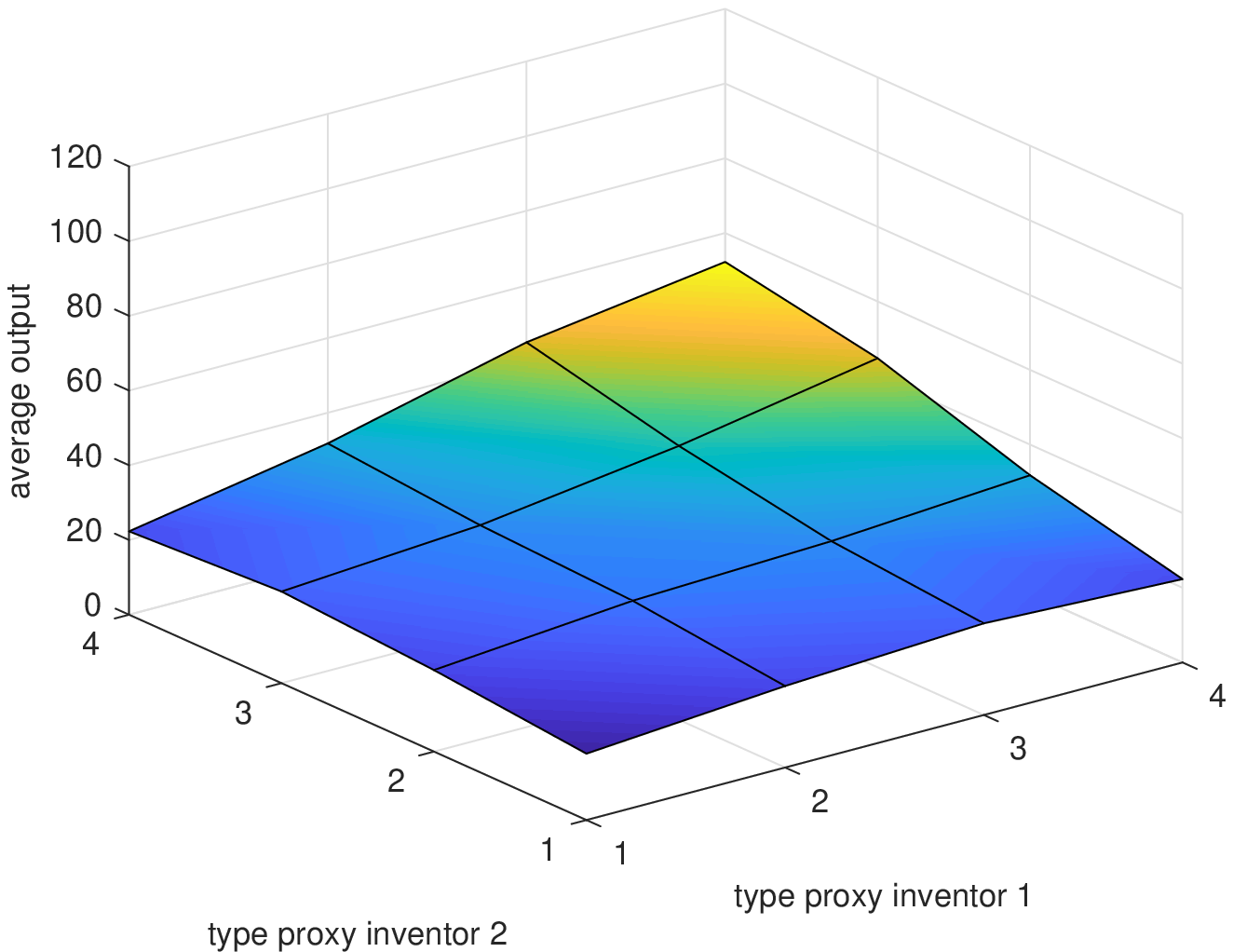}. \\\end{tabular}
 	\end{center}
 	\par
 	\textit{{\footnotesize Notes: Subsample of inventors who produce at least 5 patents on their own. I compute type proxies as quartiles of average sole-authored output. In panel (a) I show the proportions of type proxies for inventors producing together in a 2-inventor team. In panel (b) I show average output (i.e., average truncation-adjusted forward citations) for different combinations of the type proxies.}}
 \end{figure}

\begin{table}[tbp]
	\caption{Nonlinear production, patents and inventors, other specifications ($K=4$)\label{Tab_nonlin_PAT_spec}}
	\begin{center}
		\begin{tabular}{||l||cc|cc|c||}\hline\hline
			& \multicolumn{2}{c}{Correlated RE} & \multicolumn{2}{c}{Joint RE} &\multicolumn{1}{c||}{2-inventor only} \\
			& $n=1$ & $n=2$ & $n=1$ & $n=2$ & $n=2$\\\hline\hline
			Total variance &1270.4&1202.0& 1270.4&1202.0 & 1202.0\\
			Heterogeneity &  355.9&394.8&197.0 & 103.0 &401.3\\
			Sorting & -& 79.7& -&92.5& 86.2\\
			Nonlinearities  & -&  52.9& -&6.9& 47.1\\
			Other factors &914.5 & 674.5&1073.4& 999.6 & 667.3\\\hline\hline		
		\end{tabular}
	\end{center}
	{\footnotesize \textit{Notes: Estimates of variance components for different team sizes $n$ in the nonlinear model with $K$ types, for three specifications. ``Correlated RE'' comes from a model where the types follow a multinomial logit distribution given the numbers of 1-inventor and 2-inventor collaborations at the extensive and intensive margins. ``Joint RE'' comes from a model where collaborations follow independent type-specific Poisson distributions. ``2-inventor only'' only uses information from 2-inventor teams. Descriptive statistics on the sample are given in panel (b) of Table \ref{Tab_desc_pat}. For these results I net out multiplicative year and inventor-age effects from the output.}}
\end{table}

\end{document}